\documentclass[11pt,twoside]{article}
\usepackage{fancyhdr}
\PassOptionsToPackage{hyphens}{url}\usepackage[bookmarks=false]{hyperref}
\usepackage{amsfonts,epsfig,graphicx}
\usepackage{afterpage}
\usepackage{amsmath,amssymb,amsthm} 
\usepackage{fullpage}
\usepackage[T1]{fontenc} 
\usepackage{epsf} 
\usepackage{graphics} 
\usepackage{amsfonts,amsmath}
\usepackage{natbib} 
\usepackage{psfrag,xspace}
\usepackage{color,etoolbox}

\usepackage[usenames,dvipsnames]{xcolor}
\hypersetup{colorlinks,
citecolor=RoyalBlue,
urlcolor=RoyalBlue,
linkcolor=RoyalBlue}

\theoremstyle{plain}
\newtheorem{theorem}{Theorem}
\newtheorem{lemma}{Lemma}
\newtheorem{proposition}{Proposition}
\newtheorem{corollary}{Corollary}

\theoremstyle{definition}

\newtheorem{game}{Game}
\theoremstyle{remark}

\usepackage{times}
\usepackage{subcaption}
\usepackage{makecell}
\usepackage{multirow}
\usepackage{outlines}
\usepackage{cancel}
\usepackage{footnote}
\usepackage{pifont}
\usepackage{rotating}
\usepackage{dsfont}
\usepackage{enumitem}
\usepackage{accents}

\DeclareMathOperator*{\argmax}{argmax}

\newcommand{\email}[1]{\href{mailto:#1}{\textcolor{black}{\texttt{#1}}}}

\newcommand{\inparen}[1]{\left(#1\right)}
\newcommand{\incurly}[1]{\left\{#1\right\}}
\newcommand{\insquare}[1]{\left[#1\right]}

\newcommand{\floor}[1]{\left\lfloor#1\right\rfloor}

\newcommand{\absval}[1]{\left\lvert#1\right\rvert}
\newcommand{\norm}[1]{\left\lVert#1\right\rVert}

\newcommand{\indicator}[1]{\mathds{1}\inparen{#1}}

\newcommand{\N}{\mathbb{N}}

\newcommand{\R}{\mathbb{R}}

\newcommand{\calF}{\mathcal{F}}
\newcommand{\calG}{\mathcal{G}}
\newcommand{\calH}{\mathcal{H}}

\newcommand{\calN}{\mathcal{N}}

\newcommand{\calP}{\mathcal{P}}

\newcommand{\calX}{\mathcal{X}}
\newcommand{\calY}{\mathcal{Y}}

\newcommand{\bfe}{\mathbf{e}}

\newcommand{\bfp}{\mathbf{p}}

\newcommand{\bfy}{\mathbf{y}}

\newcommand{\Esymb}{\mathbb{E}}
\newcommand{\Psymb}{\mathbb{P}}

\DeclareMathOperator*{\ExpOp}{\Esymb}

\DeclareMathOperator*{\ProbOp}{\Psymb}
\renewcommand{\Pr}{\ProbOp}

\newcommand{\prob}[1]{\Pr\left({#1}\right)}

\newcommand{\ex}[1]{\ExpOp\left[{#1}\right]}
\newcommand{\Ex}[2]{\ExpOp_{{#1}}\left[{#2}\right]}

\newcommand{\frakF}{\mathfrak{F}}
\newcommand{\frakG}{\mathfrak{G}}

\newcommand{\sfA}{\mathsf{A}}

\newcommand{\sfp}{\mathsf{p}}
\newcommand{\sfq}{\mathsf{q}}

\newcommand{\sfs}{\mathsf{s}}
\newcommand{\sfw}{\mathsf{w}}

\newcommand{\cmark}{\text{\ding{51}}}%
\newcommand{\xmark}{\text{\ding{55}}}%

\let\svthefootnote\thefootnote
\newcommand\freefootnote[1]{%
  \let\thefootnote\relax%
  \footnotetext{#1}%
  \let\thefootnote\svthefootnote%
}

\title{%
\vspace{-2em}
{\bf Comparing Sequential Forecasters}\freefootnote{This manuscript is published in \emph{Operations Research}; see \url{https://doi.org/10.1287/opre.2021.0792}.}
\vspace{1em}
}
\author{%
    {\bf Yo Joong Choe}\thanks{Work done while this author was at Carnegie Mellon University.} \\
    Data Science Institute \\
    University of Chicago \\
    \email{yjchoe@uchicago.edu}
    \and
    {\bf Aaditya Ramdas} \\
    Department of Statistics and Data Science \\
    Machine Learning Department \\
    Carnegie Mellon University \\
    \email{aramdas@cmu.edu} 
    \vspace{1em}
}
\date{\normalsize\today}

\begin{document}











\maketitle

\vspace{-1em}
\begin{abstract}
    \normalsize 
    Consider two forecasters, each making a single prediction for a sequence of events over time. 
We ask a relatively basic question: how might we compare these forecasters, either online or post-hoc, while avoiding unverifiable assumptions on how the forecasts and outcomes were generated?
In this paper, we present a rigorous answer to this question by designing novel sequential inference procedures for estimating the time-varying difference in forecast scores. 
To do this, we employ confidence sequences (CS), which are sequences of confidence intervals that can be continuously monitored and are valid at arbitrary data-dependent stopping times (``anytime-valid'').
The widths of our CSs are adaptive to the underlying variance of the score differences.
Underlying their construction is a game-theoretic statistical framework, in which we further identify e-processes and p-processes for sequentially testing a weak null hypothesis --- whether one forecaster outperforms another \emph{on average} (rather than \emph{always}).
Our methods do not make distributional assumptions on the forecasts or outcomes; our main theorems apply to any bounded scores, and we later provide alternative methods for unbounded scores. 
We empirically validate our approaches by comparing real-world baseball and weather forecasters.

\end{abstract}


\tableofcontents
\clearpage

\renewcommand{\arraystretch}{1.15}

\section{Introduction}\label{sec:intro}

\begin{savenotes}
\begin{table}[t]
    \centering
    \begin{tabular}{c|ccccccc}
    \Xhline{1.1pt} 
    \bf Forecasters & \bf 1    & \bf 2    & \bf 3 & \bf 4 & \bf 5 & \bf 6    & \bf 7    \\ \hline
    FiveThirtyEight\footnote{Source: \url{https://projects.fivethirtyeight.com/2019-mlb-predictions/games/}.} 
    & 37.9\% & 41.0\% & 52.7\% & 58.7\% & 37.3\% & 40.5\% &  48.5\% \\
    Vegas-Odds.com\footnote{\smash{Source: \url{https://sports-statistics.com/sports-data/mlb-historical-odds-scores-datasets/}.}} 
    & 34.9\% & 37.7\% & 41.0\% & 50.7\% & 33.7\% & 37.4\% & 43.1\% \\ \hline
    Adjusted Win Percentage
    & 47.1\% & 47.4\% & 47.6\% & 47.4\% & 47.2\% & 47.0\% & 47.2\% \\
    K29 Defensive Forecast 
    & 50.0\% & 50.0\% & 50.9\% & 51.6\% & 50.7\% & 49.9\% & 49.1\% \\ 
    Constant Baseline
    & 50.0\% & 50.0\% & 50.0\% & 50.0\% & 50.0\% & 50.0\% & 50.0\% \\
    \hline
    Average Joe       
    & 40.0\% & 50.0\% & 60.0\% & 50.0\% & 30.0\% & 40.0\% & 50.0\% \\ 
    Nationals Fan 
    & 70.0\% & 70.0\% & 80.0\% & 70.0\% & 60.0\% & 60.0\% & 70.0\% \\ \hline
    \it Did the Nationals Win? & Yes & Yes & No & No & No & Yes & {Yes} \\ \hline
    \Xhline{1.1pt}
    \end{tabular}
    \caption{Probability forecasts (\%) on whether a baseball team (Washington Nationals) would win each game of the 2019 World Series. 
    The first two forecasters publish their forecasts online in the form of probabilities or betting odds.
    The next three forecasters are baselines computed using the 10-year win/loss records.
    The last two forecasters are imaginary (but not unrealistic) casual sports fans making their own forecasts using different heuristics. 
    All forecasts are made prior to the beginning of each game.
    See Section~\ref{sec:2019ws} for more details.}
    \label{tbl:2019ws}
\end{table}
\end{savenotes}

Forecasts of future outcomes are widely used across domains, including meteorology, economics, epidemiology, elections, and sports. 
Often, we encounter multiple forecasters making probability forecasts on a regularly occurring event, such as whether it will rain the next day and whether a sports team will win its next game.
Yet, despite the ubiquity of forecasts, it is not obvious how we can formally compare different forecasters on their predictive ability, particularly in a sequential setting where they each make a prediction on a sequence of outcomes (once for each outcome). 

As an illustrative example, consider the probability forecasts made on each game of the 2019 World Series by real-world (and fictitious) forecasters in Table~\ref{tbl:2019ws}.
It is not clear how we can effectively model the sequence of baseball game outcomes over time, and we also do not have full information on how each forecaster comes up with their predictions.
As we observe these forecasts and outcomes game-by-game, we may see one forecaster appearing to be better than the other, according to some scoring rule.
But how much of that difference can be attributed to chance or luck?
How much evidence do we have that one forecaster has been ``genuinely'' better than another, even after accounting for chance, and can we quantify this evidence without having to make assumptions about reality or how the forecasts are made?

In this work, we derive statistically rigorous procedures for \emph{sequentially} comparing forecasters via the powerful tool of \emph{confidence sequences (CS)}~\citep{darling1967confidence,lai1976confidence,howard2021timeuniform}.
CSs are sequences of confidence intervals (CIs) that provide time-uniform coverage guarantees, which allow valid sequential inference under continuous monitoring and at data-dependent stopping times.
The parameter of interest in this paper is the time-varying mean difference in forecast scores up to time $t$.
Most CSs we develop in our paper are also nonasymptotically valid, meaning that their coverage guarantee holds at every time point $t \geq 1$.

\begin{figure}[t]
    \centering
    \includegraphics[width=0.9\textwidth]{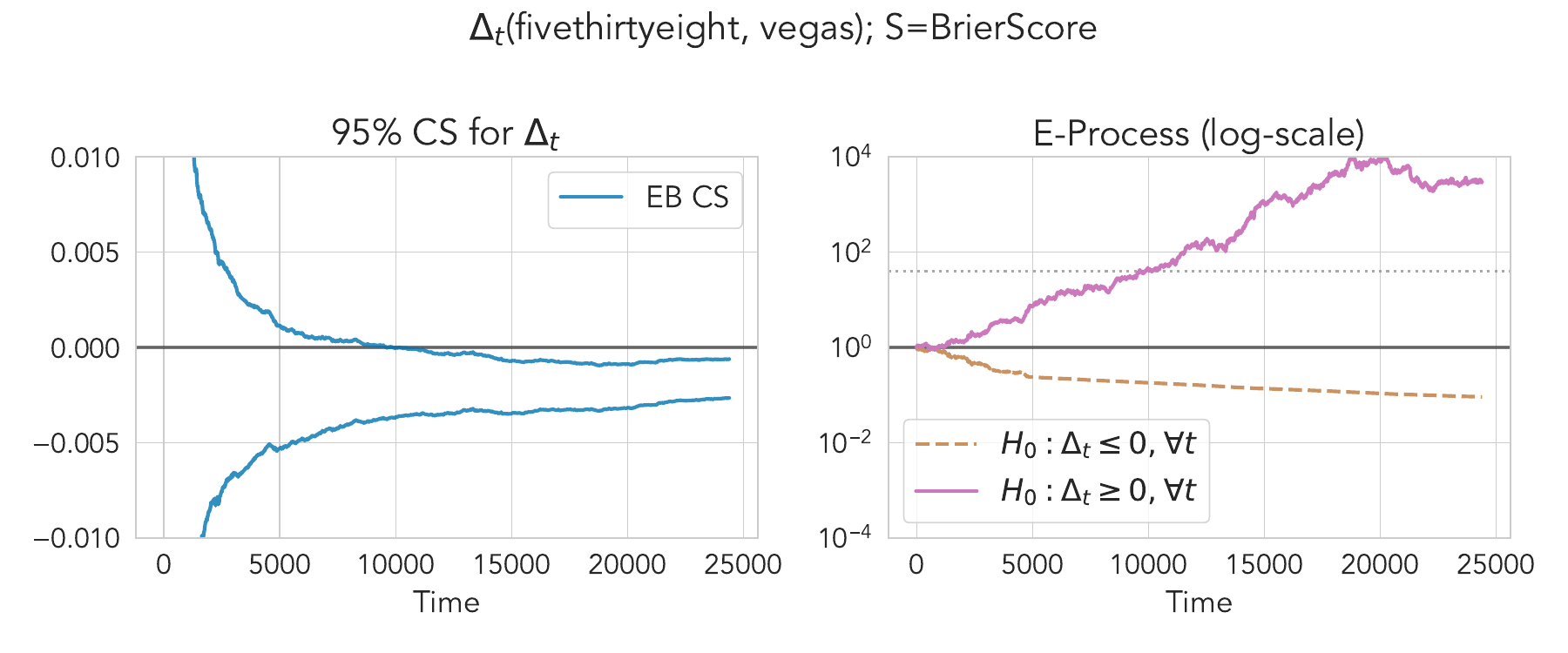}
    \caption{\emph{Left:} A 95\% CS (Theorem~\ref{thm:main}) for the average Brier score differentials $(\Delta_t)_{t=1}^T$ between \emph{FiveThirtyEight} and \emph{Vegas}, two real-world forecasters that made game-by-game probability forecasts on Major League Baseball (MLB) games from 2010 to 2019 ($T=25,165$).
    Positive values of $\Delta_t$ indicate that the first forecaster is better than the second on average.
    Unlike a classical CI, a CS covers the time-varying parameter $\Delta_t$ uniformly over all $t$ with high probability.
    In this case, we find that, with 95\% probability, the sequence $\Delta_t$ trends negative for $t \geq 10,000$, indicating that \emph{Vegas} outperformed \emph{FiveThirtyEight} on average across most of the time horizon.
    \emph{Right:} E-processes (Theorem~\ref{thm:eprocess}) for the null hypotheses, $\calH_0: \Delta_t \leq 0,\; \forall t$ (brown, dashed) and $\calH_0: \Delta_t \geq 0,\; \forall t$ (purple, solid), respectively.
    An e-process quantifies the accumulated evidence against the null, and it has a direct correspondence to the CS.
    In this example, larger values in the e-process for $\calH_0: \Delta_t \geq 0,\; \forall t$ indicate evidence of \emph{Vegas} outperforming \emph{FiveThirtyEight} on average. 
    The gray dashed line plots the value $2/\alpha = 40$, and the time at which an e-process upcrosses this line is also when the $(1-\alpha)$-CS moves entirely below or above zero.
    See Sections~\ref{sec:cs} and~\ref{sec:experiments} for details.
    }
    \label{fig:figure1}
\end{figure}

In addition, we derive \emph{e-processes} and \emph{p-processes}~\citep{ramdas2021testing} for testing whether one forecaster outperforms the other on average, which is a composite null that we formally define in Section~\ref{sec:eprocess}.
An e-process $E_t$ is a nonnegative process such that under the null, its expectation at any stopping time is at most one.
It quantifies the amount of accumulated evidence against the null up to time $t$: a larger $E_t$ is more evidence against the null. 
Further, $\sfp_t = 1/\sup_{i \leq t}E_i$ is a p-process --- its realization at any stopping time is a valid p-value, a property referred to as \emph{anytime-valid} or \emph{always-valid}~\citep{johari2021always,howard2021timeuniform}. 
These are also formally defined in Section~\ref{sec:eprocess}.
Throughout the paper, we define \emph{safe, anytime-valid inference (SAVI)} methods as ones that satisfy either the time-uniform coverage guarantee (CS) or the anytime-valid guarantee (e- or p-processes).

The setup in which we develop our methods is game-theoretic~\citep{shafer2019game}: we posit that two players participate in a forecasting game on a sequence of outcomes with an unknown distribution.
This setup naturally leads to ``distribution-free'' inference procedures --- other than requiring bounded scoring rules, we make no assumptions on the time-varying dynamics of the outcomes and forecasts, such as stationarity.
We further discuss how to relax even the assumption of bounded scores using asymptotic CSs (Section~\ref{sec:asympcs}) and normalized scores (Section~\ref{sec:winkler}).

In Figure~\ref{fig:figure1}, we show an example of a CS and its corresponding e-processes applied to a forecasting game between two real-world forecasters, \emph{FiveThirtyEight} and \emph{Vegas}, on the outcomes of Major League Baseball (MLB) games.
The CS in the left plot continuously tracks the expected average score differential over time and effectively visualizes the time-varying trend along with the uncertainty on its estimation.
The two e-processes in the right plot each measure the accumulated evidence favoring each forecaster over time.
In this example, both the CS and the e-processes show that \emph{Vegas} has outperformed \emph{FiveThirtyEight} on average.
We return to this example in Section~\ref{sec:2019ws}.

The rest of the paper is organized as follows.
After discussing related work (Section~\ref{sec:relatedwork}) and preliminaries  (Section~\ref{sec:preliminaries}), we derive CSs for the time-varying average forecast score differentials between two probabilistic forecasters in Sections~\ref{sec:game}-\ref{sec:ebcs}, with the case of binary outcomes as a working example.
In Section~\ref{sec:eprocess}, we also derive e-processes and p-processes as duals to our CSs, providing alternative sequential inference procedures for forecast comparison.
In Section~\ref{sec:simulated}, we empirically validate our CSs and compare them against fixed-time and asymptotic confidence intervals (CIs) on simulated data; in Sections~\ref{sec:2019ws} and~\ref{sec:weather}, we apply our methods to real-world forecast comparison tasks, namely comparing game-by-game predictions in Major League Baseball (MLB) and comparing statistical postprocessing methods of ensemble weather forecasts.
In addition, Section~\ref{sec:proof} contains omitted proofs; 
Section~\ref{sec:boundary_details} contains technical details about the time-uniform boundary choices;
Section~\ref{sec:asympcs} contains an alternative forecast comparison approach using an asymptotic CS; 
Sections~\ref{sec:winkler}-\ref{sec:predictable} contain extensions to normalized scores~\citep{winkler1994evaluating}, lag-$h$ forecasts, and predictable conditions/bounds, respectively; 
Section~\ref{sec:extensions} contains extensions from binary outcomes to categorical and continuous outcomes;
Section~\ref{sec:comparison} contains detailed comparisons with the methods of~\citet{henzi2021valid,diebold1995comparing,giacomini2006tests}; and
Section~\ref{sec:additional_experiments} contains additional details about our simulated, MLB, and weather experiments as well as details about experimentally fine-tuning the CS width.

\section{Related Work}\label{sec:relatedwork}

\paragraph{Evaluation and Comparison of Forecasts.}

Forecast evaluation is a well-studied subject in the literature of statistics, economics, finance, and climatology, dating back to the works of \citet{brier1950verification,good1952rational,degroot1983comparison,dawid1984present,schervish1989general}.
The primary tool for evaluating forecasts is proper scoring rules, of which the literature is extensive.
Many characterization theorems for proper scoring rules exist across different forecasting scenarios, notably including the case of probability forecasts for binary and categorical outcomes, point forecasts (e.g., mean, quantiles, and prediction intervals) for continuous outcomes, and fully probabilistic forecasts (e.g., densities and CDFs) for continuous outcomes.
See, e.g., \citet{mccarthy1956measures,savage1971elicitation,schervish1989general,winkler1996scoring,grunwald2004game,gneiting2007strictly,gneiting2011making,abernethy2012characterization,dawid2014theory,ehm2016quantiles,ovcharov2018proper,frongillo2021general,waggoner2021linear}, for both classical and recent developments.

The problem of comparing forecasts while accounting for sampling uncertainty was first popularized in the case of probability forecasts by \citet{diebold1995comparing} (DM), who proposed tests of equal (historical) forecast accuracy using the differences in forecast errors. The DM test is based on the asymptotic normality of the average forecast score differentials, and it makes stationarity assumptions about the outcomes.
\citet{giacomini2006tests} (GW) developed tests of \emph{conditional} predictive accuracy given past information, allowing for the comparison of ``which forecaster is more accurate given the information available at the time of forecasting.''
The GW test thus allows for nonstationarity, although it restricts the forecasters to a fixed window size $m$ and its validity depends on mixing assumptions.
\citet{lai2011evaluating} presented a comprehensive overview of the aforementioned methods of forecast comparison and developed a martingale-based theory of scoring rules whose differentials are linear in the outcome, such as proper scoring rules.
They proved the asymptotic normality of both forecast scores and score differentials, leading to an asymptotic and fixed-time CI that we use as a point of comparison in our work.
More recent work by \citet{ehm2018forecast,ziegel2020robust,yen2021testing} derive fixed-time tests of forecast dominance under all consistent scoring functions~\citep{gneiting2011making}.
In comparison with all of these previous methods that presuppose a fixed sample size, the key difference in our work is that we develop inference methods that are valid at arbitrary data-dependent stopping times, while making virtually no assumption on the time-varying dynamics of the data generating process.
The resulting graphical representations of CSs and e-processes also convey information about the entire time-varying trend of score differences, as in Figure~\ref{fig:figure1}, unlike classical tests and CIs that concern a single comparison at a fixed time point.

Recently, \citet{henzi2021valid} constructed sequential tests of conditional forecast dominance based on e-processes~\citep{howard2020chernoff,grunwald2019safe,shafer2019language,ramdas2021testing,vovk2021evalues}.
These methods are also anytime-valid and nonasymptotic; yet, they test a ``strong\footnote{This distinction of strong and weak nulls come from the discussion of randomized experiments in causal inference; see, e.g., \citet{lehmann1975nonparametrics,rosenbaum1995design}. Within the context of forecast comparison, \citet{ehm2018forecast} distinguish between tests of average and step-by-step conditional predictive ability, which mirrors that of weak and strong nulls.} null,'' which states that one forecaster is better than the other at \emph{every} point in time, something we rarely believe a priori. 
Thus, rejecting the strong null only suggests that there exists \emph{some} time point where the latter forecaster is better than the former, which may not come as much of a surprise. (One case where the strong null is appropriate is if we test two sets of forecasts produced by the same data scientist, with one forecaster using more features or more sophisticated models; but for two unrelated forecasters, we rarely expect the strong null to be true.)
In contrast, our e-processes test whether one forecaster dominates the other \emph{on average} over time (thus requiring consistent outperformance), and the CSs can even test such averaged nulls in a two-sided fashion (equivalently, it tests both one-sided nulls). 
We examine this distinction further in Sections~\ref{sec:eprocess} and~\ref{sec:weather}; other methodological differences are summarized in Section~\ref{sec:comparison_hz}.

\begin{table}[t]
    \centering
    \small
    \begin{tabular}{l|l|c|c|c|c|c}
    \Xhline{1.1pt}
    \bf Method \& Key Result & {\bf Null Hypothesis} $\calH_0$ & \bf Weak & \bf CI & \bf SAVI & \bf NA & \bf DF \\ \hline
    
    \citet{diebold1995comparing} & 
    $\delta = 0\;$ & \xmark & \cmark & \xmark & \xmark & \xmark \\
    $\sqrt{n}(\hat\Delta_n - \delta) \rightsquigarrow N(0, 2\pi f_d(0))$ &   & & & & \\ \hline
    
    \citet{giacomini2006tests} & 
    $\mathbb{E}_{n-1}[\hat\delta_{m,n}] = 0, \; \forall n\;$ & \xmark & \xmark & \xmark & \xmark & \xmark \\
    $T_{m}(\hat\Delta_n) \rightsquigarrow \chi^2$ & & & & & & \\ 
    ($m$: max. forecasting window) & & & & & & \\ \hline
    
    \citet{lai2011evaluating} & 
    $\frac{1}{n}\sum_{i=1}^t \mathbb{E}_{i-1}[\hat\delta_i] = 0,\; \forall n \;$ & \cmark & \cmark & \xmark & \cmark & \xmark \\
    $\sqrt{n}(\hat\Delta_n - \Delta_n)/s_n \rightsquigarrow N(0, 1)$, & & & & & & \\ 
    $s_n \leq \frac{1}{4n}\sum_{i=1}^n [\delta_i(1)-\delta_i(0)]^2 $ & & & & & & \\ \hline
    
    \citet{henzi2021valid} & 
    $\mathbb{E}_{t-1}[\hat\delta_t] \leq 0,\; \forall t \;$ & \xmark & \xmark & \cmark & \cmark & \cmark \\
    $E_t = \prod_{i=1}^t \inparen{1 + \lambda \frac{\delta_i(y_i)}{\delta_i(\indicator{p_i>q_i})}}$ & & & & & & \\
    is an e-process, $\lambda>0$ & & & & & & \\ \hline
    
    Ours & 
    $\frac{1}{t}\sum_{i=1}^t \mathbb{E}_{i-1}[\hat\delta_i] \leq 0,\; \forall t \;$ & \cmark & \cmark & \cmark & \cmark & \cmark \\
    $t(\hat\Delta_t - \Delta_t)$ is sub-exponential, & & & & & & \\
    which yields a CS \& an e-process & & & & & & \\
    \Xhline{1.1pt}
    \end{tabular}
    \caption{Inference methods for comparing probability forecasts for binary outcomes.
    This table is meant to be a quick summary only; see each referenced paper for the precise definitions, conditions, and guarantees for the method.
    The last two methods are the only ones that are anytime-valid, nonasymptotic, and distribution-free --- both of which develop e-processes. Among the two, only our method tests the weak null and provides a CS for \emph{estimating} $\Delta_t$.
    \textbf{Notations:} for each $t \in \N$, $p_t$ and $q_t$ are two probability forecasts on the outcome $y_t$; $\delta_t(y) = S(p_t, y) - S(q_t, y)$; $\hat\delta_t = \delta_t(y_t)$; $\hat\Delta_t = t^{-1}\sum_{i=1}^t \hat\delta_i$; $\Delta_t = t^{-1}\sum_{i=1}^t \mathbb{E}_{i-1}[\hat\delta_i]$.
    We also use $t$ to refer to a time index varying over time, and $n$ to denote a fixed sample size that must be determined before the experiment.
    \textbf{Weak}: whether the method tests a weak null involving a time-varying average.
    \textbf{CI}: whether the method provides a confidence interval for the score difference (as opposed to only deriving a test).
    \textbf{SAVI}: whether inference is valid at arbitrary data-dependent stopping times (as opposed to only fixed times).
    \textbf{NA}: whether the method has a nonasymptotic guarantee. 
    \textbf{DF}: whether the method has a distribution-free guarantee (as opposed to requiring distributional assumptions like stationarity/mixing/IID).}
\label{tbl:relatedwork}
\end{table}

Table~\ref{tbl:relatedwork} summarizes the aforementioned methods of forecast comparison in terms of whether they have a stopping time (or equivalently, time-uniform; see Section~\ref{sec:eprocess} for further details) guarantee, a non-asymptotic guarantee, and a distribution-free guarantee.

\paragraph{Time-Uniform Confidence Sequences.}

Confidence sequences were developed by Robbins and coauthors~\citep{darling1967confidence,robbins1970statistical,robbins1970boundary,lai1976boundary}.
Recent renewed interests on CSs are partly due to best-arm identification in multi-armed bandits~\citep{jamieson2014lilucb,jamieson2018bandit}, where CSs are sometimes referred to as always-valid or anytime confidence intervals. 
CSs are also duals to sequential hypothesis tests, analogously to CIs being dual to fixed-time hypothesis tests, and one can further derive a sequence of e-processes and p-processes given the CSs (more precisely, its underlying exponential process)~\citep{ramdas2021testing}.
In Section~\ref{sec:eprocess}, we make this connection explicit and discuss how our approach also leads to p-processes, or anytime-valid p-values~\citep{johari2021always}, for weak nulls.

The recent work by \citet{howard2021timeuniform} is of particular importance in our paper, as it develops tight CSs that are uniformly valid over time under nonparametric assumptions and has widths that shrink to zero. 
This work and its underlying technique of developing exponential test (super)martingales~\citep{howard2020chernoff,darling1967confidence,ville1939etude} have led to several interesting results, including state-of-the-art concentration inequalities for IID mean estimation~\citep{waudbysmith2020estimating} and sequential quantile estimation~\citep{howard2022sequential}. 
Our work makes the connection between the empirical Bernstein (EB) CSs derived in \citet{howard2021timeuniform} and the martingale property of forecast score differentials~\citep{lai2011evaluating}, leading to a novel sequential inference procedure for forecaster comparison.

\section{Preliminaries}\label{sec:preliminaries}

\subsection{Test Supermartingales, Ville's Inequality, and Confidence Sequences}\label{sec:martingales}

The theory of martingales and their interpretation as a gambler's wealth in a betting game are instrumental in deriving SAVI methods.
See \citet{ramdas2022game} for a comprehensive introduction. 
Let $(\calX, \calG)$ be a measurable space equipped with a filtration $\frakG := (\calG_t)_{t=0}^\infty$, where each $\calG_t$ represents the accumulated information up to time $t$.
Given any probability distribution $P$ on $(\calX, \calG)$, a sequence of random variables $(X_t)_{t=0}^\infty$ is called a \emph{process} if it is \emph{adapted} to $\frakG$, meaning that $X_t$ is $\calG_t$-measurable for all $t$.
A process is also \emph{predictable} w.r.t.~$\frakG$ if $X_t$ is $\calG_{t-1}$-measurable for all $t \geq 1$.
A \emph{stopping time} $\tau$ w.r.t.~$\frakG$ is a nonnegative integer random variable that satisfies $\{\tau \leq t\} \in \calG_t$ for all $t \geq 1$.

Let $\Esymb_{t-1}[\cdot] = \Esymb_P[\cdot \mid \calG_{t-1}]$ denote the conditional expectation w.r.t.~$\calG_{t-1}$ under $P$. 
A process $(L_t)_{t=0}^\infty$ is a \emph{supermartingale} if $\Esymb_P[|L_t|] < \infty$ and $\Esymb_{t-1}[L_t] \leq L_{t-1}$ for each $t \geq 1$, 
and a \emph{martingale} if ``$\leq$'' is replaced with ``$=$''.
A nonnegative supermartingale $(L_t)_{t=0}^\infty$ that starts at one ($L_0=1$) is called a \emph{test supermartingale} (for $P$)~\citep{shafer2011test}.
If $(L_t)_{t=0}^\infty$ is a test supermartingale for $P$, then Ville's inequality~\citep{ville1939etude} states that, for any $\alpha \in (0, 1)$,
\begin{equation}\label{eqn:ville}
    P\inparen{\exists t \geq 1: L_t \geq 1/\alpha} \leq \alpha.
\end{equation}
Ville's inequality is the primary tool for constructing confidence sequences, as illustrated in, e.g., \citet{howard2021timeuniform}; in fact, it is the only admissible way to construct them~\citep{ramdas2020admissible}.
Given $\alpha \in (0, 1)$, a $(1-\alpha)$-\emph{confidence sequence (CS)} for a time-varying sequence of target parameters $(\theta_t)_{t=1}^\infty$ is a sequence of confidence intervals (CIs) $(C_t)_{t=1}^\infty$ such that
\begin{equation}\label{eqn:cs_definition}
    P\inparen{\exists t \geq 1 : \theta_t \notin C_t} \leq \alpha,\quad \text{or equivalently,}\quad P\inparen{\forall t \geq 1 : \theta_t \in C_t} \geq 1 - \alpha.
\end{equation}
In particular, the guarantee remains valid at arbitrary stopping times and without a prespecified sample size, so that collecting additional data over time does not invalidate it~\citep[Lemma 3]{howard2021timeuniform}:
\begin{equation}\label{eqn:cs_anytime_valid}
\text{ for all stopping times $\tau$, possibly infinite, } \quad P\inparen{\theta_\tau \in C_\tau} \geq 1 - \alpha.
\end{equation}
This coverage guarantee at stopping times is sometimes referred to as being \emph{anytime-valid}.
This crucially differentiates a CS from a fixed-time CI, $C_n$, which only has the following weaker guarantee: 
\begin{equation}
    \forall n \geq 1,\; P\inparen{\theta_n \notin C_n} \leq \alpha, \quad\text{or equivalently,}\quad \forall n \geq 1,\; P\inparen{\theta_n \in C_n} \geq 1- \alpha.
\end{equation}
In short, CSs, as opposed to CIs, are the appropriate tools for sequential inference.

\subsection{Forecast Evaluation via Scoring Rules}\label{sec:proper}

Let $\calY$ be the space of all possible outcomes equipped with a $\sigma$-field $\calG$.
Let $\Delta(\calY)$ be the set of all probability distributions on $(\calY, \calG)$ and $\calP \subseteq \Delta(\calY)$.
To facilitate our discussion, the primary working example in this paper will be the space of binary outcomes $\calY = \{0, 1\}$ and probability forecasts parametrized by their means in $\calP = [0, 1]$. 
But our setup can be generalized to any finite sample space $\calY = \{1, \dotsc, K\}$ with $K$-dimensional probability forecasts $\calP = \Delta^{K-1}$, for $K \geq 2$, and $d$-dimensional sample space $\calY \subseteq \R^d$, for $d \geq 1$, with point (e.g., mean and quantile) or probabilistic (e.g., CDF) forecasts.
(We defer our discussion of these general cases to Section~\ref{sec:extensions}.)

A \emph{scoring rule} is any extended real-valued function\footnote{More formally, the scoring rule $S$ is required to be \emph{$\calP$-quasi-integrable} in its second argument, meaning that for every $p \in \calP$, $S(p, \cdot)$ is measurable and, for all $q \in \calP$, the integral $\int_\calY S(p, y) dq(y)$ exists as a possibly infinite but not indeterminate value~\citep{bauer2001measure,abernethy2012characterization}.} $S : \calP \times \calY \to \overline{\R}$ and can be used to evaluate the performance of a (probabilistic) forecast $p \in \calP$ given an observation $y \in \calY$.
Following \citet{gneiting2007strictly}, we take scoring rules to be \emph{positively oriented}, meaning that higher scores reflect better forecasts.
A prominent example is the Brier score~\citep{brier1950verification}, which in the binary case can be expressed as $S(p, y) = 1-(p - y)^2$ for $p \in [0, 1]$ and $y \in \{0, 1\}$.

Given a forecast $p \in \calP$ and a probability distribution $q \in \Delta(\calY)$, we can naturally extend the definition of a scoring rule $S$ to its \emph{expected score} w.r.t.~$y \sim q$ (conditional on $p$):
\begin{equation}\label{eqn:expected_score}
    S(p; q) = \Ex{y \sim q}{S(p, y)}.
\end{equation}
Here, we make the distinction between the scoring rule $S$ on $\calP \times \calY$ and its expected score $S$ defined on $\calP \times \Delta(\calY)$ by the notations $S(p, y)$ and $S(p; q)$, respectively.
We can recover the scoring rule from the expected score definition via $S(p, y) = S(p; \delta_y)$, where $\delta_y$ is a point measure on $y$.

A scoring rule $S$ is \emph{proper} if any probability $q \in \Delta(\calY)$ maximizes the expected score $S(\cdot; q)$:
\begin{equation}\label{eqn:proper}
q \;\in\; \argmax_{p \in \calP}\; S(p; q).
\end{equation}
$S$ is \emph{strictly proper} if the $\argmax$ in \eqref{eqn:proper} is unique.
Intuitively, a proper scoring rule encourages forecasters to be honest, because if a forecaster believes that the outcome follows the distribution $q \in \calP$, then they are incentivized to honestly forecast $q$, instead of any other distribution $p \neq q$, as $q$ maximizes the expected score (uniquely, if $S$ is strictly proper) according to their belief.
Proper scoring rules are often considered as the primary means of evaluating probabilistic forecasts, as they assess both calibration and sharpness~\citep{winkler1996scoring,gneiting2007probabilistic}.

Classical examples of proper scoring rules for probability forecasts $p \in \calP = [0, 1]$ on binary outcomes $y \in \calY = \{0, 1\}$ include the following:
\begin{itemize}
    \item The Brier score or the quadratic score~\citep{brier1950verification}:
    $ S(p, y) = 1-(p - y)^2$.
    \item The spherical score~\citep{good1971comment}:
        $S(p, y) = \frac{py + (1-p)(1-y)}{\sqrt{p^2 + (1-p)^2}}$.
    \item The logarithmic score~\citep{good1952rational}:
    $S(p, y) = y\log(p) + (1-y)\log(1-p)$.
    \item The zero-one score or the success rate:
    $S(p, y) = y \indicator{p \geq 0.5} + (1-y) \indicator{p < 0.5}$.
\end{itemize}

The Brier, spherical, and logarithmic scores are examples of strictly proper scoring rules, while the zero-one score is an example of a proper but not strictly proper scoring rule.
An example of an improper scoring rule for probability forecasts is the absolute score, $S(p, y) = 1 - |p-y|$.
Also note that all of the examples except the logarithmic score are bounded for $p \in [0, 1]$ and $y \in \{0, 1\}$.

\section{Anytime-Valid Inference for Average Forecast Score Differentials}\label{sec:cs}

In this section, we derive CSs and e-processes, as well as their corresponding sequential tests and p-processes, for the time-varying average difference in the quality of forecasts, as measured by a scoring rule.
Our intuition comes from the extensive literature on evaluating and comparing probability forecasts via scoring rules~\citep{winkler1996scoring, gneiting2007strictly, degroot1983comparison, schervish1989general, gneiting2011making, lai2011evaluating}, combined with the powerful tool of time-uniform CSs~\citep{darling1967confidence,howard2021timeuniform}.
For now, our working example in this section will be the case of comparing probability forecasts on binary outcomes; we further discuss extensions to categorical and certain continuous outcomes in Section~\ref{sec:extensions}.

\subsection{A Game-Theoretic Formulation}\label{sec:game}

The intuition behind our SAVI methods for forecast score differentials comes from the game-theoretic statistical framework~\citep{shafer2019language,ramdas2022game}.
Consider a forecasting game where two players make probabilistic forecasts on an event that happens over time (e.g., whether it will rain on each day, whether a sports team will win its game each week, and more) and an unknown player named reality chooses a sequence of distributions that generates the outcomes that the forecasters are trying to predict.
Let $t = 1, 2, \dotsc$ denote each round of the game.
Though not required, we can also optionally allow having any historical data $y_{-(H-1)}, \dotsc, y_{-1}, y_0$ for some $H \geq 0$.
The forecasting game can be formulated in general as follows --- the case of probability forecasts on binary outcomes is obtained by setting $\calP = \Delta(\calY) = [0, 1]$ ($y_t \sim r_t$ would refer to $y_t \sim \mathrm{Bernoulli}(r_t)$).
\begin{game}[Comparing Sequential Forecasters]\label{game:general}
For rounds $t = 1, 2, \dotsc$:
\begin{enumerate}
    \setlength\itemsep{0em}
    \item Forecasters 1 and 2 make their forecasts, $p_t, q_t \in \calP$, respectively. 
    \emph{The order in which the forecasters make their forecasts is not specified.}
    \item Reality chooses $r_t \in \Delta(\calY)$. 
    \emph{$r_t$ is not revealed to the forecasters.}
    \item $y_t \sim r_t$ is sampled and revealed to the forecasters. 
\end{enumerate}
\end{game}
We now elaborate on the role of each player in Game~\ref{game:general}.

\paragraph{Forecasters 1 \& 2.} 
At each round $t$, the two forecasters can make their forecasts using any information available to them.
This includes historical and previous outcomes $y_{-(H-1)}, \dotsc, y_0, y_1, \dotsc, y_{t-1}$, any of the previous forecasts made, $p_1, \dotsc, p_{t-1}$, $q_1, \dotsc, q_{t-1}$, as well as any other side information available to either forecaster.
They cannot, however, make their predictions using any of $r_1, \dotsc, r_t$'s (or information from the future).
For example, when predicting the outcome of the next baseball game, the forecasters' filtration may include not only all of previous games' results but also any side information that either forecaster may have, such as which players are starting the game and whether there are injuries. 
The setup also allows for the case where two forecasters have different side information, as our results are completely agnostic to such details.

This game-theoretic framework for forecast comparison is \emph{prequential}~\citep{dawid1984present}, in the sense that we put no restrictions on how these forecasts are generated, and we only evaluate forecasters based on the forecasts they did make and the outcomes that did occur, as opposed to forecasts they would have made had the outcomes been different.

\paragraph{Reality.} 
In our game, Reality is the player that determines the unknown distribution $r_t$ of the eventual outcome $y_t$ conditioned on its past, which notably includes the forecasters' choices $p_t$ and $q_t$. 
In the binary case, for example, Reality chooses the conditional mean sequence of the outcomes $y_t$ given everything it has seen.
Reality can essentially choose $r_t$ ``however they want,'' and they can even choose $r_t$ after seeing $p_t$ or $q_t$.
Put differently, the framework is agnostic to what information Reality sees: Reality may only see its past choices $r_1, \dotsc, r_{t-1}$ and (optionally)  the past outcomes $y_1, \dotsc, y_{t-1}$, or it may act adversarially after seeing $p_t$ and $q_t$. 
In particular, $r_t$ could also be a point distribution at $y_t$.

We note that the distribution-free property of our methods corresponds to the fact that the game places no distributional assumptions on the time-varying dynamics of $(r_t)_{t=1}^\infty$, such as stationarity, Markovian or other conditional independence assumptions.

\paragraph{The Statistician.}
The statistician, who stands outside of the game, has the goal of comparing the predictive performance of the two forecasters according to a chosen scoring rule and based only on the observed data $(p_t,q_t,y_t)_{t=1}^\infty$, without making any assumptions about the behavior of any player involved.\footnote{Specifically, we do not explicitly consider strategic issues arising from (say) the choice of the scoring rule or the method of comparison. In other words, we consider the comparison problem separately from the elicitation problem (how to elicit honest forecasts). A separate line of work considers these important, but orthogonal, issues.}
The statistician may choose to update their inferential conclusions as the game progresses.
How the statistician achieves such a goal will be the focus of the subsequent sections.

\subsection{The Measure-Theoretic Setup}\label{sec:setup}

We now formalize Game~\ref{game:general} in the context of comparing the two probabilistic forecasters over time.
Let $(p_t)_{t=1}^\infty$ and $(q_t)_{t=1}^\infty$ be two sequences of forecasts in $\calP$, for a sequence of outcomes $(y_t)_{t=1}^\infty$ in $\calY$.
In the binary case, the forecasts will take values in $\calP = [0, 1]$ and the outcomes in $\calY = \{0, 1\}$.
We can define Game~\ref{game:general} in a measure-theoretic sense by specifying the associated filtrations, i.e., a sequence of ``information sets'' with which we perform inference.
Our formulation is closely related to the setup of \citet{lai2011evaluating}, although we make the game-theoretic intuitions explicit.

\paragraph{The ``Observable'' Forecaster Filtration $\frakF$.}
We first define the filtration with which the two forecasters generate their forecasts, denoted as $\frakF := (\calF_t)_{t=0}^\infty$.
For each $t \geq 1$, let $\calF_{t-1}$ represent \emph{any} information available to the forecasters before making their predictions at time $t$, as described in the previous subsection.
Mathematically, this means that $(p_t)_{t=1}^\infty$, $(q_t)_{t=1}^\infty$, and $(y_t)_{t=1}^\infty$ are adapted w.r.t.~$\frakF$.
Note that $\frakF$ also includes the information available to the statistician, making this the ``observable'' filtration that contrasts with the ``oracle'' filtration (defined below).

\paragraph{The ``Oracle'' Game Filtration $\frakG$.}
The game filtration, denoted as $\frakG := (\calG_t)_{t=0}^\infty$, represents \emph{all} sets of information associated with Game~\ref{game:general}.
The parameter of interest (unknown to the statistician) is defined w.r.t.~this ``oracle'' filtration.
More precisely, for each $t \geq 1$, $\calG_{t-1}$ includes not only everything in $\calF_{t-1}$ but also any information available to Reality before the outcome $y_t$ is realized, including Reality's choice $r_t$.
Mathematically, this implies that $(p_t)_{t=1}^\infty$, $(q_t)_{t=1}^\infty$, and $(r_t)_{t=1}^\infty$ are \emph{predictable} w.r.t.~$\frakG$, while $(y_t)_{t=1}^\infty$ is adapted w.r.t.~$\frakG$.
The setup allows for the flexible choices of Reality described in the previous subsection, as it does not preclude Reality's actions in any way.

In the remainder of the paper, we use the notation $\Esymb_{t-1}[\cdot] = \ex{\cdot \mid \calG_{t-1}}$ to denote the conditional expectation with respect to the game filtration for each $t$.
In the case of binary (and categorical) outcomes, because the outcome distribution is completely specified by their mean, we simply let $r_t$ denote the (unknown) conditional mean of the outcome $y_t$ given $\calG_{t-1}$ for each $t$, with a slight abuse of notation.
In such cases, we have that 
\begin{equation}
    r_t = \Esymb_{t-1}[y_t] \quad \forall t = 1, 2, \dotsc,
\end{equation}
where $\Esymb_{t-1}$ refers to the conditional expectation over $y_t \sim r_t \mid \calG_{t-1}$.

\paragraph{Comparing Sequential Forecasters via Average Forecast Score Differentials.}
With the aforementioned setup, we can now use scoring rules to assess and compare the quality of the two forecasters over time. 
We define the \textbf{\emph{average (forecast) score differential}} $\Delta_t$ between the sequences of forecasts $(p_i)_{i=1}^\infty$ and $(q_i)_{i=1}^\infty$, up to time $t$, as the average difference in \emph{expected scores}:
\begin{equation}\label{eqn:fsd}
    \Delta_t := \frac{1}{t}\sum_{i=1}^t \Esymb_{i-1}\insquare{ S\inparen{p_i, y_i} - S\inparen{q_i, y_i} }, \quad t \geq 1,
\end{equation}
where $\Esymb_{i-1}$ denotes the expectation over $y_i \sim r_i$ \emph{conditioned on} the game filtration $\calG_{i-1}$, which includes both forecasts $p_i$ and $q_i$ as well as $r_i$.
The time-varying parameter $\Delta_t$ provides an intuitive way of quantifying the difference in the quality of forecasts made up to time $t$. 
We highlight that $\Delta_t$ helps us infer whether one forecaster is better than the other \emph{on average} (over time), as opposed to one strictly dominating the other~\citep{giacomini2006tests,henzi2021valid}.
This estimand is also used in \citet{lai2011evaluating}'s asymptotic CI.

The parameter $\Delta_t$ is not observable to the statistician or the forecasters, because reality's moves $r_1, \dotsc, r_t$ are unknown and never observed. 
We thus define the \textbf{\emph{empirical average (forecast) score differential}} $\hat\Delta_t$ as the unbiased estimate of each summand in \eqref{eqn:fsd}, also averaged over time:
\begin{equation}\label{eqn:empirical_fsd}
    \hat\Delta_t := \frac{1}{t}\sum_{i=1}^t \insquare{ S\inparen{p_i, y_i} - S\inparen{q_i, y_i} }, \quad t \geq 1.
\end{equation}
$\hat\Delta_t$ is completely observable to the statistician after time $t$.

The statistician's goal then becomes quantifying how far $\hat\Delta_t$ is from $\Delta_t$, while accounting for the uncertainty associated with sampling $y_t$ at each time $t$. 
To this end, we define the \emph{pointwise (forecast) score differential} $\delta_i := \Esymb_{i-1}[S(p_i, y_i) - S(q_i; y_i)]$ and its empirical counterpart $\hat\delta_i := S(p_i, y_i) - S(q_i, y_i)$. 
Then, it is immediate that the cumulative sums of deviations, defined by $S_0=1$ and
\begin{equation}\label{eqn:cumulative_sums}
    S_t := t\inparen{\hat\Delta_t - \Delta_t} = \sum_{i=1}^t \inparen{\hat\delta_i - \delta_i}, \quad t \geq 1,
\end{equation}
forms a martingale, i.e., $\Esymb_{t-1}[S_t] = S_{t-1},\; \forall t \geq 1$.
Previous work including \citet{seillier1993testing,lai2011evaluating} use this property to derive the asymptotic normality of empirical average score differentials.
In the following sections, we illustrate how $(S_t)_{t=0}^\infty$ can further be uniformly and non-asymptotically bounded by constructing \emph{exponential} test supermartingales.
As a result, we will be able to estimate and cover $\Delta_t$ using CSs and also test its sign using e-processes.

\subsection{Time-Uniform Confidence Sequences for Average Score Differentials}\label{sec:ebcs}

\subsubsection{Time-Uniform Boundaries and Exponential Test Supermartingales}\label{sec:subpsi}

We now show that we can uniformly bound the difference between $\hat\Delta_t$ and $\Delta_t$ over time using uniform boundaries and test supermartingales. 
To do this, we start with a \emph{cumulative sum} process $S_t := \sum_{i=1}^t (\hat\delta_i - \delta_i)$ as well as its \emph{intrinsic time} $\hat{V}_t$, which is the variance process for $S_t$ (to be defined later).
Our goal is then to uniformly bound the sum $S_t$ over the intrinsic time $\hat{V}_t$, which corresponds to bounding the difference between $\hat\Delta_t$ and $\Delta_t$ over time due to~\eqref{eqn:cumulative_sums}.

Following~\cite{howard2020chernoff}, for any sum process $(S_t)_{t=0}^\infty$ and its intrinsic times $(\hat{V}_t)_{t=0}^\infty$, we define a \emph{(one-sided) uniform boundary $u = u_\alpha$ with crossing probability $\alpha \in (0, 1)$} as any function of the intrinsic time that gives a time-uniform bound on the sums: 
\begin{equation}\label{eqn:uniform_boundary}
    P\inparen{\forall t \geq 1: S_t \leq u_\alpha(\hat{V}_t)} \geq 1 - \alpha ,
\end{equation}
that is, with probability at least $1-\alpha$, the sums $S_t$ are upper-bounded by $u(\hat{V}_t)$ \emph{at all times $t$}.
By similarly computing a uniform boundary to $(-S_t, \hat{V}_t)_{t=0}^\infty$, we can also obtain a time-uniform lower bound on $S_t$.
(Alternatively, we can directly define a \emph{two-sided} sub-$\psi$ uniform boundary, which satisfies $P(\forall t \geq 1: -u_\alpha(\hat{V}_t) \leq S_t \leq u_\alpha(\hat{V}_t)) \geq 1-\alpha$. An example is \citet{robbins1970statistical}'s two-sided normal mixture that we describe in Section~\ref{sec:uniform_boundary}.)
The upper and lower bounds then jointly form a time-uniform CS on $(\Delta_t)_{t=1}^\infty$ by rearranging the terms.

How do we show that there exists such a uniform boundary for our definitions of $(S_t, \hat{V}_t)_{t=0}^\infty$? 
\citet{howard2020chernoff,howard2021timeuniform} show that there exists such a uniform boundary if, for each $\lambda \in [0, \lambda_\mathrm{max})$, the \emph{exponential process} defined by $L_0(\lambda)=1$ and
\begin{equation}\label{eqn:expm}
    L_t(\lambda) = \exp\incurly{ \lambda S_t - \psi(\lambda) \hat{V}_t }, \quad t \geq 1,
\end{equation}
is a test supermartingale w.r.t.~$\frakG$.
Here, $\psi: [0, \lambda_\mathrm{max}) \to \R$ is a ``CGF-like'' function~\citep{howard2020chernoff}, with a scale parameter $c > 0$, that controls how fast $S_t$ can grow relative to the intrinsic time $\hat{V}_t$. 
It is called a ``CGF-like'' function because it closely resembles (or equals) a cumulant generating function (CGF) of a mean-zero random variable.
In this paper, we use two $\psi$ functions:
\begin{itemize}
    \item $\psi_{N,c}(\lambda) = {c^2\lambda^2}/{2}, \; \forall \lambda \in [0, \infty)$, which is the CGF of a centered Gaussian with variance $c^2$;
    \item $\psi_{E,c}(\lambda) = c^{-2}(-\log(1 - c\lambda) - c\lambda), \; \forall \lambda \in [0, 1/c)$, which is a rescaled CGF of a centered Exponential with scale $c$.
\end{itemize}
If $L_t(\lambda)$ is a test supermartingale for each $\lambda \in [0, \lambda_\mathrm{max})$ for some $\psi$, then we say that $(S_t)_{t=0}^\infty$ is \emph{sub-$\psi$ with variance process $(\hat{V}_t)_{t=0}^\infty$}.
In particular, we say that $(S_t)_{t=0}^\infty$ is sub-Gaussian or sub-exponential, with variance process $(\hat{V}_t)_{t=0}^\infty$ and scale $c$, if it is sub-$\psi_{N,c}$ or sub-$\psi_{E,c}$ respectively; these generalize the definitions of sub-Gaussian and sub-exponential random variables to cumulative sums w.r.t.~intrinsic time.
The uniform boundary $u$ defined using $\psi$ is then called a \emph{sub-$\psi$ uniform boundary}.

Our goal is now to identify the conditions with which $(L_t(\lambda))_{t=0}^\infty$ is indeed a test supermartingale and use different $\psi$ functions to obtain different uniform boundaries and hence CSs.

\subsubsection{Warmup: Hoeffding-Style Confidence Sequences}

We first derive an illustrative example of a CS for $\Delta_t$ solely based on the sub-Gaussianity of the empirical pointwise score differentials $(\hat\delta_i)_{i=1}^\infty$. 
While the resulting CS is not the tightest one in our case, its derivation is simple enough to showcase the general pipeline for deriving CSs.

Recall the problem setup in Section~\ref{sec:setup}, and for each $i \geq 1$, consider two probability forecasts $p_i, q_i \in [0, 1]$ on a binary outcome $y_i \in \{0, 1\}$ with unknown mean $r_i \in [0, 1]$.
Since $p_i$, $q_i$, and $y_i$ are all bounded, we know that the pointwise score differentials $\hat\delta_i$ for $i \geq 1$ are also bounded for many of the scoring rules we've discussed (e.g., $|\hat\delta_i| \leq 1$ for the Brier, spherical, and zero-one scores).
If $|\hat\delta_i| \leq c$ for some $c > 0$, we know that $\hat\delta_i$ is $c$-sub-Gaussian~\citep{hoeffding1963probability}
conditioned on the game filtration $\calG_{i-1}$, meaning that $\mathbb{E}_{i-1}[e^{\lambda(\hat\delta_i - \delta_i)}] \leq e^{\lambda^2c^2/2} = \exp\{\psi_{N,c}(\lambda)\}$ for all $\lambda \in \R$. 

Now, for each $t$, define the cumulative sum $S_t = \sum_{i=1}^t (\hat\delta_i - \delta_i)$ and the intrinsic time $\hat{V}_t = \sum_{i=1}^t 1 = t$.
It then follows that, for each $\lambda \in [0, \infty)$, the exponential process $(L_t(\lambda))_{t=0}^\infty$ given by $L_t(\lambda) = \exp\{\lambda S_t - \psi_{N,c}(\lambda)\hat{V}_t\}$ is a test supermartingale:
\begin{align}
    \mathbb{E}_{t-1}[L_t(\lambda)]
    &= L_{t-1}(\lambda) \cdot \mathbb{E}_{t-1}\insquare{\exp\incurly{ \lambda \inparen{\hat\delta_t - \delta_t} - \psi_{N,c}(\lambda)}} \leq L_{t-1}(\lambda) .
\end{align}
Hence, there exists a sub-Gaussian uniform boundary for $(S_t, \hat{V}_t)$ such that the time-uniform guarantee in \eqref{eqn:uniform_boundary} holds.
By rearranging terms and also using the analogous argument for $(-S_t, \hat{V}_t)$, we arrive at our first CS.
Hereafter, the notation $(a \pm b)$ denotes the interval $(a - b, a + b)$.

\begin{theorem}[Hoeffding-style confidence sequences for $\Delta_t$]\label{thm:hoeffding}
Suppose that $\hat\delta_i$ is $c$-sub-Gaussian conditioned on $\calG_{i-1}$ for $i \geq 1$, for some $c \in (0, \infty)$.
Then, for any $\alpha \in (0, 1)$,
\begin{equation}\label{eqn:hoeffding}
    C_t^{\mathsf{H}} := \inparen{\hat\Delta_t \pm \frac{u(t)}{t}} \indent \text{forms a $(1-\alpha)$-CS for $\Delta_t$,}
\end{equation}
where $u = u_{\alpha/2, c}$ is any (one-sided) sub-Gaussian uniform boundary with crossing probability $\frac{\alpha}{2}$ and scale $c$ (or alternatively, a two-sided version with crossing probability $\alpha$ and scale $c$).
\end{theorem}

The statement~\eqref{eqn:hoeffding} is equivalent to saying that, with probability at least $1-\alpha$, $\Delta_t$ is contained in $C_t^{\mathsf{H}}$ \emph{for all time $t$}, or that $P(\forall t \geq 1: \Delta_t \in C_t^{\mathsf{H}}) \geq 1 - \alpha$.
This CS is called a Hoeffding-style CS, as it extends \citet{hoeffding1963probability}'s inequality for the sums of independent sub-Gaussian random variables to the sequential case.
In the sub-Gaussian case, it is also possible to construct a two-sided boundary without separately constructing a one-sided boundary.
This is due to a classical result by~\citet{robbins1970statistical} that we restate later in~\eqref{eqn:cm_normalmixture}, so the upper and lower confidence bounds need not be constructed separately; in practice, the one-sided and two-sided variants are nearly identical~\citep{howard2021timeuniform}.
We further discuss the possible choices of the uniform boundary in Section~\ref{sec:uniform_boundary}.

The condition for Theorem~\ref{thm:hoeffding} (and for Theorem~\ref{thm:main} that will follow shortly) is satisfied by many scoring rules for probability forecasts on binary or categorical outcomes, including the Brier, spherical, and zero-one scores. 
For the unbounded logarithmic score, one can use its truncated variant $S(p, y) = y\log(p \vee \epsilon) + (1-y) \log ((1 - p) \vee \epsilon)$ for some small $\epsilon > 0$; although the score is no longer proper, our methods remain valid.
The condition is also satisfied for scoring rules on bounded continuous outcomes, such as Brier and quantile scores on $[0,1]$-valued outcomes (See Section~\ref{sec:extensions}).

\subsubsection{Main Result: Empirical Bernstein Confidence Sequences}

Now we are ready to present our main result, which is the derivation of a tight CS for $\Delta_t$.
The key difference from the Hoeffding-style CS is that we now use an empirical estimate of the variance process for the cumulative sums, leading to a variance-adaptive CS that is often much tighter in practice.\footnote{The improvement from a Hoeffding-style CS to an empirical Bernstein CS mirrors the improvement from Hoeffding's inequality to empirical Bernstein's inequality for bounded random variables in the fixed-sample case.}
Recall the problem setup in Section~\ref{sec:setup} once again.
\begin{theorem}[Empirical Bernstein confidence sequences for $\Delta_t$]\label{thm:main}
Suppose that $|\hat\delta_i| \leq \frac{c}{2}$ for each $i \geq 1$, for some $c \in (0, \infty)$.
Also, let $\hat{V}_t = \sum_{i=1}^t (\hat\delta_i - \gamma_i)^2$, where $(\gamma_i)_{i=1}^\infty$ is any $\insquare{-\frac{c}{2}, \frac{c}{2}}$-valued predictable sequence w.r.t.~$\frakG$.
Then, for any $\alpha \in (0, 1)$,
\begin{equation}\label{eqn:main}
    C_t^{\mathsf{EB}} := \inparen{\hat\Delta_t \pm \frac{u(\hat{V}_t)}{t}} \indent \text{forms a $(1-\alpha)$-CS for $\Delta_t$,}
\end{equation}
where $u = u_{\alpha/2, c}$ is any sub-exponential uniform boundary with crossing probability $\frac{\alpha}{2}$ and scale $c$.
\end{theorem}

As before, the statement~\eqref{eqn:main} is equivalent to saying that, with probability at least $1-\alpha$, $\Delta_t$ is contained in $C_t^\mathsf{EB}$ \emph{for all time $t$}, or that $P\inparen{\forall t \geq 1: \Delta_t \in C_t^\mathsf{EB}} \geq 1 - \alpha$. 
The proof is provided in Section~\ref{sec:proof_main}.
Theorem~\ref{thm:main} (and its proof) can be viewed as an extension of Theorem~4 in~\citet{howard2021timeuniform} to our setup of sequential forecast comparison.

Like the Hoeffding-style CS in Theorem~\ref{thm:hoeffding}, the EB CS estimates the conditional predictive ability in an anytime-valid and distribution-free manner.
The EB CS is further variance-adaptive because its width is a function of the empirical variance process $(\hat{V}_t)_{t=0}^\infty$, and we illustrate this empirically in Section~\ref{sec:experiments}.
As before, we can use any bounded scoring rules, which in the binary and categorical cases include the Brier, spherical, and zero-one scores (proper), as well as the truncated logarithmic score (improper); scoring rules for bounded continuous outcomes can similarly be used.
In addition, for unbounded \emph{proper} scores for binary forecasts, such as the logarithmic score, we show in Section~\ref{sec:winkler} that a normalized version of the average score differential, due to \citet{winkler1994evaluating}, can be used.

The choice of the uniform boundary $u$ is discussed in the following subsection.
A reasonable choice for the predictable sequence $(\gamma_i)_{i=1}^\infty$ is the average of previous score differentials, i.e., $\gamma_i = \hat\Delta_{i-1}$, although a smarter choice may lead to tighter CS.
For the rest of this paper, our default choice of CS for $\Delta_t$ will be that of Theorem~\ref{thm:main}, using $\hat{V}_t = \sum_{i=1}^t (\hat\delta_i - \hat\Delta_{i-1})^2$, unless specified otherwise.

\subsubsection{Choosing the Uniform Boundary via the Method of Mixtures}\label{sec:uniform_boundary}

The specific choice of the uniform boundary $u$ controls the tightness of the CS across time, and an extensive list of choices for $u$ is covered in detail in \citet{howard2021timeuniform}. 
While the simplest uniform boundaries are given as linear functions of the intrinsic time~\citep{howard2020chernoff}, curved uniform boundaries can produce CSs that are tighter across time.
Here, we focus on a type of curved boundaries called the conjugate-mixture boundary; another option, called the polynomial stitching boundary, is also discussed in Section~\ref{sec:stitching}.
Either boundary type is applicable to both Theorems~\ref{thm:hoeffding} and~\ref{thm:main}.

The conjugate-mixture (CM) boundary~\citep{howard2021timeuniform}, denoted as $u_\alpha^{\mathsf{CM}}$, represents a class of uniform boundaries arising from the method of mixtures, the first instance of which was derived by \citet{darling1967confidence}.
The key idea is summarized as follows. 
Since $L_t(\lambda) = \exp\{\lambda S_t - \psi(\lambda) \hat{V}_t\}$ is a test supermartingale for every $\lambda \in [0, \lambda_\text{max})$, it follows that for any distribution $F$ on $[0, \lambda_\text{max})$, the mixture $L_t^\mathsf{mix} := \int L_t(\lambda) dF(\lambda)$ is also a test supermartingale.
Choosing $F$ to be conjugate (in the Bayesian sense) to $\psi$ then gives a closed-form expression for $L_t^\mathsf{mix}$.
For example, if $(S_t)_{t=0}^\infty$ is sub-Gaussian with $(\hat{V}_t)_{t=0}^\infty$ (Theorem~\ref{thm:hoeffding}), then choosing $F$ to be a Gaussian results in the \emph{normal mixture} boundary~\citep{robbins1970statistical}; if $(S_t)_{t=0}^\infty$ is sub-exponential with $(\hat{V}_t)_{t=0}^\infty$ (Theorem~\ref{thm:main}), then choosing $F$ as a Gamma results in a \emph{gamma-exponential mixture} boundary.

To elaborate, by Lemma 2 of~\citet{howard2021timeuniform}, if $L_t(\lambda)= \exp\{\lambda S_t - \psi(\lambda) \hat{V}_t\}$ is a test supermartingale for each $\lambda \in [0, \lambda_\text{max})$ and $F$ is any probability distribution on $[0, \lambda_\text{max})$, then the following function is a sub-$\psi$ uniform boundary with crossing probability $\alpha \in (0, 1)$:
\begin{equation}\label{eqn:cm_boundary}
   u_\alpha^{\mathsf{CM}}(v) := \sup \incurly{
   s \in \R: m(s, v) < \frac{1}{\alpha} 
   }, \quad v \geq 0,
\end{equation}
where $m(s, v) := \int \exp\incurly{\lambda s - \psi(\lambda) v} dF(\lambda)$.
Because $m(S_t, \hat{V}_t) = L_t^\textsf{mix}$ is a test supermartingale, Ville's inequality says that $P(\forall t \geq 1: m(S_t, \hat{V}_t) < 1/\alpha) \geq 1-\alpha$, which in turn implies that $P(\forall t \geq 1: S_t \leq u_\alpha^{\mathsf{CM}}(\hat{V}_t)) \geq 1 - \alpha$.
Similarly, if $(-S_t, \hat{V}_t)_{t=0}^\infty$ is also sub-$\psi$, then the above procedure also gives the lower bound on $S_t$.

Importantly, the uniform boundary~\eqref{eqn:cm_boundary} can be used for both Theorems~\ref{thm:hoeffding} and~\ref{thm:main}, with the choice of $F$ differing in each case.
For the Hoeffding-style CS in Theorem~\ref{thm:hoeffding}, a two-sided normal mixture boundary can be computed directly in closed-form by choosing $F$ to be $\calN(0, \rho^{-1})$~\citep{robbins1970statistical}:
\begin{equation}\label{eqn:cm_normalmixture}
    u_\alpha^{\mathsf{CM}}(v; \psi_N) = \sqrt{(v+\rho) \log \inparen{\frac{v+\rho}{\alpha^2\rho}}}
\end{equation}
where $\rho > 0$ is a free parameter. 
In practice, $\rho$ can be chosen to optimize the width of the resulting CS at a pre-specified intrinsic time.
A one-sided normal mixture boundary can also be derived in closed-form~\citep{howard2021timeuniform}.

For the EB CS in Theorem~\ref{thm:main}, a one-sided gamma-exponential mixture boundary $u_\alpha^{\mathsf{CM}}(v; \psi_E)$, with $F$ as a Gamma, can be computed efficiently using a numerical root finder ($m(s,v)$ has a closed form, and the boundary $u_\alpha^{\mathsf{CM}}$ is obtained numerically; see Section~\ref{sec:gm_mixture} for details).
The one-sided boundary can be used for computing both the upper and lower confidence bounds of the EB CS.
If a closed-form boundary is needed, then the polynomial stitching boundary (Section~\ref{sec:stitching}) can be used.
Also, while the CM boundary has an asymptotic rate of $O(\sqrt{v \log v})$ as illustrated in \eqref{eqn:cm_normalmixture}, it is usually tighter than the polynomial stitched boundary in practice. 
In fact, the CM boundary is unimprovable in the case of sub-Gaussian random variables without additional assumptions~\citep[Proposition 4]{howard2021timeuniform}.

\begin{table}[t]
    \centering
    \begin{tabular}{l|c|c|c}
        \Xhline{1.1pt} 
        \bf Type & \bf CS $C_t$ & \bf Intrinsic Time $\hat{V}_t$ & \bf Uniform Boundary $u$ \\ \hline
        \bf Hoeffding-Style & \multirow{2}{*}{$\inparen{\hat\Delta_t \pm \frac{u(\hat{V}_t)}{t}}$} & \multirow{2}{*}{$t$} & Normal Mixture \\
        (Theorem~\ref{thm:hoeffding}) & & & Polynomial Stitching \\
        \hline
        \bf Emp. Bernstein & \multirow{2}{*}{$\inparen{\hat\Delta_t \pm \frac{u(\hat{V}_t)}{t}}$} & $\sum_{i=1}^t (\hat\delta_i - \gamma_i)^2$, & Gamma-Exponential Mixture \\
        (Theorem~\ref{thm:main})   & & $(\gamma_i)_{i=1}^\infty$ predictable & Polynomial Stitching \\
        \Xhline{1.1pt} 
    \end{tabular}
    \caption{Summary of confidence sequences and their uniform boundary choices.}
    \label{tbl:cs_boundary}
\end{table}

Table~\ref{tbl:cs_boundary} summarizes the choice of uniform boundaries and the CSs we derived for estimating $\Delta_t$.
In our experiments, we use the conjugate-mixture uniform boundary by default, although we also perform an empirical comparison between the different choices as well as their hyperparameters in Section~\ref{sec:iid_mean}.
We use the publicly available implementation of the polynomial stitching and CM uniform boundaries by \citet{howard2021timeuniform}.\footnote{\url{https://github.com/gostevehoward/confseq}}

\subsection{Sequential Tests, e-Processes and p-Processes}\label{sec:eprocess}

While our derivation so far has focused on CSs, we can also derive e-processes and p-processes~\citep{shafer2019game,vovk2021evalues,grunwald2019safe,ramdas2020admissible}.
In particular, an e-process can be derived as a lower bound on the exponential test supermartingale~\eqref{eqn:expm} that we used to construct the CS in the previous section. 
This correspondence is general to any exponential process upper-bounded by a test supermartingale, as noted in, e.g., \citet{ramdas2020admissible,howard2021timeuniform}; our work utilizes this fact to introduce alternative sequential inference procedures with the same anytime-valid and distribution-free guarantees.

\paragraph{Weak and Strong Null Hypotheses.} Before deriving e- and p-processes, we first make clear the null hypotheses that correspond to the CS derived in Theorem~\ref{thm:main}.
We define the \emph{weak one-sided null} $\calH_0^\sfw(p, q)$ as
\begin{equation}\label{eqn:weak_null}
    \calH_0^\sfw(p, q): \Delta_t = \frac{1}{t} \sum_{i=1}^t \delta_i \leq 0, \quad\forall t = 1, 2, \dotsc.
\end{equation}
$\calH_0^\sfw(p, q)$ implies that, across all times $t$, the first forecaster ($p$) is no better than the second forecaster ($q$) \emph{on average}.
Note that $\calH_0^\sfw(p, q)$ is a composite null, in the sense that it consists of all joint distributions $P$ on $\frakG$ such that $\Delta_t \leq 0$ for all $t \geq 1$ under $P$.
$\calH_0^\sfw(q, p)$ is analogously defined as $\calH_0^\sfw(q, p): \Delta_t = \frac{1}{t}\sum_{i=1}^t \delta_i \geq 0$.

We now illustrate how the CSs derived in Theorem~\ref{thm:hoeffding} and Theorem~\ref{thm:main} would correspond to sequential tests of the weak one-sided nulls $\calH_0^\sfw(p, q)$ and $\calH_0^\sfw(q, p)$, drawing from the duality between CSs and sequential tests~\citep{johari2021always,howard2021timeuniform,ramdas2020admissible}.
Specifically, because the upper and lower confidence bounds are often constructed separately, the $(1-\alpha)$-level CS for $\Delta_t$ denoted as $C_t = (L_t, U_t)$ satisfies $\Delta_t \leq U_t$ with probability at least $1 - \frac{\alpha}{2}$ \emph{and} that $\Delta_t \geq L_t$ with probability at least $1 - \frac{\alpha}{2}$.
Thus, if for any time $t$ we find that $L_t > 0$ or $U_t < 0$, then we can reject either $\calH_0^\sfw(p, q)$ or $\calH_0^\sfw(q, p)$ with high probability.
More generally, the CSs readily provide a valid stopping rule for rejecting $\calH_0^\sfw$, a fact that we summarize in the following corollary. Below, we follow Robbins' power-one testing framework which uses one-sided stopping rules that only stop on rejecting the null (and do not stop otherwise).

\begin{corollary}[A sequential test for $\calH_0^\sfw$ using a CS]\label{cor:sequential_test}
Given a $(1-\alpha)$-CS $C_t = (L_t, U_t)$ obtained using either Theorem~\ref{thm:hoeffding} or~\ref{thm:main}, the following stopping rule provides a valid level-$\alpha$ sequential test for $\calH_0^\sfw(p,q)$ and $\calH_0^\sfw(q, p)$ (jointly):
\begin{align}\label{eqn:cs_stopping}
    \textit{
    Reject $\calH_0^\sfw(p, q)$ if $L_t > 0$; reject $\calH_0^\sfw(q, p)$ if $U_t < 0$.
    }
\end{align}
This means that:
\begin{align}
\sup_{P \in \calH_0^\sfw(p, q)} P(\exists t \geq 1: \text{Reject }\calH_0^\sfw(p, q)) + \sup_{P \in \calH_0^\sfw(q, p)} P(\exists t \geq 1: \text{Reject }\calH_0^\sfw(q, p))\leq \alpha.
\end{align}
\end{corollary}
The stopping rule~\eqref{eqn:cs_stopping} is equivalent to \emph{deciding that $p$ has been better (worse) than $q$ if $C_t$ is entirely above (below) zero.}
The anytime-validity of this rule implies that the statistician can, e.g., periodically perform the test as $t$ increases and update their decision accordingly.
On one extreme, the statistician can choose to perform the test after every round $t$, or on the other extreme, they can test just once at a designated time $t^*$ (while leaving open the possibility of revisiting the experiment some time later).
Compared to a standard hypothesis test for a stationary mean, the underlying $\Delta_t$ can change its course over time, so in general it may not be sufficient to test once at $t^*$ in order to have power against the weak null. 
See Section~\ref{sec:experiments} for an illustration and Section~\ref{sec:discussion} for a further discussion.

We note that separately testing for both $\calH_0^\sfw(p,q)$ and $\calH_0^\sfw(q, p)$ is not equivalent to simply testing for $\Delta_t = 0,\: \forall t$, which is equivalent to $\delta_t = 0,\: \forall t$.
Rather, the sequential test~\eqref{eqn:cs_stopping} is the combination of two separate sequential tests in~\eqref{eqn:cs_stopping} for $\calH_0^\sfw(p, q)$ and $\calH_0^\sfw(q, p)$, each at the significance level $\alpha/2$.
The interpretation of the CS as two simultaneous sequential tests allows the user to continuously monitor the score differential on both sides via the CS-based stopping rule~\eqref{eqn:cs_stopping}.

For the sake of comparison, we also define the \emph{strong one-sided null} $\calH_0^\sfs = \calH_0^\sfs(p, q)$ as
\begin{equation}\label{eqn:strong_null}
    \calH_0^\sfs(p, q): \delta_t \leq 0, \quad \forall t = 1, 2, \dotsc.
\end{equation}
$\calH_0^\sfs(q, p)$ is defined analogously as $\calH_0^\sfs(q, p) : \delta_t \geq 0, \; \forall t = 1, 2, \dotsc$.
The recent work by~\citet{henzi2021valid} develops e-processes (defined in the next paragraph) and sequential tests for this null.
In contrast to $\calH_0^\sfw$, $\calH_0^\sfs$ corresponds to saying that the first forecaster ($p$) is no better than the second forecaster ($q$) at \emph{every} time step $t = 1, 2, \dotsc$. 
Thus, the strong null $\calH_0^\sfs$ implies the weak null $\calH_0^\sfw$, but not vice versa.
The critical distinction here is that rejecting $\calH_0^\sfs$ only tells us that $p$ outperformed $q$ at \emph{some} time step $t$, but it does not tell us if either was better on average over time. 
To give a concrete example, fix $k>2$ (say, $k=7$ indicating Sundays), and define
\begin{equation}\label{eqn:strong_vs_weak_example}
    \delta_t = +0.1 \text{ if } t = k, 2k, 3k, \dotsc; \quad \delta_t = -1 \text{ otherwise. }
\end{equation}
In other words, $p$ is generally worse than $q$ but marginally better than $q$ every $k$th time step (e.g., every Sunday).
Because the strong null is false, any (powerful) sequential test for the strong null will reject it, and yet this may be a confusing conclusion as $q$ is generally a better forecaster.

\paragraph{Sub-exponential E-processes for the Weak Null.} 

We now show that the exponential test supermartingale underlying the CS in Theorem~\ref{thm:main} can also be transformed to directly measure evidence against the weak one-sided null (rather than make a decision at a level $\alpha$).
Formally, an \emph{e-process}~\citep{ramdas2021testing} for a (possibly composite) null hypothesis $\calH_0$ is defined as a nonnegative process $(E_t)_{t=0}^\infty$, starting at one ($E_0=1$), such that:
\begin{equation}\label{eqn:eprocess_defn}
    \text{for any $P \in \calH_0$ and any arbitrary stopping time $\tau$,} \quad \mathbb{E}_{P}[E_\tau] \leq 1 ,
\end{equation}
where we define $E_\infty := \limsup_{t\to\infty} E_t$.
The larger the value of $E_t$, the more the evidence against the null. In particular, if the null is true, then it is unlikely to observe large values of the process at any stopping times (by Markov's inequality, $P(E_\tau 
\geq 1/\alpha) \leq \alpha$).
An e-process is anytime-valid by definition~\eqref{eqn:eprocess_defn} (validity at arbitrary stopping times), analogous to the anytime-validity of a CS in Equation~\ref{eqn:cs_anytime_valid}, and the term `process' is also used to emphasize this property.
An e-process can also be interpreted in a fully game-theoretic statistical sense: an e-process for a composite null measures the \emph{minimum} wealth among bets against each member of the null~\citep{ramdas2021testing}, such that it only grows large when there is evidence against all members.
At a fixed $t$, $E_t$ is also called an e-variable, and its realization is called an e-value~\citep{vovk2021evalues,grunwald2019safe}.

We can now define and show an e-process that corresponds to Theorem~\ref{thm:main}.
(We can also define an analogous e-process corresponding to Theorem~\ref{thm:hoeffding}, but this is omitted due to space constraints.)
The following e-process is for the weak one-sided null $\calH_0^\sfw(p,q)$ and is related to the lower confidence bound of the CS from Theorem~\ref{thm:main}; the e-process for $\calH_0^\sfw(q,p)$ is analogous and related to the upper confidence bound of the CS.
Recall once again the problem setup in Section~\ref{sec:setup}.
\begin{theorem}[Sub-exponential E-processes for $\calH_0^\sfw$]\label{thm:eprocess}
Assume the same conditions as Theorem~\ref{thm:main}. 
Then, for each $\lambda \in [0, 1/c)$, 
\begin{equation}\label{eqn:eprocess}
    E_t(\lambda) := \exp\incurly{ \lambda \sum_{i=1}^t \hat\delta_i - \psi_{E,c}(\lambda) \hat{V}_t } \indent \text{is an e-process for $\calH_0^\sfw(p, q)$}.
\end{equation}
Furthermore, given a probability distribution $F$ on $[0, 1/c)$, the mixture process $E^\textsf{mix}_t := \int E_t(\lambda) dF(\lambda)$ is an e-process for $\calH_0^\sfw(p, q)$.
\end{theorem}
The proof, provided in Section~\ref{sec:proof_eprocess}, shows that under each $P \in \calH_0^\sfw$, $E_t(\lambda)$ is upper-bounded by a exponential test supermartingale for $P$, namely $L_t(\lambda)$ in \eqref{eqn:expm}.
Because a process is upper-bounded by a test supermartingale for $P \in \calH_0$ if and only if it is an e-process for $\calH_0$~\citep{ramdas2020admissible}, this establishes that $E_t(\lambda)$ is an e-process in the sense of~\eqref{eqn:eprocess_defn}.
It then follows that $E^\textsf{mix}_t \leq \int L_t(\lambda) dF(\lambda) = L_t^\textsf{mix} \; \forall t$, so $E_t^\textsf{mix}$ is also an e-process.

The e-process of Theorem~\ref{thm:eprocess} is an anytime-valid inference procedure that provides a measure of accumulated evidence against the weak one-sided null $\calH_0^\sfw(p, q)$ at any stopping time.
By definition, it is expected to be small under the weak null, and we only expect to see it grow large when the weak null does not hold.
In comparison with \citet{henzi2021valid}'s e-process for the \emph{strong} null, we see that our e-process provides a more useful notion of evidence for saying that one forecaster outperforms another.
In the example of~\eqref{eqn:strong_vs_weak_example}, an e-process for the strong null can grow large, even though $q$ is generally a better forecaster; in contrast, our e-process~\eqref{eqn:eprocess} for the weak null is expected to remain small.
In Section~\ref{sec:weather}, we provide an empirical comparison of the two e-processes.

\paragraph{Choosing $\lambda$ (or $F$) for E-processes.}
Theorem~\ref{thm:eprocess} tells us that the expected value of $E_t(\lambda)$ and $E_t^\mathsf{mix}$ are bounded by 1 at all stopping times under the null, for any choice of $\lambda$ or any mixture distribution $F$.
In practice, we default to using a mixture e-process with the conjugate distribution $F$, as in Section~\ref{sec:uniform_boundary}. 
For the sub-exponential e-process, the gamma-exponential mixture as before provides a closed form for the function $m(s,v)$ in~\eqref{eqn:cm_boundary}, so that $E_t^\mathsf{mix} = m(\sum_{i=1}^t\hat\delta_i, \hat{V}_t)$ can be computed efficiently.
The expression for $m(s,v)$ is included in Section~\ref{sec:gm_mixture}.

\paragraph{P-processes.} Finally, we remark that any e-process for $\calH_0$ can also be converted into an \emph{p-process} for $\calH_0$, i.e., the sequence $(\sfp_t)_{t=0}^\infty$ that satisfies: for any $\alpha \in (0, 1)$,
\begin{equation}\label{eqn:pprocess_defn}
    \text{for any $P \in \calH_0$ and for any arbitrary stopping time $\tau$,} \quad P (\sfp_\tau \leq \alpha) \leq \alpha .
\end{equation}
A p-process evaluated at any stopping time $\tau$, i.e. $\sfp_{\tau}$, is a p-value, but unlike a classical p-value, a p-process is valid at arbitrary stopping times.

Any e-process $(E_t)_{t=0}^\infty$ can be converted into a p-process via
\begin{equation}\label{eqn:pprocess}
    \sfp_t := 1/\sup_{i \leq t} E_i , \quad \forall t,
\end{equation}
following derivations from, e.g., \citet{ramdas2020admissible,ramdas2021testing}.
We also remark that $\sfp_t$ can alternatively be defined from a CS as the smallest $\alpha$ for which the $(1-\alpha)$-level CS does not include zero~\citep{howard2021timeuniform}, so all three notions (CS, e-process, and p-process) are closely related.

\section{Experiments}\label{sec:experiments}

In this section, we run both simulated and real-data experiments for sequential forecast comparison using our CSs as well as e-processes.
All code and data sources for the experiments are made publicly available online at \url{https://github.com/yjchoe/ComparingForecasters}.

\subsection{Numerical Simulations}\label{sec:simulated}

As our first experiment, we compare our Hoeffding-style and EB CSs (Theorems~\ref{thm:hoeffding} and~\ref{thm:main}, respectively) on simulated data with the asymptotic fixed-time CIs due to Theorem~2 of~\citet{lai2011evaluating}.
The main goal is to confirm that the CSs cover time-varying average score differentials uniformly, unlike the fixed-time CI, and are also nearly as tight as the CI.

In our simulated experiments, we also include an asymptotic CS for time-varying means, recently developed by \citet{waudbysmith2021doubly}, as an additional tool for anytime-valid inference.
Asymptotic CSs can be viewed as alternatives to their non-asymptotic counterparts, including the ones we introduced in Section~\ref{sec:cs}, and they trade off non-asymptotic validity to achieve versatility and also comparatively smaller widths at smaller sample sizes.
A formal review of asymptotic CSs in the context of sequential forecast comparison is included in Section~\ref{sec:asympcs}.

\begin{figure}[t]
    \centering
    \includegraphics[width=\textwidth]{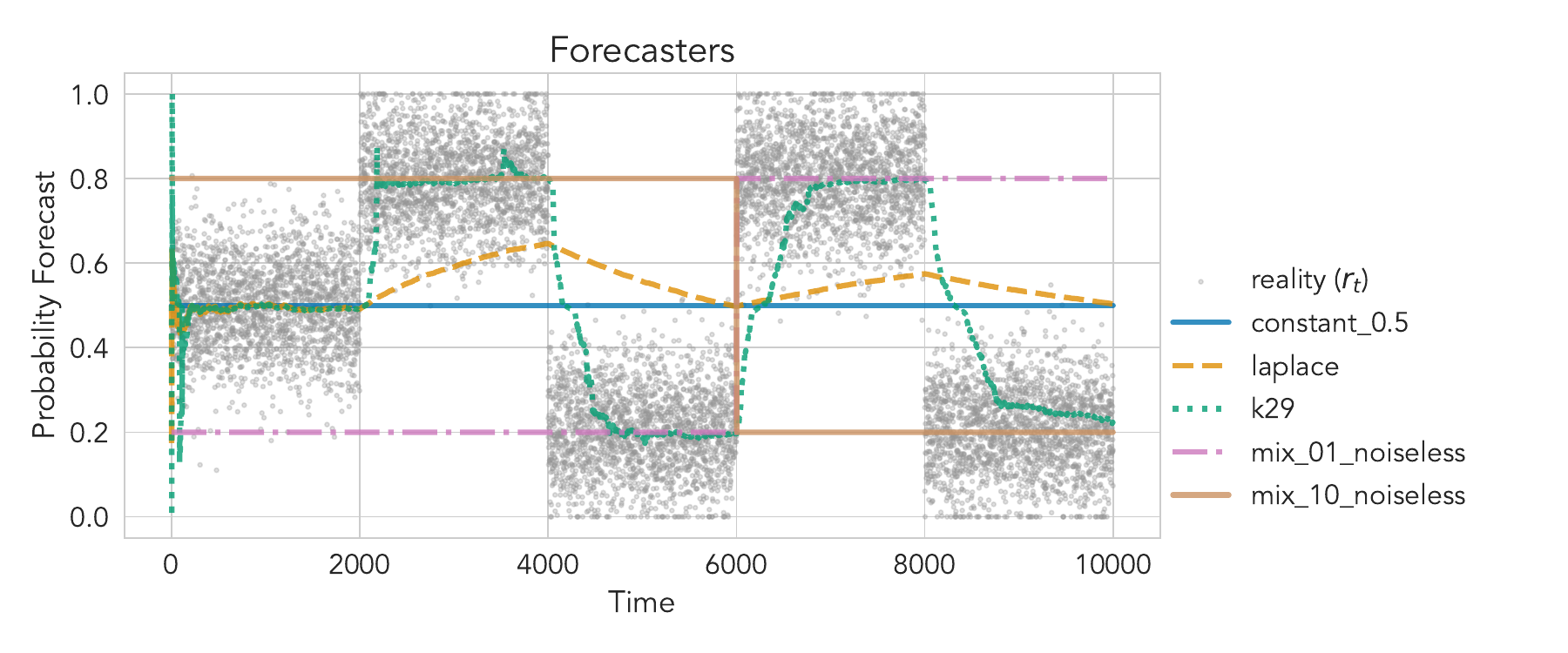}
    \caption{Various forecasters on a simulated non-IID data ($T=10^4$) with sharp changepoints across time. Note that, instead of plotting the binary outcomes $y_t \in \{0, 1\}$, we plot the Reality's choices $(r_t)_{t=1}^T$ that generates the outcome sequence. See text for details about the forecasters.}
    \label{fig:forecasters}
\end{figure}

As for our simulated data, we generate a sequence of non-IID binary outcomes and compare different forecasters using our CSs.
The overall simulation pipeline closely follows Game~\ref{game:general}, with $\calP = \Delta(\calY) = [0, 1]$, $\calY = \{0, 1\}$, and $T = 10^4$.
At each round $t = 1, \dotsc, T$, each forecaster makes a probability forecast $p_t, q_t \in \calP$, then reality chooses $r_t$, and finally $y_t \sim \mathrm{Bernoulli}(r_t)$ is sampled.
The forecasts $p_t$ and $q_t$ are made only using the previous outcomes, i.e., $y_1, \dotsc, y_{t-1}$.
The Reality's choices $(r_t)_{t=1}^T$ is specifically chosen to be non-IID and contain sharp changepoints, as shown in Figure~\ref{fig:forecasters}.
This serves as a challenging test case for the EB CS, as the sharp changepoints make it difficult to quickly adapt to the underlying variance.
See Section~\ref{sec:simulated_details} for further details.

At the end of each round $t = 1, \dotsc, T$, we compute the 95\% Hoeffding-style and EB CS for $\Delta_t$, using Theorems~\ref{thm:hoeffding} and~\ref{thm:main} respectively.
We use the Brier score $S(p, q) = 1 - (p - q)^2$ as our default scoring rule, but we also explore other scoring rules later in the section.
As for the hyperparameter choices for sub-$\psi$ uniform boundaries, we are guided by preliminary experiments in Section~\ref{sec:iid_mean}.

We consider several forecasters, which are drawn with lines in Figure~\ref{fig:forecasters}. 
These include the constant baseline, i.e., $p_t = 0.5$ (\texttt{constant\_0.5}), as well as the Laplace forecasting algorithm (\texttt{laplace}) $p_t = \frac{k+0.5}{t+1}$, where $k = \#\{i \in [t]: y_i = 1\}$. 
We further add predictions using the K29 defensive forecasting algorithm  (\texttt{k29})~\citep{vovk2005defensive}, which is a game-theoretic forecasting method that yields calibrated forecasts.
The method depends on the choice of a kernel function, and here we use the Gaussian RBF $K(p, q) = \exp\inparen{-\frac{(p-q)^2}{2\sigma^2}}$ with bandwidth $\sigma=0.01$.
The \texttt{mix\_01\_noiseless} forecaster is defined as $p_t = 0.8$ for $t \leq 6000$ and $p_t = 0.2$ for $t > 6000$; the \texttt{mix\_01} forecaster is a noisy version that adds an independent noise to $p_t$ by $\tilde{p_t} = p_t + 0.5 \cdot \epsilon_t$ (clipped at 0 and 1), where $\epsilon_t$ is drawn IID from Student's $t$-distribution with 1 degree of freedom. 
The \texttt{mix\_10\_noiseless} forecaster is defined as $q_t = 1-p_t$ and the \texttt{mix\_10} forecaster $\tilde{q}_t$ is analogously defined.

The choices of forecasters and Reality are made in such a way that the unknown parameter $\Delta_t$, for $t = 1, \dotsc, T$, can not only change its sign but also have different variances over time.
For example, the \texttt{mix\_10} forecaster outperforms ($\Delta_t > 0$) the \texttt{mix\_01} forecaster on average during $t \in (2000, 6000)$, while the sign then reverses ($\Delta_t < 0$) for $t \in (6000, 10000)$.
Among the algorithmic forecasters, the K29 variants consistently perform better than the Laplace algorithm, especially when using sharper kernels, because they are better at modeling the sharp changepoints over time.

\begin{figure}[t]
    \centering
    \includegraphics[width=0.9\textwidth]{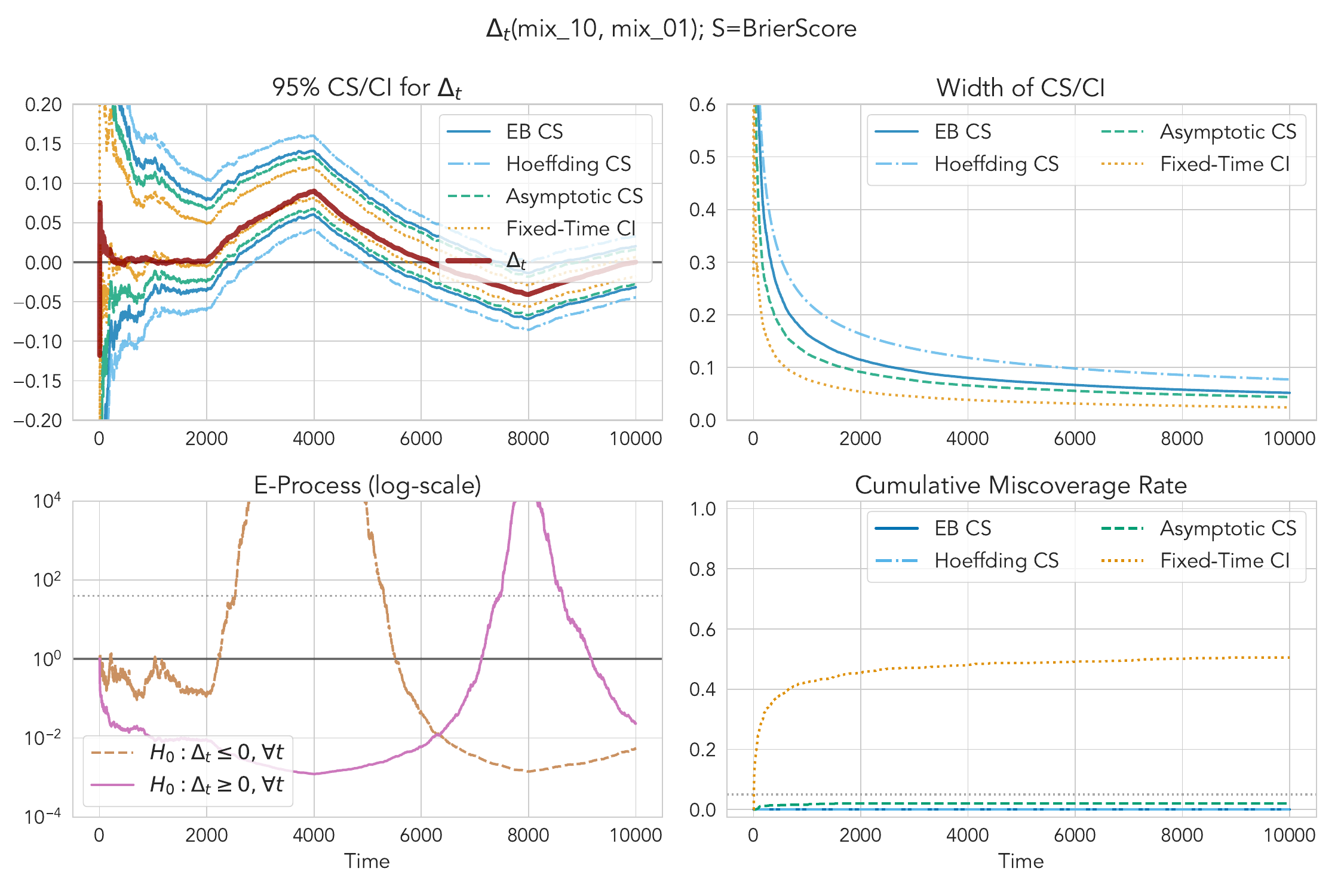}
    \caption{\emph{Top Left:} 95\% EB CS (blue, solid), Hoeffding-style CS (skyblue, dash-dotted), asymptotic CS (green, dashed; Section~\ref{sec:asympcs}), and a fixed-time asymptotic CI (orange, dotted) for simulated time-varying average score differentials $(\Delta_t)_{t=1}^T$ between the \texttt{mix\_10} and \texttt{mix\_01} forecasters ($T = 10^4$).
    The Brier score is used.
    \emph{All CSs, but not the CI, uniformly cover the true score differential sequence, which changes signs sharply multiple times across the horizon.}
    \emph{Top Right:} Widths of the CSs and the CI across time steps.
    The variance-adaptive EB CS is tighter than the Hoeffding CS and slightly looser than the asymptotic CS; the fixed-time CI is the tightest, but it does not have the time-uniform guarantee.
    \emph{Bottom Left:} Sub-exponential e-processes (Theorem~\ref{thm:eprocess}) that measure the accumulated evidence against either forecaster (first forecaster: brown, dashed; second: purple, solid). 
    Testing whether the e-process exceeds the dashed gray line at $2/0.05 = 40$ corresponds to a sequential test at $\alpha=0.05$ (Corollary~\ref{cor:sequential_test}).
    \emph{Bottom Right:} The cumulative miscoverage rate, which estimates $\alpha_t = P\inparen{\exists i \leq t: \Delta_i \notin C_i}$ over repeated sampling of $y_1, \dotsc, y_t$ under $P$, of the CSs/CIs.
    For a 95\% CS, this rate is controlled at 0.05 by definition; it is in fact always zero for the non-asymptotic CSs in our experiments.
    For the fixed-time CI, this rate exceeds well above $\alpha$ and continues to increase (in log-scale of time).
    }
    \label{fig:confseq_illustrative}
\end{figure}

In Figure~\ref{fig:confseq_illustrative}, we plot the 95\% Hoeffding-style CS (Theorem~\ref{thm:hoeffding}), EB CS (Theorem~\ref{thm:main}), and a fixed-time CI for $\Delta_t$ (top left), as well as their widths (top right), the corresponding e-process (bottom left), and the cumulative miscoverage rates (bottom right).
First, both CSs successfully cover $\Delta_t$ at any given time point, and their widths decrease as more outcomes are observed. 
As expected, the width of the EB CS decays more quickly than the width of the Hoeffding CS due to its use of the empirical variance term ($\hat{V}_t$) but more slowly than the fixed-time CI, matching the patterns observed in \citet{howard2021timeuniform,waudbysmith2021doubly}.
As noted before, the fixed-time CI is only valid at a fixed time $t$ and not uniformly over time, despite its tighter width, and this is illustrated by its large cumulative miscoverage rate, i.e., $\alpha_t = P\inparen{\exists i \leq t: \Delta_i \notin C_i}$ (estimated over the repeated sampling of $y_1, \dotsc, y_t$ under $P$).
In contrast, the EB CS\footnote{The EB CS is computed with the polynomial stitching bound for computational efficiency.} keeps its cumulative miscoverage rate well below $\alpha$ (it is in fact zero, as it is constructed using supermartingales and not martingales).
In Section~\ref{sec:dmgw}, we also include an analogous plot comparing our methods with other classical tests~\citep{diebold1995comparing,giacomini2006tests}.

The sub-exponential e-processes for $\calH_0(p, q)$ (solid green) and $\calH_0(q, p)$ (dotted purple) show how they accurately track the accumulated evidence for/against each forecaster over time. 
For example, the e-process for $\calH_0(p, q)$ stays below 1 during $t < 2000$, when neither forecaster outperforms the other, and grows large during $t \in (2000, 6000)$ when data shows more evidence against the null hypothesis that $\Delta_t \leq 0,\; \forall t$ because the true $\Delta_t$ in fact becomes positive.
It then decreases back to values below 1 during $t \in (6000, 10000)$, when the true $\Delta_t$ becomes negative.
We note that the gray dotted line indicates the value $2/\alpha = 40$; testing whether an e-process exceeds $2/\alpha$ corresponds to a level-$(\alpha/2)$ sequential test equivalent to the one stated in Corollary~\ref{cor:sequential_test}.
In fact, the plots show that the points at which the $(1-\alpha)$-level EB CS excludes zero (on either side) are precisely when either e-process exceeds $2/\alpha$, illustrating the duality between the CS and the e-process.

\begin{figure}[t]
    \centering
    \includegraphics[width=0.9\textwidth]{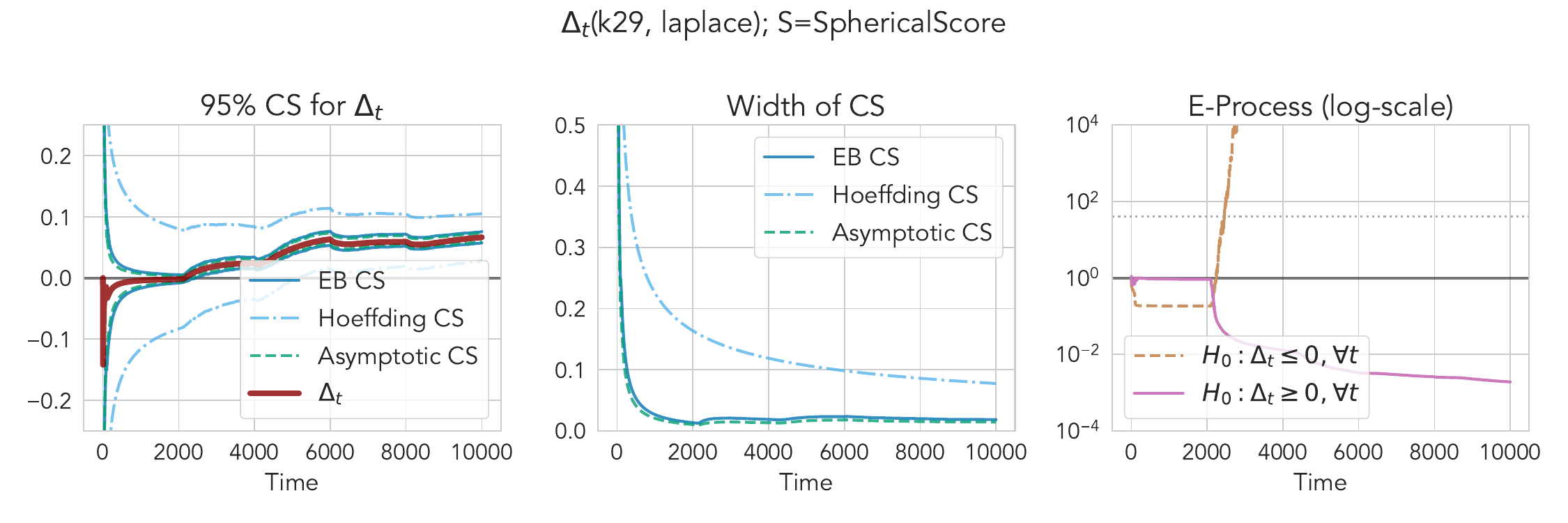}
    \includegraphics[width=0.9\textwidth]{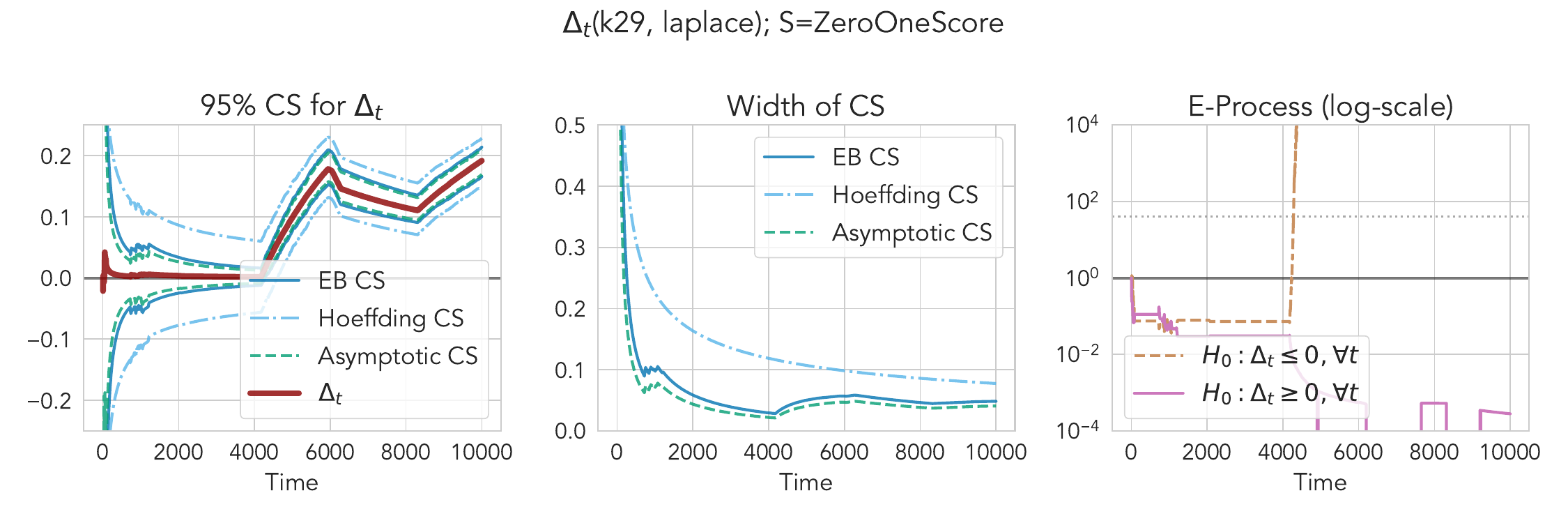}
    \includegraphics[width=0.9\textwidth]{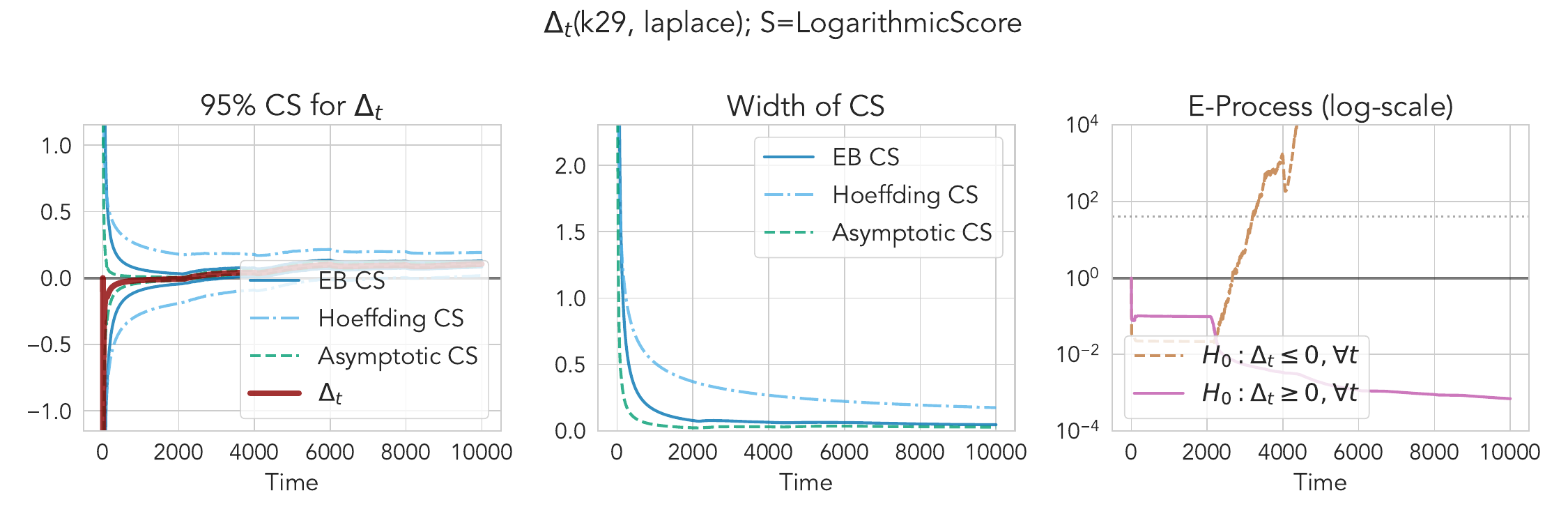}
    \caption{95\% EB (blue, solid), Hoeffding-style (skyblue, dash-dotted), and asymptotic (green, dashed) CSs (left), their widths (middle), and the sub-exponential e-processes (right) between the K29 forecaster and the Laplace forecaster.
    Three different scoring rules are used here: the spherical (top), the zero-one (middle), and the $\epsilon$-truncated logarithmic ($\epsilon=0.01$) (bottom) scores. 
    All scoring rules are positively oriented, such that positive values of $\Delta_t$ indicate that the first forecaster is better than the second.
    Even when the scoring rule is not strictly proper (zero-one) or not proper at all (truncated logarithmic), all CSs still cover $\Delta_t$ uniformly, and in general the width of the EB CS shrinks close to the asymptotic CS than the Hoeffding-style CS, which is wider.
    The e-processes for $\calH_0^\sfw: \Delta_t \leq 0$ (brown, dashed) cross the $2/\alpha$ line (gray, dotted) as the lower confidence bound of the EB CS crosses zero.
    }
    \label{fig:scoring_rules}
\end{figure}

In Figure~\ref{fig:scoring_rules}, we now plot the 95\% CSs (left), their widths (middle), and also the corresponding e-processes (right) for comparing the \texttt{k29\_poly3} forecaster against the \texttt{laplace} baseline, using the spherical score (strictly proper), zero-one score (proper), the $\epsilon$-truncated logarithmic score ($\epsilon=10^{-8}$) (improper).
We observe that all variants of CSs always cover the true $\Delta_t$ over time, at $\alpha=0.05$, and its width decreases similarly to the case of Brier scores and eventually approaches that of the asymptotic CS. 
In terms of the width comparison between EB and Hoeffding CSs, we see that the EB CS is generally much tighter than the Hoeffding CS, and it decreases more slowly around time steps when there are sharp changepoints in $\Delta_t$.
This can be explained by the variance-adaptive nature of the EB CS, which would use larger values of intrinsic time $\hat{V}_t$ at sharp changepoints, whereas the Hoeffding CS simply uses $\hat{V}_t = t$ irrespective of the variance process. 
The sub-exponential e-processes for $\calH_0^\sfw(p,q)$ and $\calH_0^\sfw(q, p)$ illustrate the accumulated evidence for the first forecaster in all three cases around the same time the CS moves entirely above zero, illustrating the duality between the two methods.

We include a plot of all pairwise comparisons between four of the forecasters in Section~\ref{sec:pairwise}.

\clearpage

\clearpage  

\subsection{Comparing Forecasters on Major League Baseball Games}\label{sec:2019ws}

As our first real-world application of the CSs, we consider the problem of predicting wins and losses for baseball games played in the Major League Baseball (MLB).
Sports game prediction is particularly suitable for our setting, because there are multiple publicly available probability forecasts on the outcome of each game (e.g., FiveThirtyEight, betting odds, and pundits/experts), that are frequently updated across time.
There is also no obvious assumption to be reasonably made about the outcome of the games, such as stationarity or assumptions of parametric models.
Recall Table~\ref{tbl:2019ws} for an illustration of various probability forecasts made on MLB games.

We specifically focus on predicting the outcome of MLB games over ten years (2010-2019), culminating in the 2019 World Series between the Houston Astros and the Washington Nationals.
We use every regular season and postseason MLB game from 2010 to 2019 as our dataset.
We convert each game as a single time point in chronological order, leading to a total of $T=25,165$ games. 
As for the forecasters, we consider the following:
\begin{itemize}
    \setlength\itemsep{0em}
    \item \texttt{538}: Game-by-game probability forecasts by FiveThirtyEight on every MLB game since 1871, available at \url{https://data.fivethirtyeight.com/#mlb-elo}. 
    \item \texttt{vegas}: Pre-game closing odds made on each game by online sports bettors, converted and scaled to probabilities, as reported by \url{https://Vegas-Odds.com}.\footnote{\url{https://sports-statistics.com/sports-data/mlb-historical-odds-scores-datasets/}}
    \item \texttt{constant}: a constant baseline corresponding to $p_t = 0.5$ for each $t$.
    \item \texttt{laplace}: A seasonally adjusted Laplace algorithm, representing the season win percentage for each team. 
    The final adjust win percentage from the previous season, reverted to the mean by one-third, is used as the baseline probability for the next season.
    The final probability forecast for a game between two teams is rescaled to sum to 1.
    \item \texttt{k29}: The K29 algorithm applied to each team, using the Gaussian kernel with $\sigma=0.1$, computed using data from the current season only. 
    The final probability forecast for a game between two teams is rescaled to sum to 1.
\end{itemize}
In Section~\ref{sec:mlb_forecasters}, we give further details about the five forecasters and also plot their forecasts on the last 200 games of 2019. 

We perform all pairwise comparisons of the five aforementioned forecasters on the 10-year win/loss predictions. 
See Sections~\ref{sec:iid_mean} for details on tuning the free hyperparameter on the uniform boundary.
First, as we showed in Figure~\ref{fig:figure1}, we compare the two publicly available forecasters in \texttt{538} ($p$) and \texttt{vegas} ($q$), finding that the \texttt{vegas} forecaster has marginally outperformed the \texttt{538} forecaster: after $T=25,165$ games, 95\% EB CS for $\Delta_T$ is $(-0.00265, -0.00062)$, and the e-value for $\calH_0^\sfw(q,p): \Delta_t \geq 0,\;\forall t$ is $2979.0$.
The fact that the \texttt{vegas} forecaster (marginally) outperformed the \texttt{538} forecaster is interesting, especially given that the primary goal of sports bettors is not to maximize predictive accuracy but their overall profit.\footnote{\url{https://fivethirtyeight.com/features/the-imperfect-pursuit-of-a-perfect-baseball-forecast/}}
Yet, given the relatively small score difference and also the inherent uncertainty in sports game outcomes,\footnote{\url{https://projects.fivethirtyeight.com/checking-our-work/mlb-games/}} more fine-grained comparisons between real-world sports forecasters (e.g., regular season vs. playoffs, team-specific comparisons, and comparisons with or without specific side information) remain interesting future work.

In Table~\ref{tbl:mlb_comparison}, we further compare every other forecaster against the \texttt{vegas} forecaster by estimating the average Brier score differential $\Delta_T$ using the 95\% EB CS.
We also show the corresponding sub-exponential e-processes (Theorem~\ref{thm:eprocess}) for the null of $\calH_0^\sfw(q, p): \Delta_t \geq 0, \; \forall t$, which translates to saying that \texttt{vegas} is not assumed to be better under the null, evaluated at time $T$.
Furthermore, we include comparisons involving the logarithmic score, namely via the average Winkler score $W_T(p,q)$ (Proposition~\ref{ppn:winkler}, Section~\ref{sec:winkler}) that quantifies the relative ``skill'' of forecasters~\citep{winkler1994evaluating,lai2011evaluating} as measured by a scoring rule (the logarithmic score, in this case).
The Winkler score approach allows us to utilize unbounded proper scoring rules, such as the logarithmic score, when dealing with binary outcomes.
Because the score is normalized and thus always maximized at 1, we can construct a one-sided CS with an upper confidence bound (UCB), and also construct an e-process against the null $\calH_0^{\sfw\sfw}: W_t \geq 0,\;\forall t$.
A negative UCB or a high value in the e-process indicates that $p$ is significantly worse than $q$ in relative skill.

Our results show that none of the other forecasters, including the \texttt{538} forecaster, have outperformed \texttt{vegas}, both in terms of the Brier score and the Winkler-logarithmic score.

We include a plot of all pairwise comparisons between the five forecasters in Section~\ref{sec:mlb_comparisons}.

\begin{table}[t]
    \centering
    \begin{subtable}[c]{0.45\textwidth}
        \begin{tabular}{l|c|c}
            \Xhline{1.1pt}
            \bf Forecaster & \bf $C_T^\mathsf{EB}$ & \bf $E_T$ \\ \hline
            \texttt{538} & (-0.00265, -0.00061) &  2979.0 \\
            \texttt{laplace} & (-0.00980, -0.00596) & $> 10^4$ \\
            \texttt{k29} & (-0.01392, -0.00905) & $> 10^4$ \\
            \texttt{constant} & (-0.01115, -0.00713) & $> 10^4$ \\
            \Xhline{1.1pt}
        \end{tabular}
        \subcaption{$\Delta_T$ (Brier) against \texttt{vegas}}
    \end{subtable}
    \hspace{0.05\textwidth}
    \begin{subtable}[c]{0.45\textwidth}
        \begin{tabular}{l|c|c}
            \Xhline{1.1pt}
            \bf Forecaster & \bf $C_T^\mathsf{EB}$ & \bf $E_T$ \\ \hline
            \texttt{538} & (    $-\infty$, -0.01012) & $> 10^4$ \\
            \texttt{laplace} & (    $-\infty$, -0.04723) & $> 10^4$ \\
            \texttt{k29} & (    $-\infty$, -0.14684) & $> 10^4$ \\
            \texttt{constant} & (    $-\infty$, -0.05165) & $> 10^4$ \\
            \Xhline{1.1pt}
        \end{tabular}
        \subcaption{$W_T$ (Winkler-logarithmic) against \texttt{vegas}}
    \end{subtable}
    \caption{Comparing forecasters against the \texttt{vegas} forecaster. 
    In (a), we present 95\% EB CSs for the average Brier score differential $(\Delta_t)_{t=0}^\infty$, evaluated at time $T = 25,165$ (i.e., $C_T^\mathsf{EB}$), as well as the e-process for the null of $\calH_0^\sfw(q,p): \Delta_t \geq 0,\;\forall t$, also evaluated at time $T$ (i.e., $E_T$).
    In (b), we present the analogous table for the average Winkler score $W_T$ (Section~\ref{sec:winkler}), which is a normalized difference in a proper score (the logarithmic score, in this case). 
    Note that $C_T^\mathsf{EB}$ is one-sided due to the one-sided boundedness of $W_T$.
    Positive (negative) values of $\Delta_T$ and $W_T$ indicate that the forecaster is better (worse) than the baseline.
    We find that none of the other forecasters, including \texttt{538}, have outperformed \texttt{vegas} from 2010 to 2019.
    }
    \label{tbl:mlb_comparison}
\end{table}

\subsection{Comparing Statistical Postprocessing Methods for Weather Forecasts}\label{sec:weather}

As our second real-data experiment, we compare a set of statistical postprocessing methods for weather forecasts~\citep{vannitsem2021statistical}, following the recent work by \citet{henzi2021valid}.
Statistical postprocessing here refers to the process of correcting for biases and dispersion errors in ensemble weather forecasts, which are produced by perturbing the initial conditions of numerical weather prediction (NWP) methods.
As ensemble forecasts are commonly used in state-of-the-art weather forecasting systems as a means of producing probabilistic forecasts, statistical postprocessing is considered a key component of modern weather forecasting.

Given 24-hour precipitation data from 2007 to 2017 at four locations (Brussels, Frankfurt, London Heathrow, and Zurich), our goal is to compare three postprocessing methods over time: isotonic distributional regression (IDR; \citet{henzi2021isotonic}), heteroscedastic censored logistic regression (HCLR; \citet{messner2014extending}), and a variant of HCLR without its scale parameter (HCLR\_).
We use the Brier score throughout this section.
See Section~\ref{sec:weather_forecast} for details regarding data as well as a plot of the three forecasting methods.

Our main goal here is to sequentially compare the three statistical postprocessing methods using the EB CS and the sub-exponential e-process.
As noted in Sections~\ref{sec:relatedwork} and~\ref{sec:eprocess}, the inferential conclusions drawn from the sub-exponential e-process (Theorem~\ref{thm:eprocess}) are different from \citet{henzi2021valid}'s e-process, which provides a test of conditional forecast dominance at all times (i.e., the strong null), instead of average (i.e., the weak null).
Given that the weak null is larger than the strong null, we would generally expect the sub-exponential e-process for the weak null to be smaller than \citet{henzi2021valid}'s e-process for the strong null.
On the other hand, the two methods are similar in that they are both valid at arbitrary (data-dependent) stopping times.

\begin{figure}[t]
    \centering
    \includegraphics[width=\textwidth]{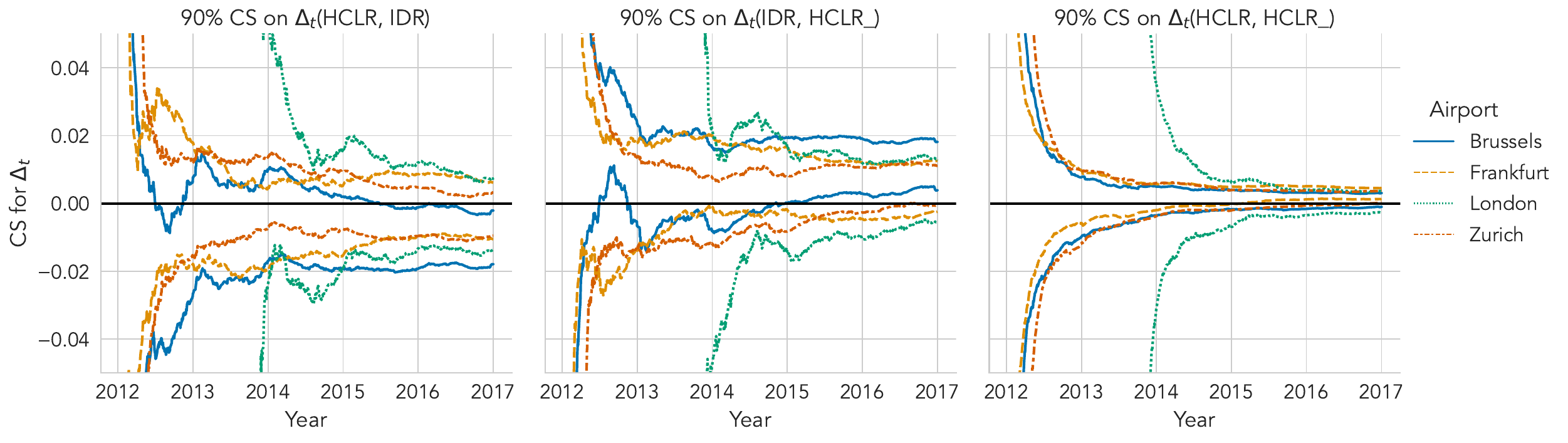}
    \includegraphics[width=\textwidth]{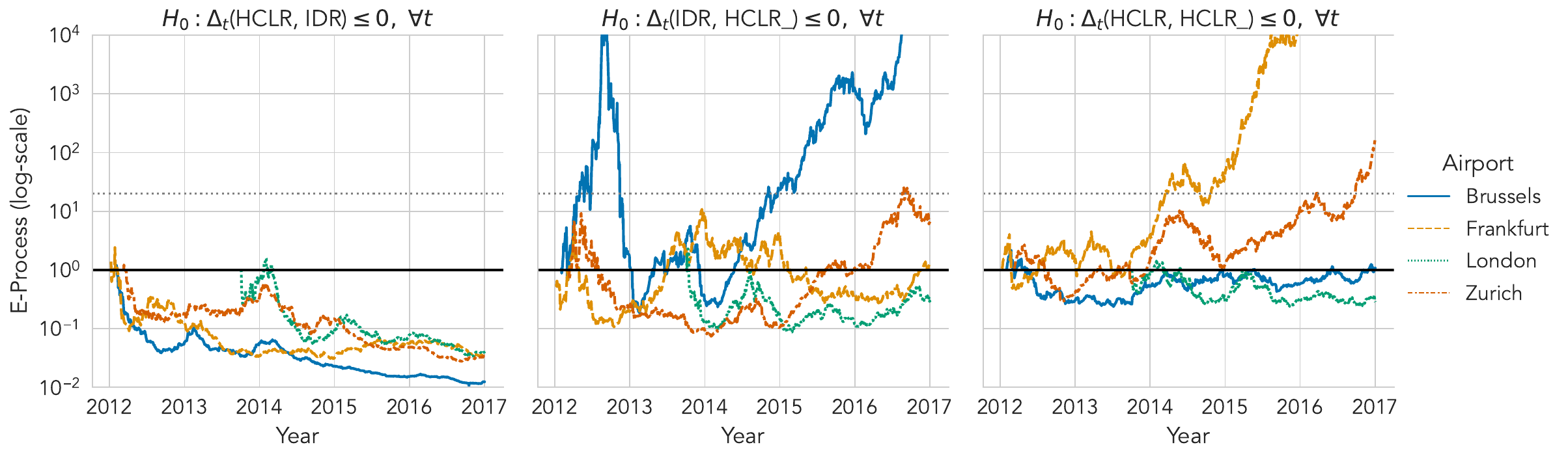}
    \caption{\emph{Top:} 90\% EB CSs for $\Delta_t$ between pairs of statistical postprocessing methods (HCLR and IDR; IDR and HCLR\_; HCLR and HCLR\_) for 1-day ensemble forecasts using Theorem~\ref{thm:main}, computed and plotted separately for each airport: Brussels ($T=1,703$), Frankfurt ($T=1,809$), London ($T=1,128$), and Zurich ($T=1,621$).
    Positive (negative) scores of $\Delta_t(p, q)$ indicate that forecaster $p$ is better (worse) than forecaster $q$.
    Overall, the CSs capture the time-varying score gap on average between the two forecasters across the years.
    \emph{Bottom:} E-processes for the null that $\calH_0^\sfw: \Delta_t \leq 0,\; \forall t$, corresponding to (the lower bound of) the 90\% CSs above. These e-processes are the \emph{weak} (average-based) counterpart to \citet{henzi2021valid}'s e-processes for the \emph{strong} (step-by-step) null that $\calH_0^\sfs: \delta_t \leq 0 \;\forall t$. 
    Note that the e-processes exceed 20 approximately when the lower bound of the 90\% CS exceeds 0.
    Both procedures use the Brier score as the scoring rule.
    }
    \label{fig:weather_plots}
\end{figure}

In Figure~\ref{fig:weather_plots}, we plot both the 90\% EB CS on $\Delta_t$ (top) as well as the sub-exponential e-processes for the weak one-sided null $\calH_0^\sfw$ (bottom), between HCLR and IDR, IDR and HCLR\_, and HCLR and HCLR\_ on 1-day PoP forecasts at the four airport locations.
Note that we compare the same three pairs as \citet{henzi2021valid}, who compare e-processes for the strong one-sided null $\calH_0^\sfs$.
The EB CS is computed using Theorem~\ref{thm:main} and the gamma-exponential mixture boundary~\eqref{eqn:cm_boundary}; the analogous mixture e-processes are then computed using Theorem~\ref{thm:eprocess}.
We use the significance level of $\alpha = 0.1$ for the EB CS, corresponding the threshold of $2/\alpha=20$ for each one-sided e-process.

We first note from Figure~\ref{fig:weather_plots} that the lower bound of our 90\% EB CS on $\Delta_t(p, q)$ and the e-process for $\calH_0^\sfw: \Delta_t(p, q) \leq 0$ share a similar trend over time, where the e-process grows large when the lower bound grows significantly larger than zero, implying that the forecaster $p$ is better than the forecaster $q$, using the stopping rule~\eqref{eqn:cs_stopping}. 
Whereas the CS provides a (two-sided) estimate of $\Delta_t(p, q)$ with uncertainty, the e-process explicitly gives the amount of evidence for whether one is better than the other. 
This illustrates how the two procedures complement each other for anytime-valid inference on $\Delta_t$.
We also remark that, although we only plot the e-processes for one-sided null $\calH_0^\sfw(p, q)$, we can further compute the e-processes for $\calH_0^\sfw(q, p): \Delta_t(q, p) \leq 0$, and they would correspond to the upper confidence bounds of the EB CSs.

Based on these results, we find from the 90\% EB CSs that IDR forecasts are found to outperform both HCLR and HCLR\_ 1-day forecasts for Brussels and that HCLR forecasts outperform HCLR\_ forecasts for Frankfurt and Zurich, but we do not find significant differences at other locations between other pairs. 
The e-processes (thresholded at 20) lead to the same conclusions, and they clearly visualize at which point in time is one forecaster first found to outperform the other and how that pattern changes.
For example, when comparing IDR to HCLR\_ for Brussels, IDR is found to be better as early as 2012, and it also shows the period between late 2012 and late 2015 where it is no longer found to be better, before eventually regaining evidence favoring IDR starting 2016.

When we compare the sub-exponential e-processes for the weak null $\calH_0^\sfw$ with the e-processes for the strong null $\calH_0^\sfs$, which are drawn in Figure 3 of \citet{henzi2021valid}, we find that e-processes for the strong null are large whenever e-processes for the weak null are also large, but not vice versa. 
For example, the comparison of IDR against HCLR\_ in Frankfurt is only found to have strong evidence against the strong null, but not the weak null.
This is consistent with our previous discussion in Section~\ref{sec:eprocess} that the strong null implies the weak null and thus is easier to ``reject'' (or gather evidence against). 
For example, in Frankfurt, we can infer we only have strong evidence that IDR has outperformed HCLR\_ \emph{at some point in time} between 2012 and 2017, but we do not have sufficient evidence that IDR has outperformed HCLR\_ \emph{on average} in the same time period.

In Section~\ref{sec:lagh}, we include e-processes for comparing lag-$h$ forecasts in the same setting.

\section{Extensions and Discussion}\label{sec:discussion}

In the following, we discuss some related points that were not highlighted in previous sections.

\paragraph{On the use of unbounded scoring rules.}
Our main results in Theorems~\ref{thm:main} and~\ref{thm:eprocess} require the use of bounded scoring rules, which may be restrictive in certain use cases.
If the score differentials are unbounded, a general solution would be to use the asymptotic CS (Section~\ref{sec:asympcs}), which assumes that only $2+\delta$ moments are bounded.
When it comes to unbounded proper scores for binary outcomes, such as the logarithmic score, the Winkler score (Section~\ref{sec:winkler}), which we used in Section~\ref{sec:2019ws}, offers a nonasymptotic and anytime-valid solution.

\paragraph{Comparing forecasts of lag $h > 1$.}
In general forecasting scenarios, we may encounter forecasts that are made $h > 1$ rounds ahead of when the outcome is revealed at time $t$.
In these cases, the expected score differential we seek to estimate should be conditioned on the filtration available at the time of forecasting, rather than the filtration at round $t-1$. 
We formally derive methods for comparing lag-$h$ forecasts in Section~\ref{sec:lagh}.
These include lagged sequential e-values~\citep{arnold2021sequentially}, which are not e-processes themselves but can nevertheless quantify the evidence against the weak null (and a ``less weak'' variant), as well as p-processes and e-processes that are more conservative.
The technical details follow the recent discussions by \citet{arnold2021sequentially,henzi2021valid}.
Constructing a more powerful e-process and also a CS for the lagged weak null remains a challenging problem.

\paragraph{On ``looking ahead'' in distribution-free sequential inference on time-varying means.}
Our methods are valid without any assumptions about the time-varying dynamics of the forecast score differentials $(\hat\delta_i)_{i=1}^\infty$, and in particular we avoid conditions involving stationarity or mixing. A large e-value against $\calH_0: \Delta_t(p,q) \leq 0, \; \forall t$ at some stopping time $\tau$ tells us that $p$ has achieved a better conditional predictive performance than $q$ up to $\tau$ on average.
The utility of comparing forecasters in such a descriptive sense is often significant in the real world: determining a winner in real-world forecasting competitions can often land significant cash prizes (e.g., financial forecasting\footnote{\url{https://m6competition.com}}) and/or media attention (e.g., election and sports forecasting).

This also means that the inferential conclusions drawn from our methods need not extrapolate to \emph{future} time steps, because hypothetically the forecasters or Reality (from Game~\ref{game:general}) can completely change their behaviors going forward.
Indeed, there is a distinction between saying that one \emph{has done} better than the other and that one \emph{is going to be} better than the other in the future --- the former is descriptive, while the latter is predictive. 
All our methods provide evidence and uncertainty related to the former statement.
Because we do not make any assumption that says ``the future will resemble the past,''  no method can make conclusive statements about the latter without clairvoyance. Our setup highlights that past performance can be compared in a distribution-free manner, while predictions of future performance will require nontrivial distributional assumptions.

Ultimately, the decision to take the inferential conclusion and extrapolate it toward the future is (and should be) left to the practitioner's own beliefs.
If a practitioner opts to make additional assumptions about Reality, then in principle, the conclusions drawn from our methods can extend to settings that the assumptions allow. 
If one is willing to assume, say, that the score differentials are constant, then the inferential conclusions will straightforwardly extrapolate to future time steps (in the assumed setting).
Furthermore, the variance-adaptive EB CS will remain tight, because the underlying variance remains constant.
It should be noted that, even under such assumptions, which are often made by classical methods like the \citet{diebold1995comparing} test, anytime-valid approaches avoid the ``p-hacking'' problem that the classical methods are susceptible to.

\clearpage

\subsection*{Acknowledgements}
YJC and AR thank Alexander Henzi, Johanna F. Ziegel, Rafael M. Frongillo, and the anonymous reviewers for their valuable feedback on this work.
AR acknowledges funding from NSF DMS 1916320. Research reported in this paper was sponsored in part by the DEVCOM Army Research Laboratory under Cooperative Agreement W911NF-17-2-0196 (ARL IoBT CRA). The views and conclusions contained in this document are those of the authors and should not be interpreted as representing the official policies, either expressed or implied, of the Army Research Laboratory or the U.S. Government. The U.S. Government is authorized to reproduce and distribute reprints for Government purposes notwithstanding any copyright notation herein.

\bibliography{contents/references}
\bibliographystyle{apalike}

\clearpage

\appendix

\section{Main Proofs}\label{sec:proof}

\subsection{Sub-exponential Test Supermartingales for Time-Varying Means}\label{sec:proof_bbelow}

The proofs of Theorems~\ref{thm:main} and~\ref{thm:eprocess} are both based on a variance-adaptive test supermartingale that uniformly bounds sums of random variables that are bounded from below.
We first derive this test supermartingale (which, by definition, is also an e-process itself) and use the result for the proofs of the main theorems in the following subsections.

We start by revisiting a useful lemma for the sub-exponential processes.
Recall from Section~\ref{sec:subpsi} that $\psi_{E,c}(\lambda) = c^{-2}(-\log(1 - c\lambda) - c\lambda), \; \forall \lambda \in [0, 1/c)$ is the exponential CGF-like function. 
By the proof of Lemma 4.1 in \citet{fan2015exponential}, for any $\lambda \in [0, 1/c)$ and any $\xi \geq -c$,
\begin{equation}\label{eqn:fan_lemma}
    \exp\incurly{\lambda \xi - \psi_{E,c}(\lambda) \xi^2} \leq 1 + \lambda \xi.
\end{equation}
Note that the original proof uses $c=1$, but it straightforwardly generalizes to any value of $c > 0$.
To see this, for any $c>0$, set $\tilde{\lambda} = c\lambda \in [0, 1)$ and $\tilde{\xi} = c^{-1}\xi \geq -1$.
Then, applying the lemma with $c=1$ using $(\tilde{\lambda}, \tilde{\xi})$ gives the desired result.

Now, we show a time-uniform sub-exponential boundary that is generally applicable to sums of random variables that are bounded from below.
This is an extension of Lemma 3(e) from \citet{howard2020chernoff}, which also utilizes~\eqref{eqn:fan_lemma}.
We note that a similar extension is utilized in the recent work of \citet{waudbysmith2022anytime} but without the predictable bounds $(c_i)_{i=1}^\infty$.

In the following, let $(X_i)_{i=1}^\infty$ be any process whose conditional means $\mu_i := \mathbb{E}_{i-1}[X_i]$ exist.
Let $(S_t)_{t=0}^\infty$ be its cumulative deviations from the conditional means, i.e., $S_0 = 0$ and $S_t = \sum_{i=1}^t (X_i - \mu_i)$. 
Note that $S_t$ is a martingale, i.e., $\mathbb{E}_{t-1}[S_t] = S_{t-1}$.
Also, let $(\hat{V}_t)_{t=0}^\infty$ be a nondecreasing variance process of the form $\hat{V}_0=0$ and $\hat{V}_t = \sum_{i=1}^t (X_i - \gamma_i)^2$, where $(\gamma_i)_{i=1}^\infty$ is a predictable process.
Also, we take $1/\infty=0$ and, with a slight abuse of notation, $[0, 0) = \{0\}$.
\begin{proposition}[Sub-exponential test supermartingales for time-varying means]\label{ppn:subexp_bdd_below}
Suppose that there exists a predictable positive sequence $(c_i)_{i=1}^\infty$ such that $X_i - \gamma_i \geq -c_i$ a.s. for all $i \geq 1$.
Then,
\begin{align}\label{eqn:nsm_bounded_below}
    L_t(\lambda) 
    = \prod_{i=1}^t \exp \incurly{ \lambda (X_i-\mu_i) - \psi_{E,c_i}(\lambda) \inparen{X_i - \gamma_i}^2 }
\end{align}
is a test supermartingale for each $\lambda \in [0, 1/c_0)$, where $c_0 = \sup_{i\geq1} c_i$.
\end{proposition}

\begin{proof}
For each $i \geq 1$, it suffices to show that
\begin{equation}
    \mathbb{E}_{i-1} \insquare{\exp\incurly{\lambda (X_i-\mu_i) - \psi_{E,c_i}(\lambda) (X_i - \gamma_i)^2}} \leq 1.
\end{equation}

Let $\tilde{X}_i = X_i - \mu_i$ and $\tilde{\gamma}_i = \gamma_i - \mu_i$.
Then, $\tilde{X}_i - \tilde{\gamma_i} = X_i - \gamma_i \geq -c_i$ a.s. by assumption.
By~\eqref{eqn:fan_lemma},
\begin{equation}
    \exp\incurly{\lambda (\tilde{X}_i - \tilde\gamma_i) - \psi_{E,c_i}(\lambda) (\tilde{X}_i - \tilde{\gamma}_i)^2} \leq 1 + \lambda(\tilde{X}_i - \tilde{\gamma}_i).
\end{equation}
Multiplying each side by $\exp\{\lambda \tilde{\gamma}_i\}$ and rearranging terms, we get
\begin{equation}
    \exp\incurly{\lambda \tilde{X}_i - \psi_{E,c_i}(\lambda) (\tilde{X}_i - \tilde{\gamma}_i)^2} \leq e^{\lambda \tilde{\gamma}_i}(1 - \lambda \tilde{\gamma}_i) + e^{\lambda \tilde{\gamma}_i}\lambda \tilde{X}_i \leq 1 + e^{\lambda \tilde{\gamma}_i}\lambda \tilde{X}_i,
\end{equation}
where in the second inequality we used the fact that $1-x \leq e^{-x}$ for all $x \in \R$.

Finally, we take the conditional expectation $\mathbb{E}_{i-1}$ on each side.
Because $\mathbb{E}_{i-1}[\tilde{X}_i] = \mathbb{E}_{i-1}[X_i - \mu_i] = 0$, and also because $(\gamma_i)_{i=1}^\infty$ and $(c_i)_{i=1}^\infty$ are predictable, we get
\begin{align}
    \mathbb{E}_{i-1}\insquare{\exp\incurly{\lambda \tilde{X}_i - \psi_{E,c_i}(\lambda) (\tilde{X}_i - \tilde{\gamma}_i)^2}}
    &\leq 1 + e^{\lambda \tilde{\gamma}_i} \lambda \mathbb{E}_{i-1}\insquare{\tilde{X}_i} = 1.
\end{align}
Substituting back in $\tilde{X}_i = X_i - \mu_i$ and $\tilde{X}_i - \tilde\gamma_i = X_i - \gamma_i$, we get the desired result.
\end{proof}

Proposition~\ref{ppn:subexp_bdd_below} is stated for a general setting in which bounds on the pointwise score differentials can vary across time, as long as they form a predictable sequence.
If there is a constant $c \in (0, \infty)$ such that $|\hat\delta_i| \leq \frac{c}{2}$, such as in Theorems~\ref{thm:main} and~\ref{thm:eprocess}, then we can simply choose $c_i = c$ for all $i$ and further simplify the expression~\eqref{eqn:nsm_bounded_below} to
\begin{equation}
    L_t(\lambda) = \exp \incurly{ \lambda S_t - \psi_{E,c}(\lambda) \hat{V}_t }, \quad \forall \lambda \in [0, 1/c).
\end{equation}
We return to the case of using non-constant predictable bounds in Section~\ref{sec:predictable_bounds}.

\subsection{Proof of Theorem~\ref{thm:main}}\label{sec:proof_main}

The proof is a direct consequence of Proposition~\ref{ppn:subexp_bdd_below}, applied once each to the lower and upper confidence bounds.

The stated conditions imply that $\hat\delta_i - \gamma_i \geq -c$ a.s. for all $i \geq 1$.
Define $S_t = \sum_{i=1}^t (\hat\delta_i - \delta_i)$.
Then, by Proposition~\ref{ppn:subexp_bdd_below}, the process
\begin{equation}
    L_t^\textsf{lcb}(\lambda) = \exp\incurly{ \lambda S_t - \psi_{E, c}(\lambda)\hat{V}_t}
\end{equation}
is a test supermartingale for $\lambda \in [0, 1/c)$.
By definition, this implies that $(S_t)_{t=0}^\infty$ is sub-$\psi_{E,c}$ (``sub-exponential with scale $c$'') with variance process $(\hat{V}_t)_{t=0}^\infty$, and thus we have 
\begin{equation}
    \prob{\exists t \geq 1: S_t \geq u_{\alpha/2}(\hat{V}_t)} \leq \alpha/2,
\end{equation}
for any sub-exponential uniform boundary~\eqref{eqn:uniform_boundary} with crossing probability $\alpha/2$ and scale $c$, denoted here as $u_{\alpha/2}$.
Using the fact that $\frac{1}{t}S_t = \frac{1}{t}\sum_{i=1}^t \hat\delta_i - \frac{1}{t} \sum_{i=1}^t \delta_i = \hat\Delta_t - \Delta_t$, we can divide each side of the inequality by $t$ to obtain the lower confidence bound (LCB).

Similarly, the conditions also imply that $-\hat\delta_i + \gamma_i \geq -c$, so Proposition~\ref{ppn:subexp_bdd_below} also implies that the process
\begin{equation}
    L_t^\textsf{ucb}(\lambda) = \exp\incurly{ \lambda (-S_t) - \psi_{E, c}(\lambda)\hat{V}_t}
\end{equation}
is also a test supermartingale for $\lambda \in [0, 1/c)$, or equivalently, $(-S_t)_{t=0}^\infty$ is sub-$\psi_{E,c}$ with the same variance process $(\hat{V}_t)_{t=0}^\infty$.
Applying the same argument to $L_t^\textsf{ucb}(\lambda)$ gives the analogous upper confidence bound (UCB) using the \emph{same} uniform boundary $u_{\alpha/2}$.

Finally, combining the lower and upper confidence bounds with a union bound, we obtain the CS:
\begin{equation}
    \prob{\forall t \geq 1 : \absval{\hat\Delta_t - \Delta_t} < \frac{u(\hat{V}_t)}{t}} \geq 1 - \alpha .
\end{equation}

\subsection{Proof of Theorem~\ref{thm:eprocess}}\label{sec:proof_eprocess}

We state and prove a slightly more general version of Theorem~\ref{thm:eprocess} that only assumes the empirical score differentials $\hat\delta_i$ are bounded from \emph{below} and the predictable estimates $\gamma_i$ are bounded (or truncated) from \emph{above}. 
Theorem~\ref{thm:eprocess} assumes that the score differentials are bounded from below \emph{and} above, so applying the following proposition twice to $(\hat\delta_i, \gamma_i)_{i=1}^\infty$ and $(-\hat\delta_i, -\gamma_i)_{i=1}^\infty$ will give us the result.

\begin{proposition}\label{ppn:eprocess_bdd_below}
Suppose that $\hat\delta_i \geq -\frac{c}{2}$ for each $i \geq 1$, for some $c \in (0, \infty)$.
Also, let $(\gamma_i)_{i=1}^\infty$ be any predictable sequence and $\hat{V}_t = \sum_{i=1}^t (\hat\delta_i - \overline\gamma_i)^2$, where $\overline\gamma_i = \gamma_i \wedge \frac{c}{2}$.
Then, for each $\lambda \in [0, 1/c)$, the process $(E_t(\lambda))_{t=0}^\infty$ defined as $E_0(\lambda) = 1$ and
\begin{equation}\label{eqn:eprocess_bdd_above}
    E_t(\lambda) := \exp\incurly{ \lambda \sum_{i=1}^t \hat\delta_i - \psi_{E,c}(\lambda) \hat{V}_t } \indent \text{is an e-process for $\calH_0^\sfw(p, q)$}.
\end{equation}
\end{proposition}

Proposition~\ref{ppn:eprocess_bdd_below} tells us that, if the pointwise empirical score differentials are bounded from below (or above), then we can derive a sub-exponential e-process for $\calH_0(p,q)$ (or $\calH_0(q,p)$).
An important use case for the more general scenario is when using the Winkler score~\citep{winkler1994evaluating}, which is bounded from above by 1 but unbounded from below, as we describe in Section~\ref{sec:winkler}.

\begin{proof}[Proof of Proposition~\ref{ppn:eprocess_bdd_below}]
First, note that $(E_t(\lambda))_{t=0}^\infty$ is an adapted process w.r.t.~$\frakG$ (and also consists of empirical quantities only).
Let $S_t = \sum_{i=1}^t (\hat\delta_i - \delta_i) = t(\hat\Delta_t - \Delta_t)$.
Since $\hat\delta_i - \overline\gamma_i \geq -c$ for all $i \geq 1$, Proposition~\ref{ppn:subexp_bdd_below} implies that
\begin{equation}
    L_t(\lambda) := \exp\incurly{ \lambda S_t - \psi_E(\lambda) \hat{V}_t }
\end{equation}
is a test supermartingale for each $\lambda \in [0, 1/c)$.

Now, under any $P \in \calH_0^\sfw(p, q)$, we have that $\exp \incurly{ -\lambda \sum_{i=1}^t \delta_i } \geq 1$, so for any $t \geq 1$,
\begin{align}
    L_t(\lambda) 
    &= \exp\incurly{ \lambda \sum_{i=1}^t \hat\delta_i - \psi_E(\lambda) \hat{V}_t } \exp \incurly{ -\lambda \sum_{i=1}^t \delta_i}  \nonumber \\
    &\geq \exp\incurly{ \lambda \sum_{i=1}^t \hat\delta_i - \psi_E(\lambda) \hat{V}_t } 
    = E_t(\lambda) .
\end{align}
In other words, for each $P \in \calH_0^\sfw(p, q)$, the process $(E_t(\lambda))_{t=0}^\infty$ is upper-bounded by the test supermartingale $(L_t(\lambda))_{t=0}^\infty$ at all times $t$.
This implies that $(E_t(\lambda))_{t=0}^\infty$ is an e-process for $\calH_0^\sfw(p, q)$, by Corollary 22 of \citet{ramdas2020admissible}.
\end{proof}

\section{Details on Time-Uniform Boundary Choices}\label{sec:boundary_details}

\subsection{Computing the Gamma-Exponential Mixture}\label{sec:gm_mixture}

Here, we derive a closed-form expression (up to efficiently computable gamma functions) for the gamma-exponential mixture, which is used in both the mixture boundary for the CS (Equation~\eqref{eqn:cm_boundary}) and in the mixture e-process for the weak null (Theorem~\ref{thm:eprocess}).
The mixture takes the following form:
\begin{equation}
     m(s, v) := \int \exp\incurly{\lambda s - \psi_{E,c}(\lambda)v} f_\rho(\lambda)d\lambda,
\end{equation}
where $f_\rho$, for any $\rho > 0$, is a reparametrized Gamma density $f_\rho(\lambda) = C(\rho) (1-\lambda)^{\rho-1} e^{-\rho(1-\lambda)}$, $\lambda \in [0, 1/c)$, where $C(\rho) = \frac{\rho^\rho}{\underline{\gamma}(\rho, \rho) \Gamma(\rho)}$ is the normalization constant, $\Gamma(a, z) := \int_z^\infty u^{a-1}e^{-u}du$ is the upper incomplete gamma function, $\Gamma(a) := \Gamma(a, 0)$ is the gamma function, and $\underline{\gamma}$ is the regularized lower incomplete gamma function:
\begin{equation}\label{eqn:lower_gamma}
    \underline{\gamma}(a, z) := \frac{1}{\Gamma(a)} \int_0^z u^{a-1}e^{-u}du, \quad \forall a,z>0.
\end{equation} 
Both $\Gamma$ and $\underline{\gamma}$ can be computed efficiently in standard scientific computing software. (E.g., $\underline{\gamma}$ can be computed using \texttt{boost::math::gamma\_p} in C++ and \texttt{scipy.special.gammainc} in Python.)

We note here that all time-uniform boundaries have a ``tradeoff of tightness'' across different (intrinsic) times~\citep{howard2021timeuniform}, so that it is natural to have a hyperparameter that controls at what intrinsic time we want the resulting CS width to be optimized.
In the above, the single hyperparameter, $\rho > 0$, can be related to the user-specified optimal intrinsic time $v_\text{opt}$ (and the significance level $\alpha$) via the mapping $\rho = -v_\text{opt}(W_{-1}(-\alpha^2/e) + 1)$, where $W_{-1}$ is the lower branch of the Lambert $W$ function.
As described in Proposition 3 of \citet{howard2021timeuniform}, this choice of $\rho$ uniquely minimizes the width function $v \mapsto u(v)/\sqrt{v}$, when $u$ is the two-sided normal mixture boundary, and it is also known to also provide a good approximation for the (one-sided) gamma-exponential mixture boundary in practice.

The first part of the following proposition is essentially a restatement of Proposition 9 in \citet{howard2021timeuniform}; the second part additionally provides an upper bound for the mixture when $s \ll 0$ (e.g., the mixture e-process when data supports the null).
\begin{proposition}[Gamma-exponential mixture for e-processes]\label{ppn:ge_mixture}
Fix $c>0$ and $\rho>0$. 
Consider any values of $s \in \R$ and $v \geq 0$.
If $\frac{cs+v+\rho}{c^2} > 0$, then
\begin{equation}\label{eqn:ge_mixture}
    m(s, v) = 
    C\inparen{\frac{\rho}{c^2}}
    \frac{\Gamma\inparen{\frac{v+\rho}{c^2}} \underline{\gamma}\inparen{\frac{v+\rho}{c^2}, \frac{cs+v+\rho}{c^2}}}{\inparen{\frac{cs+v+\rho}{c^2}}^{\frac{v+\rho}{c^2}}} \exp\incurly{\frac{cs+v}{c^2}};
\end{equation}
otherwise, if $\frac{cs+v+\rho}{c^2} < 0$, then
\begin{equation}\label{eqn:ge_mixture_ub}
    m(s, v) \leq C\inparen{\frac{\rho}{c^2}}\frac{\exp\incurly{-\frac{\rho}{c^2}}}{\frac{v+\rho}{c^2}} \leq 1.
\end{equation}
\end{proposition}
This is precisely the formula for the sub-exponential mixture e-process in Theorem~\ref{thm:eprocess}: $E_t^{\textsf{mix}} = m(\sum_{i=1}^t \hat\delta_i, \hat{V}_t)$ with $f_\rho$ being the mixture density.
It makes sense that $m(s, v)$ is upper-bounded by 1 when $\frac{cs+v+\rho}{c^2} < 0$, because $s < -\frac{v+\rho}{c} < 0$ would imply that the sum of score differentials is negative, supporting the weak null.
In our implementation, we use the first upper bound in~\eqref{eqn:ge_mixture_ub}, which can be computed efficiently and get substantially smaller than 1 when $v \gg 0$.

\begin{proof}[Proof of Proposition~\ref{ppn:ge_mixture}]
For simplicity, we assume $c=1$. The proof is analogous for any $c>0$.

Recall that $\psi_E(\lambda) = -\log(1-\lambda) - \lambda$ for $\lambda \in [0, 1)$. 
For any $\rho > 0$,
\begin{align}
    m(s, v) &= C(\rho) \int_0^1 \exp\incurly{\lambda s - \psi_E(\lambda) v} \cdot  (1-\lambda)^{\rho-1} e^{-\rho(1-\lambda)} d\lambda  \nonumber  \\
    &= C(\rho) \int_0^1 e^{\lambda(s + v)} (1-\lambda)^v \cdot (1-\lambda)^{\rho-1} e^{-\rho(1-\lambda)} d\lambda  \nonumber \\
    &= C(\rho) \int_0^1 (1-\lambda)^{v+\rho-1} e^{\lambda(s + v) - \rho(1-\lambda)} d\lambda  \nonumber \\
    &= C(\rho) \left(\int_0^1 (1-\lambda)^{v+\rho-1} e^{-(s+v+\rho)(1-\lambda)} d\lambda \right) e^{s+v} ,
\end{align}
where in the last equality we used 
\[
\lambda(s+v) - \rho(1-\lambda) = (s+v) - (1-\lambda)(s+v) - (1-\lambda)\rho = -(s+v+\rho)(1-\lambda) + (s+v) .
\]
Now, let $a = v + \rho$ and $z = s + v + \rho$, and note that $a > 0$.

\paragraph{Case 1: $z = s+v+\rho > 0$.} 
Using the change-of-variable formula $u = (s+v+\rho)(1-\lambda) = z(1-\lambda)$, we have that
\begin{align}
    m(s, v) &= C(\rho)  \inparen{ \int_z^0 \inparen{\frac{u}{z}}^{a-1} e^{-u} \frac{du}{-z} } e^{s+v}  \nonumber \\
    &= C(\rho) \cdot \frac{1}{z^a} \inparen{ \int_0^{z} u^{a-1} e^{-u} du } e^{s+v} \label{eqn:m_lower_gamma} \\
    &= C(\rho) \frac{\Gamma(a)\underline{\gamma}(a,z)}{z^a} e^{s+v} , \label{eqn:m_pos_z}
\end{align}
where we use the fact that the integral in \eqref{eqn:m_lower_gamma} corresponds to the numerator of the lower incomplete gamma function $P(a, z)$ in \eqref{eqn:lower_gamma}.
The expression~\eqref{eqn:m_pos_z} can be computed in closed-form.

\paragraph{Case 2: $z = s+v+\rho < 0$.} 
Using the change-of-variable formula $u = -(s + v + \rho)(1-\lambda) = -z(1-\lambda)$, we obtain
\begin{align}
    m(s, v) &= C(\rho)  \inparen{ \int_{-z}^0 \inparen{\frac{u}{-z}}^{a-1} e^{u} \frac{du}{z} } e^{s+v}  \nonumber \\
    &= C(\rho) \cdot \frac{1}{(-z)^a} \inparen{ \int_0^{-z} u^{a-1} e^{u} du } e^{s+v}  \nonumber \\
    &= C(\rho) \cdot \frac{1}{|z|^a} \inparen{ \int_0^{|z|} u^{a-1} e^{u} du } e^{s+v} . \label{eqn:m_neg_z}
\end{align}
Although the integral in \eqref{eqn:m_neg_z} is no longer a regularized lower incomplete gamma function, we can still show that $m(s, v)$ is upper-bounded by $1$. 
Since $e^u \leq e^{|z|} = e^{-z}$ for $u \leq |z|$, we have that
\begin{align}
    m(s, v) &\leq C(\rho) \cdot \frac{1}{|z|^a} \inparen{ \int_0^{|z|} u^{a-1} du } e^{-z} \cdot e^{s+v}  \nonumber \\
    &= C(\rho) \cdot \frac{1}{|z|^a} \inparen{ \int_0^{|z|} u^{a-1} du }  e^{-\rho} \label{eqn:z_s_v} \\
    &= C(\rho)  \cdot \frac{1}{|z|^a} \left. \inparen{\frac{u^a}{a}} \right|_0^{|z|} e^{-\rho}  \nonumber \\
    &= \frac{C(\rho) e^{-\rho}}{v+\rho} , \label{eqn:m_neg_z_upper}
\end{align}
where in \eqref{eqn:z_s_v} we used $-z + (s + v) = -(s + v + \rho) + (s + v) = -\rho$, and in \eqref{eqn:m_neg_z_upper} we substituted in $a = v + \rho$. 
We can further bound this value, using the fact that $v > 0$ and substituting back in $C(\rho)$:
\begin{align}
    m(s, v) \leq \frac{C(\rho)e^{-\rho}}{v+\rho} 
    &\leq \frac{C(\rho)e^{-\rho}}{\rho} \nonumber \\
    &= \rho^{\rho-1} e^{-\rho} \cdot \inparen{\int_0^\rho u^{\rho-1} e^{-u} du }^{-1} \nonumber \\
    &\leq \rho^{\rho-1} e^{-\rho} \cdot  \inparen{e^{-\rho} \int_0^\rho u^{\rho-1} du }^{-1} \label{eqn:ub_e} \\
    &= \rho^{\rho-1} \cdot \insquare{ \left. \inparen{\frac{u^\rho}{\rho}} \right|_0^{\rho} }^{-1} \nonumber \\
    &= 1 ,
\end{align}
where in \eqref{eqn:ub_e} we used the fact that $e^{-\rho} \leq e^{-u}$ for $u \in [0, \rho]$.
\end{proof}

\subsection{The Polynomial Stitching Boundary}\label{sec:stitching}

The \emph{polynomial stitched boundary} (Theorem 1, \citet{howard2021timeuniform}) provides a fully closed-form (without any gamma functions) alternative to the aforementioned gamma-exponential mixture boundary. 
It is constructed by finding a smooth analytical upper bound on a sequence of linear uniform bounds across different timesteps.
The boundary asymptotically grows with $O(\sqrt{v \log\log v})$ rate, matching the form of the law of the iterated logarithm (LIL). 
For example, a 95\% EB CS for $\Delta_t$ (Theorem~\ref{thm:main}) using the polynomial stitching boundary is given as follows (assuming $|\hat\delta_i| \leq 1, \; \forall i$):
\begin{equation}\label{eqn:stitching_example}
    \hat\Delta_t \pm 2 \cdot \frac{1.7\sqrt{\inparen{\hat{V}_t \vee 1} \inparen{\log\log\inparen{2\inparen{\hat{V}_t \vee 1}} + 3.8}} + 3.4\log\log\inparen{2\inparen{\hat{V}_t \vee 1}} + 13}{t}
\end{equation}
where $\hat{V}_t$ is the intrinsic time.

The polynomial stitched boundary can be applied to both Theorems~\ref{thm:hoeffding} and~\ref{thm:main} by setting $\hat{V}_t = t$ and $\hat{V}_t = \sum_{i=1}^t (\hat\delta_i - \gamma_i)^2$ respectively.
Previous work showed that the polynomial stitched boundary is a sub-gamma uniform boundary (Theorem 1, \citet{howard2021timeuniform}), which is also a ``universal'' sub-$\psi$ uniform boundary for any CGF-like function $\psi$ (Proposition 1, \citet{howard2020chernoff}).
We omit a full restatement of \citet{howard2021timeuniform}'s Theorem 1, which establishes the validity of the polynomial stitching boundary, but rather, we list its three hyperparameters for practical use:
\begin{itemize}
    \item $v_\text{opt} > 0$ determines the value of the intrinsic time at which the boundary is tightest;
    \item $s > 1$ controls how the crossing probability is distributed over intrinsic time;
    \item $\eta > 1$ controls the geometric spacing of the intrinsic time.
\end{itemize}
Throughout this paper, we fix $s=1.4$ and $\eta=2$, as recommended by the original paper, and only adjust $v_\text{opt}$, which serves the analogous role as the hyperparameter of the same name for the gamma-exponential boundary in Section~\ref{sec:gm_mixture}.

Although the stitching boundary is computed in closed form and matches the LIL rate, it is usually not as tight as the CM boundary in practice, and thus we use the CM boundary as our default in all of our main experiments.

\section{Asymptotic CSs for Sequential Forecast Comparison}\label{sec:asympcs}

In their recent work, \citet{waudbysmith2021doubly} introduce a new class of time-uniform CSs called asymptotic CSs, which trade the nonasymptotic guarantee of a standard CS~\eqref{eqn:cs_definition} for applicability to a wider variety of scenarios, e.g., estimating the average treatment effect in causal inference (for which a nonasymptotic CS is not known).
Formally, a sequence of confidence intervals $(\hat\theta_t \pm R_t^\sfA)_{t=1}^\infty$ is a \emph{$(1-\alpha)$-asymptotic CS (AsympCS)} for $(\theta_t)_{t=1}^\infty$ if there exists a nonasymptotic $(1-\alpha)$-CS $(\hat\theta_t \pm R_t^{\mathsf{NA}})_{t=1}^\infty$, for $(\theta_t)_{t=1}^\infty$, such that
\begin{equation}\label{eqn:asympcs_defn}
    {R_t^{\mathsf{NA}}}/{R_t^\sfA} \overset{a.s.}{\longrightarrow} 1.
\end{equation}
Furthermore, the AsympCS has an \emph{approximation rate} of $r(t)$ if $R_t^{\mathsf{NA}} - R_t^\sfA = O_{a.s.}(r(t))$.
Definition~\eqref{eqn:asympcs_defn} says that, as $t \to \infty$, the AsympCS is an ``arbitrarily precise approximation'' of the nonasymptotic CS, and it can be viewed as approximately satisfying the time-uniform coverage property when $t$ is large.

\citet{waudbysmith2021doubly} describes an asymptotic CS for time-varying means that can be applied to our setting of estimating $(\Delta_t)_{t=1}^\infty$ under Lyapunov CLT-type conditions.
For the sake of completeness, we include the (simplified) assumptions and the resulting closed form of the asymptotic CS, adapted to our setting and notations. 

Let $\sigma_t^2 = \mathbb{E}_{t-1}[(\hat\delta_t-\delta_t)^2]$ denote the conditional variance, $V_t = \sum_{i=1}^t \sigma_i^2$ be the cumulative conditional variance, and $\tilde{\sigma}_t^2 = t^{-1}V_t$ be the average.
Let $\hat{\sigma}_t^2$ be any estimator of $\sigma_t^2$, such as $\hat\sigma_t^2 = t^{-1}\sum_{i=1}^t(\hat\delta_i - \hat\Delta_{i-1})^2$. (Notice that, in the setting of Theorem~\ref{thm:main}, $\hat\sigma_t^2 = t^{-1}\hat{V}_t$ with $\gamma_i$ set to $\hat\Delta_{i-1}$.)
Now, we assume the following:
\begin{enumerate}[label=(\alph*)]
    \setlength\itemsep{0em}
    \item $\tilde\sigma_t^2 \overset{a.s.}{\longrightarrow} \sigma_*^2$ for some $\sigma_*^2>0$;
    \item there exists $q > 2$ such that the $q$\textsuperscript{th} moments of $\hat\delta_t$ is uniformly bounded (a.s.) for all $t \geq 1$; and
    \item $\hat\sigma_t^2/\tilde\sigma_t^2 \overset{a.s.}{\longrightarrow} 1$.
\end{enumerate}
As noted in the paper, these conditions can be substantially more general than either sub-Gaussianity or boundedness.
Given these assumptions, we know by Theorem 2.3 of \citet{waudbysmith2021doubly} that, for any $\rho > 0$ and any $\alpha \in (0, 1)$,
\begin{equation}\label{eqn:asympcs}
    C_t^{\mathsf{A}} := \inparen{\hat\Delta_t \pm \sqrt{ \frac{2(t\hat\sigma_t^2\rho^2 + 1)}{t^2\rho^2} \log\inparen{\frac{\sqrt{t\hat\sigma_t^2\rho^2 + 1}}{\alpha}} }} 
\end{equation}
forms a $(1-\alpha)$-AsympCS for $(\Delta_t)_{t=1}^\infty$ with an approximation rate of $o(\sqrt{V_t\log V_t}/t)$. 
$\rho > 0$ is a hyperparameter that affects the relative tightness of the CS across time, analogous to the hyperparameter $\rho$ in Section~\ref{sec:boundary_details}.
In our experiments, we follow \citet{waudbysmith2021doubly} (Equation 74) and use the choice that approximately optimizes the width at a pre-specified time $t^* \geq 1$:
\begin{equation}
    \rho(t^*) = \sqrt{\frac{2\log(1/\alpha) + \log(1+2\log(1/\alpha))}{t^*}}.
\end{equation}
Unless specified otherwise, $t^*$ is chosen to be $100$ in our experiments.

As illustrated in Figures~\ref{fig:confseq_illustrative} and~\ref{fig:scoring_rules}, the AsympCS is typically tighter than the EB CS (Theorem~\ref{thm:main}) for smaller values of $t$, and as $t$ grows large the widths of the two CSs become close to one another.

\section{Comparing Relative Forecasting Skills Using the Winkler Score}\label{sec:winkler}

In a typical forecast comparison scenario, we are often interested in comparing a newly developed forecasting algorithm (say, $p$) with an existing baseline (say, $q$).
For example, a company that already deploys a daily forecasting algorithm may want to A/B test if its newly developed method is at least as good as the existing one. 
In such settings, we may be interested in the \emph{relative} improvement of a forecaster over a baseline, and early work by~\citet{murphy1988skill} and~\citet{winkler1994evaluating} propose using normalized scoring rules that better reflect the relative ``skill'' of the new forecaster.

In this section, we show how our main results can be extended in a unique way to construct time-uniform CSs and e-processes for the \emph{average Winkler score}~\citep{winkler1994evaluating}, which is a normalized version of the average score differentials between probability forecasts on binary outcomes. 
Interestingly, these results yield SAVI approaches that are valid \emph{without} a boundedness or sub-Gaussianity assumption on the underlying scoring rule, and instead, they are valid whenever the scoring rule is proper~\citep{gneiting2007strictly}. 
The Winkler score is particularly useful when comparing probability forecasters based on the logarithmic score, which is a strictly proper but unbounded score, as we showcased in Section~\ref{sec:2019ws}.
We remark that \citet{lai2011evaluating} first showed the asymptotic normality of the average Winkler score.
In contrast to their work, the methods we develop here are nonasymptotic and anytime-valid, depending only on the natural upper bound (of 1) on the Winkler score; we also allow the baseline forecaster to be nonconstant.

Formally, we first define the \emph{(pointwise) Winkler score} $w(p, q, y)$ with a base scoring rule $S$ as follows:
\begin{align}\label{eqn:winkler}
    w(p, q, y) &:= \frac{S(p, y) - S(q, y)}{S(p, \indicator{p > q}) - S(q, \indicator{p > q})}, \quad p, q \in (0, 1), \; y \in \{0,1\},
\end{align}
where we set $0/0 := 0$.
We note that \eqref{eqn:winkler} is equivalent to the increment in the e-process of~\citet{henzi2021valid} (details in Section~\ref{sec:comparison_hz}), and thus we can interpret~\citet{henzi2021valid}'s e-process for the strong null as betting directly proportionally to the relative forecasting skill between the forecasters.
We also define the \emph{expected (pointwise) Winkler score} as
\begin{equation}\label{eqn:expected_winkler}
    w(p, q; r) := \mathbb{E}_{y \sim r}\insquare{w(p, q, y)} 
    = \frac{\mathbb{E}_{y \sim r}\insquare{S(p, y)} - \mathbb{E}_{y \sim r}\insquare{S(q, y)}}{S(p, \indicator{p > q}) - S(q, \indicator{p > q})},
\end{equation}
for $p, q \in (0, 1)$ and $r \in [0, 1]$. 
As before, $y \sim r$ denotes $y \sim \mathrm{Bernoulli}(r)$ (conditional on $p$ and $q$).
\citet[Section 4]{winkler1994evaluating} showed that, given any constant forecaster $q \in (0,1)$, the scoring rule $S_q'(p, y) = w(p, q, y)$ is (strictly) proper for $p$ whenever $S$ itself is (strictly) proper.
The score is also standardized in the following sense.
Suppose that $p$ is a calibrated forecaster and $q$ is the ``least skillful'' calibrated forecaster, i.e., the constant forecaster that predicts the historical average (\emph{climatology} in weather forecasting).
Then, the expected Winkler score $w(p,q;r)$ is zero (minimum) when $p = q$ and one (maximum) when $p \in \{0,1\}$. 
The \emph{empirical} Winkler score $w(p,q,y)$ can take negative values, which would suggest that $p$ is worse than $q$ on forecasting the outcome $y$ under $S$.

In the following lemma, we summarize the characteristics of the Winkler score that are useful for both its interpretation and the proofs that will follow shortly.
\begin{lemma}[\citet{winkler1994evaluating}]\label{lem:winkler}
Let $S$ be a proper scoring rule. 
Then, for any $p, q \in (0, 1)$ and $y \in \{0, 1\}$,
\begin{equation}
    w(p, q, y) = \begin{cases}
    1 & \mbox{if\;\;} y = \indicator{p > q}; \\
    \leq 0 & \mbox{otherwise.}
    \end{cases}
\end{equation}
In the case that $y \neq \indicator{p > q}$, the denominator is non-negative and the numerator is non-positive.
\end{lemma}

See~\citet{winkler1994evaluating,winkler1977rewarding} for a proof.
Lemma~\ref{lem:winkler} establishes that $p$ gets a positive score of $1$ if it is at least as good as $q$, but otherwise, it does not get a positive score. 
Two implications are: (i) the Winkler score is bounded from above by $1$, and (ii) when we take the average of pointwise Winkler scores over $t$ forecasts and outcomes, we can read off the sign of the average to tell whether $p$ has better or worse forecasting skills than $q$.

Returning to the sequential setup in Game~\ref{game:general}, we now treat the pointwise Winkler scores between $(p_t)_{t=1}^\infty$ and $(q_t)_{t=1}^\infty$ as the analogs of pointwise score differentials from Section~\ref{sec:cs}.
Because $(p_t)_{t=1}^\infty$ and $(q_t)_{t=1}^\infty$ are predictable w.r.t.~$\frakG$, we replace the expectation in \eqref{eqn:expected_winkler} with the conditional expectation w.r.t.~$\calG_{t-1}$.
Then, for each $t$, we can define the \emph{(expected) average Winkler score} up to $t$:
\begin{equation}\label{eqn:average_expected_winkler}
    W_t := \frac{1}{t}\sum_{i=1}^t \mathbb{E}_{t-1}[w(p_i, q_i, y_i)], \quad t \geq 1.
\end{equation}
This is the time-varying sequence of parameters that we seek to estimate; we also analogously define the \emph{weak Winkler (WW)} null
\begin{equation}\label{eqn:weak_null_winkler}
    \calH_0^{\sfw\sfw,\geq}(p,q): W_t \geq 0, \quad \forall t \geq 1.
\end{equation}
For this null, the sign is the opposite of~\eqref{eqn:weak_null}: we assert that $p$ is at least as good as $q$ as our null, and rejecting $\calH_0^{\sfw\sfw,\geq}(p,q)$ would mean that $p$ is decidedly worse than $q$ on average up to some time $t$.
Note also that we slightly generalize the average score from \citet{winkler1994evaluating}'s to allow the baseline forecaster to be any predictable $(0,1)$-valued forecaster $(q_t)_{t=1}^\infty$. 

We are now ready to present our main result.
In the following, we denote the (empirical) pointwise Winkler scores as $\hat{w}_i = w(p_i, q_i, y_i)$ for each $i$ and their average over time as $\hat{W}_t := \frac{1}{t}\sum_{i=1}^t w(p_i, q_i, y_i)$.
\begin{proposition}[Sequential inference on the average Winkler score]\label{ppn:winkler}
Suppose that $S$ is a proper scoring rule and that $p_i, q_i \in (0, 1)$ for each $i \geq 1$.
Let $(\underline\gamma_i)_{i=1}^\infty$ be a $\left[-1, \infty \right)$-valued predictable process and let $\hat{V}_t = \sum_{i=1}^t (\hat{w}_i - \underline\gamma_i)^2$. 
\begin{enumerate}
    \item (One-sided EB CS for $(W_t)_{t=1}^\infty$.) 
    For each $\alpha \in (0, 1)$, the sequence of intervals $(C_t^\mathsf{EB})_{t=1}^\infty$ defined as
    \begin{equation}\label{eqn:ebcs_winkler_onesided}
        C_t^\mathsf{EB} := \inparen{-\infty, \hat{W}_t + t^{-1}{u_\alpha(\hat{V}_t)}} \cap (-\infty, 1]
    \end{equation}
    is a $(1-\alpha)$-CS for $(W_t)_{t=1}^\infty$, for any sub-exponential uniform boundary $u_\alpha$ with crossing probability $\alpha$ and scale $2$.
    \item (Sub-exponential e-process for $\calH_0^{\sfw\sfw,\geq}$.) 
    For each $\lambda \in [0, 1/2)$, the process $(E_t(\lambda))_{t=0}^\infty$ defined as $E_0(\lambda)=1$ and
    \begin{equation}\label{eqn:eprocess_winkler_onesided}
        E_t(\lambda) := \exp\incurly{ -\lambda \hat{W}_t - \psi_{E,2}(\lambda)\hat{V}_t }
    \end{equation}
    is an e-process for $\calH_0^{\sfw\sfw,\geq}: W_t \geq 0,\; \forall t$, and so is the mixture process $E_t^\mathsf{mix} := \int E_t(\lambda) dF(\lambda)$ for any distribution $F$ on $[0, 1/c)$.
\end{enumerate}
\end{proposition}
The proof is a direct application of Proposition~\ref{ppn:subexp_bdd_below}, using the upper bound of $1$ on the empirical pointwise Winkler scores.
Because the Winkler score is unbounded from below, the standard machinery only readily provides the upper confidence bound for $(W_t)_{t=1}^\infty$.
Thus, we derive a one-sided CS in~\eqref{eqn:ebcs_winkler_onesided} that tells us the certainty to which we know $W_t$ is away from 1. 
The sub-exponential e-process in~\eqref{eqn:eprocess_winkler_onesided} corresponds to this upper confidence bound and measures the evidence against the null that $p$ is at least as good as $q$. 
From the sequential testing point-of-view, either a large value in the e-process or a small value of the upper confidence bound suggests that $p$ \emph{under}performs $q$; conversely, either a small value in the e-process or a value close to 1 for the upper confidence bound (i.e., a vacuous CS) tells us that there is no such evidence.
Note that, to satisfy the constraint on the predictable process $(\gamma_i)_{i=1}^\infty$ to be bounded from below by $-1$, we can choose as default the running average as in Theorem~\ref{thm:main}, but cap it from below at $-1$, i.e., $\gamma_i = -1 \vee \hat{W}_{i-1}$.

\begin{proof}[Proof of Proposition~\ref{ppn:winkler}]
We first use Lemma~\ref{lem:winkler} to obtain an upper bound of 1 on the pointwise empirical Winkler scores, $w_i = w(p_i, q_i, y_i)$.
Then, the rest of the proof follows similarly from the proofs of Proposition~\ref{ppn:subexp_bdd_below} as well as Theorem~\ref{thm:main} and Theorem~\ref{thm:eprocess}.

Specifically, define the process $(L_t(\lambda))_{t=0}^\infty$ as $L_0(\lambda)=1$ and
\begin{align}
    L_t(\lambda) &:= \exp \incurly{ \lambda \inparen{-\hat{W}_t + W_t} - \psi_{E,2}(\lambda) \hat{V}_t },
\end{align}
which is a test supermartingale an w.r.t.~$\frakG$ for each $\lambda \in [0, 1/2)$ by Proposition~\ref{ppn:subexp_bdd_below} and Lemma~\ref{lem:winkler}.
By definition, the process $(t(W_t -\hat{W}_t))_{t=0}^\infty$ is sub-exponential with scale $2$ (i.e., sub-$\psi_{E,2}$) having the variance process $(\hat{V}_t)_{t=0}^\infty$.
The results then follow analogously to Theorems~\ref{thm:main} and~\ref{thm:eprocess}.
\end{proof}

We close with the note that, if the main goal is rather to tightly estimate $(W_t)_{t=1}^\infty$ from both sides or to test the null $\calH_0^{\sfw\sfw,\leq}: W_t \leq 0,\; \forall t$, then there is a way to use either the sub-Gaussianity or the boundedness assumption on scoring rules (rather than propriety) and apply any of our main Theorems; the proof would be analogous for each application. 
The caveat with the Winkler score is that it is unbounded from below even when using a bounded base scoring rule, such as the Brier score, because the lower bound depends on how close $q$ can get to $0$ or $1$.
If $q_t = q \in (0,1)$ is the climatology forecaster, then this is not an issue, and the two-sided approach can also be useful.
We summarize the analogs of Theorem~\ref{thm:main} and Theorem~\ref{thm:eprocess} for the average Winkler score as a corollary.
\begin{corollary}[Two-sided sequential inference on the average Winkler score.]
Suppose there exists some $c>0$ such that $\hat{w}_i \geq 1-c$ for any $i \geq 1$.
Let $(\gamma_i)_{i=1}^\infty$ be a $\insquare{1-c, 1}$-valued predictable process and let $\hat{V}_t = \sum_{i=1}^t(\hat{w}_i - \gamma_i)^2$.
Then,
\begin{enumerate}
    \item (Two-sided EB CS for $(W_t)_{t=1}^\infty$.) 
    For each $\alpha \in (0, 1)$, the sequence of intervals $(C_t^\mathsf{EB})_{t=1}^\infty$ defined as
    \begin{equation}\label{eqn:ebcs_winkler_twosided}
        C_t^\mathsf{EB} := \inparen{ \hat{W}_t \pm t^{-1}{u_{\alpha/2}(\hat{V}_t)} } \cap (-\infty, 1]
    \end{equation}
    is a $(1-\alpha)$-CS for $(W_t)_{t=1}^\infty$, for any sub-exponential uniform boundary $u_{\alpha/2}$ with crossing probability $\alpha/2$ and scale $c$.
    \item (Sub-exponential e-process for $\calH_0^{\sfw\sfw,\leq}$.) 
    For each $\lambda \in [0, 1/c)$, the process $(E_t(\lambda))_{t=0}^\infty$ defined as $E_0(\lambda)=1$ and
    \begin{equation}\label{eqn:eprocess_winkler_twosided}
        E_t(\lambda) := \exp\incurly{ \lambda \hat{W}_t - \psi_{E,c}(\lambda)\hat{V}_t }
    \end{equation}
    is an e-process for $\calH_0^{\sfw\sfw,\leq}: W_t \leq 0,\; \forall t$, and so is the mixture process $E_t^\mathsf{mix} := \int E_t(\lambda) dF(\lambda)$ for any distribution $F$ on $[0, 1/c)$.
\end{enumerate}
\end{corollary}
The value of $c$ may depend on both the choice of $S$ and how close $q_i$ can get to either $0$ or $1$.
For example, if $S$ is the Brier score and $q_i \in [q_0, 1-q_0]$ for some constant $q_0 \in (0,1)$, then $c = 2/q_0$.

\section{Comparing Lagged Forecasts}\label{sec:lagh}

Given an integer lag $h \geq 1$, if $p_i$ and $q_i$ were lag-$h$ forecasts made at round $i$ for the eventual outcome $y_{i+h-1}$, then we would be interested in the following time-varying parameter:
\begin{equation}\label{eqn:fsd_lag}
    \Delta_t^{(h)} := \frac{1}{t-h+1} \sum_{i=1}^{t-h+1} \mathbb{E}_{i-1}\insquare{S(p_i, y_{i+h-1}) - S(q_i, y_{i+h-1})}, \quad \forall t \geq h.
\end{equation}
For each $t \geq h$, we take the average up to the $(t-h+1)$th round, because the forecasts made beyond that round can only be evaluated after the $t$th round.
The conditional expectation is taken in such a way that the forecasters ($p_i$ and $q_i$) are evaluated based on the information they had at the time of forecasting ($\calG_{i-1}$) and not the one right before the outcome is realized ($\calG_{i+h-1}$).

The case of $h=1$ corresponds to the setting we considered in Section~\ref{sec:cs}, but extending the construction to the case of $h>1$ is not straightforward. 
For example, the sequence $(E_t(\lambda))_{t=0}^\infty$ defined analogously to the one in Theorem~\ref{thm:eprocess} would \emph{not} be an e-process w.r.t.~the game filtration $\frakG$, let alone a process, because the $t$th term would include future outcomes that are not realized at time $t$.
Rather, the process $(E_t(\lambda))_{t=0}^\infty$ now only satisfies the weaker property that $\mathbb{E}_{t-h}[E_t] \leq 1$ for all (non-stopping) times $t \geq h$ under $\calH_0$.
In their recent work, \citet{arnold2021sequentially} refer to such processes as \emph{sequential e-values for $\calH_0$ at lag $h$} and propose to combine $h$ subsequences of the original process that are each test supermartingales w.r.t.~different sub-filtrations of $\frakG$.

Although lag-$h$ sequential e-values are not e-processes themselves, the recent preprints of \citet{arnold2021sequentially,henzi2021valid} show that there is a workaround to turn them into an e-process possessing anytime-validity.
Here, we adapt their approach and develop e- and p-processes for weaker nulls similar to the weak null in the lag-1 case; developing a tight CS for estimating $\Delta_t^{(h)}$ remains an open problem.

To proceed, we define two weak nulls related to the sequence of parameters $(\Delta_t^{(h)})_{t=h}^\infty$. 
The first is a straightforward generalization of the lag-1 weak null~\eqref{eqn:weak_null} to any $h \geq 1$: 
\begin{equation}\label{eqn:weak_null_lagh}
    \calH_0^\sfw(p,q;h): \Delta_t^{(h)} \leq 0, \quad \forall t \geq h.
\end{equation}
This recovers $\calH_0^\sfw(p,q)$ when $h=1$.
We refer to~\eqref{eqn:weak_null_lagh} as the \emph{lag-$h$ weak null} between $p$ and $q$.

Because of the aforementioned challenge in the $h>1$ case, we also define a null hypothesis for which we can derive a more powerful e-process. 
The \emph{lag-$h$ period-wise (PW) weak null}, which we denote as $\calH_0^{\sfp\sfw}(p,q; h)$, asserts that the weak null holds at every $h$th step for all periods $k \in \{1, \dotsc, h\}$, making it (slightly) stronger than the weak null but weaker than the strong null. 

Formally, define the index set
\begin{equation}
    I_t^{[k]} = \incurly{ k+1+hs : s = 0, 1, \dotsc, \floor{\frac{t-k}{h}} - 1 },
\end{equation}
which includes every $h$th round of the game starting at $k+1$ up to (at most) $t-h+1$. 
(For $t < h+k$, $I_{t}^{[k]} = \emptyset$.)
Now, for each $k = 1, \dotsc, h$, we define $\Delta_t^{[k]} := \frac{1}{t-h+1}\sum_{i \in I_t^{[k]}} \delta_i$, so that $\sum_{k=1}^h \Delta_t^{[k]} = \Delta_t^{(h)}$.
Then, the lag-$h$ PW weak null is defined as
\begin{equation}\label{eqn:weak_null_lagh_sub}
    \calH_0^{\sfp\sfw}(p,q;h): \Delta_t^{[k]} \leq 0,\quad \forall t \geq h,\; \forall k = 1, \dotsc, h.
\end{equation}
It is clear from their definitions that the following inclusion relationships hold between the three null hypotheses:
\begin{equation}\label{eqn:null_inclusions}
\calH_0^\sfw(h) \supseteq \calH_0^{\sfp\sfw}(h) \supseteq \calH_0^\sfs(h)
\end{equation}
for any $h \geq 1$.
When $h$ is a small integer (say, 5 or 10) and $t$ grows large, the lag-$h$ PW weak null is still much weaker than the lag-$h$ strong null.

Having defined the two nulls, we first present an e-process and a p-process for the lag-$h$ PW null~\eqref{eqn:weak_null_lagh_sub}. 
Because we cannot straightforwardly derive an e-process for $h > 1$, we start with a p-process constructed using the lag-$h$ sequential e-values and then use a p-to-e calibrator~\citep{shafer2011test} to obtain an e-process that remains valid at arbitrary stopping times.
An analogous proposition for~\eqref{eqn:weak_null_lagh} is shown later and relies on similar proof techniques.

Let $\hat\delta_i^{(h)} = S(p_i, y_{i+h-1}) - S(q_i, y_{i+h-1})$ be the empirical pointwise score differential for lag-$h$ forecasts.
Note that $\delta_i^{(h)} = \mathbb{E}_{i-1}[\hat\delta_i^{(h)}]$.
In addition, we say that a function $f: [0,1] \to [0, \infty)$ is a \emph{p-to-e calibrator} if it is non-increasing and satisfies $\int_0^1 f(u)du = 1$.
\begin{proposition}[Sequential inference for $\calH_0^{\sfp\sfw}(h)$]\label{ppn:lagh_pw}
Suppose that $|\hat\delta_i^{(h)}| \leq \frac{c}{2}$ for all $i \geq 1$, for some $c \in (0, \infty)$.
Let $(\gamma_i)_{i=1}^\infty$ be a $[-\frac{c}{2}, \frac{c}{2}]$-valued predictable process w.r.t.~$\frakG$.
Also, for each $k \in \{1, \dotsc, h\}$ and $\lambda \in [0, 1/c)$, define
\begin{align}
    E_t^{[k]}(\lambda) &= \prod_{i \in I_{t}^{[k]}} \exp\incurly{\lambda \hat\delta_i^{(h)}  - \psi_{E,c}(\lambda)\inparen{\hat\delta_i^{(h)} - \gamma_i}^2}, \quad \forall t \geq 0,
\end{align}
where $\prod_{i \in \emptyset}(\cdot) = 1$.
Then, for each $\lambda \in [0, 1/c)$, the following statements are true:
\begin{enumerate}
    \item (Averaged sequential e-values.) The process 
        \begin{equation}\label{eqn:seq_evalues_pw}
            \bar{E}_t^{\sfp\sfw}(\lambda) := \frac{1}{h}\sum_{k=1}^h E_t^{[k]}(\lambda), \quad \forall t \geq 0,
        \end{equation}
        is adapted w.r.t.~$\frakG$ and satisfies $\mathbb{E}_P[\bar{E}_{\tau+h-1}^{\sfp\sfw}(\lambda)] \leq 1$ for any $\frakG$-stopping time $\tau$ and any $P \in \calH_0^{\sfp\sfw}(p,q;h)$.
    \item (P-process.) The process $(\sfp_t^{\sfp\sfw})_{t=1}^\infty$ defined by
        \begin{equation}\label{eqn:pprocess_pw}
            \sfp_t^{\sfp\sfw} := \frac{he\log h}{\sum_{k=1}^h \inparen{1/\sfp_t^{[k]}}}, \quad\text{where}\quad \sfp_t^{[k]} := 1 \wedge \inparen{1/\sup_{i\leq t} E_i^{[k]}(\lambda)}, \quad \forall t \geq 0,
        \end{equation}
        is a p-process for $\calH_0^{\sfp\sfw}(p,q;h)$ w.r.t.~$\frakG$.
    \item (Calibrated e-process.) Let $f: [0,1] \to [0, \infty)$ be any p-to-e calibrator. Then, the process $(E_t^{\sfp\sfw})_{t=0}^\infty$ defined by $E_0^{\sfp\sfw} = 1$ and
        \begin{equation}
            E_t^{\sfp\sfw} := f(\sfp_t^{\sfp\sfw}), \quad \forall t \geq 1
        \end{equation}
        is an e-process for $\calH_0^{\sfp\sfw}(p,q;h)$ w.r.t.~$\frakG$.
    \end{enumerate}
\end{proposition}
The structure of the index set ensures that $E_{t}^{[k]}(\lambda)$ for each $k$ is adapted and non-increasing under the null.
For example, with lag-3 forecasts, $E_{t}^{[k]}(\lambda)$ for each $k$ is computed using each of the subsequences $(1, 4, 7, \dotsc)$, $(2, 5, 8, \dotsc)$, and $(3, 6, 9, \dotsc)$. 
As for the choice of a p-to-e calibrator $f$, we follow \citet{vovk2021evalues,ramdas2021testing} and use (as our default)
\begin{equation}
    f(p) = \frac{1-p + p \log p}{p(\log p)^2},\quad p \in [0,1].
\end{equation}

In words, sequential e-values are expected to be at most 1 at time $\tau+h-1$, where $\tau$ is any stopping time w.r.t.~$\frakG$.
In contrast, the p-process directly yields a valid sequential test without such a condition, and it can also be calibrated to yield an e-process.

\begin{proof}[Proof of Proposition~\ref{ppn:lagh_pw}]
Our goal is to derive a p-process for $\calH_0^{\sfp\sfw}(h)$ based on ideas from the proofs of Proposition 3.4 in \citet{arnold2021sequentially} and from the validity of their proposed sequential test, and then to calibrate it into an e-process~\citep{shafer2011test,ramdas2021testing}.

\paragraph{Sub-filtrations $\frakG^{[k]}$ and processes $L_t^{[k]}$.}
Recall that $\frakG = (\calG_t)_{t=0}^\infty$, and define the $\frakG^{[1]}, \dotsc, \frakG^{[h]}$ as follows: for each $k = 1, \dotsc, h$, 
\begin{equation}\label{eqn:filtration_k}
    \frakG^{[k]} := \inparen{\calG_t^{[k]}}_{t=0}^\infty,\quad \text{where} \quad  \calG_t^{[k]} := \calG_{\floor{\frac{t-k}{h}}h+k}.
\end{equation}
Because $\floor{\frac{t-k}{h}}h+k \leq \inparen{\frac{t-k}{h}}h+k \leq t$, we have $\calG_t^{[k]} \subseteq \calG_t\; \forall t$, i.e., $\frakG^{[k]}$ is a sub-filtration of $\frakG$ for each $k$.
(Each $\calG^{[k]}$ only updates its filtration every $h$ steps.)

In the following, we fix $\lambda \in [0, 1/c)$ and omit any dependence on it for notational convenience.
For each $k = 1, \dotsc, h$, define the process $(L_t^{[k]})_{t=0}^\infty$ as follows: $L_0^{[k]} := 1$ and, for each $t \geq 1$,
\begin{align}
    L_t^{[k]} := \prod_{i \in I_t^{[k]}} l_{i-1}(y_{i+h-1}),
\end{align}
where $\prod_{i \in \emptyset}(\cdot) = 1$ and
\begin{equation}\label{eqn:lagh_term}
    l_{i-1}(y_{i+h-1}) := \exp\incurly{\lambda \inparen{\hat\delta_{i}^{(h)} - \delta_{i}^{(h)}} - \psi_{E,c}(\lambda)\inparen{\hat\delta_{i}^{(h)} - \gamma_{i}}^2}.
\end{equation}
(We index \eqref{eqn:lagh_term} by $i-1$, because it only consists of $\calG_{i-1}$-measurable terms aside from $y_{i+h-1}$. For example, $\delta_{i}^{(h)} = \mathbb{E}_{i-1}[\hat\delta_i^{(h)}]$ is $\calG_{i-1}$-measurable.)
Then, each $(L_t^{[k]})_{t=0}^\infty$ is an adapted process w.r.t.~$\frakG$, because the last index of $I_t^{[k]}$ is at most $t-h+1$, and the outcome corresponding to that index is $y_t$, which is $\calG_t$-measurable.

\paragraph{$(L_t^{[k]})_{t=0}^\infty$ is a test supermartingale w.r.t.~$\frakG^{[k]}$ for each $k$.}
Recall that $\mathbb{E}[\hat\delta_{i}^{(h)} \mid \calG_{i-1}] = \delta_{i}^{(h)}$ by definition. 
Since the score differentials are bounded by assumption, the proof of Proposition~\ref{ppn:subexp_bdd_below} (with $y_i$ replaced with $y_{i+h-1}$ in the proof) implies that
\begin{equation}\label{eqn:l_leq_1_lag}
    \mathbb{E}\insquare{l_{i-1}(y_{i+h-1}) \mid \calG_{i-1}} \leq 1 \quad \forall i \geq h.
\end{equation}
Now, if $t < h$ or $\floor{\frac{t-k}{h}} \neq \frac{t-k}{h}$ (i.e., not an integer), then $I_t^{[k]} = I_{t-1}^{[k]}$ by construction, so $L_t^{[k]} = L_{t-1}^{[k]}$.
On the other hand, if $t \geq h$ and $\floor{\frac{t-k}{h}} = \frac{t-k}{h}$, then algebra shows that $L_t^{[k]} = L_{t-1}^{[k]} \cdot l_{t-h}(y_t)$, and also that $\calG_{t-1}^{[k]} = \calG_{\floor{\frac{(t-1)-k}{h}}h+k} =\calG_{\inparen{\frac{t-k}{h}-1}h+k} = \calG_{t-h}$.
Thus,
\begin{equation}
    \mathbb{E}\insquare{L_t^{[k]} \mid \calG_{t-1}^{[k]}} = L_{t-1}^{[k]} \cdot \mathbb{E}\insquare{l_{t-h}(y_t) \mid \calG_{t-h}} \leq L_{t-1}^{[k]}.
\end{equation}
The above algebra also shows that each multiplicative increment of $L_t^{[k]}$ is either constant (1) or $\frakG_t^{[k]}$-measurable. 
Therefore, $(L_t^{[k]})_{t=0}^\infty$ is a test supermartingale w.r.t.~$\frakG^{[k]}$.

\paragraph{$(\bar{E}_t^{\sfp\sfw})_{t=0}^\infty$ is a sequential e-value of lag $h$ for $\calH_0^{\sfp\sfw}$ (w.r.t.~$\frakG$).}
Under any $P \in \calH_0^{\sfp\sfw}(p,q;h)$, we know that
\begin{equation}
     \Delta_t^{[k]} = \sum_{i \in I_t^{[k]}} \delta_i^{(h)} \leq 0, \quad \forall t \geq h.
\end{equation}
We thus have, $P$-almost surely,
\begin{align}
    E_t^{[k]} &= \prod_{i \in I_t^{[k]}} \exp\incurly{\lambda \hat\delta_i^{(h)} - \psi_{E,c}(\lambda)\inparen{\hat\delta_i^{(h)} - \gamma_i}^2} \\
    &\leq \exp\incurly{-\sum_{i \in I_t^{[k]}} \delta_i^{(h)}} \cdot \prod_{i \in I_t^{[k]}} \exp\incurly{\lambda \hat\delta_i^{(h)} - \psi_{E,c}(\lambda)\inparen{\hat\delta_i^{(h)} - \gamma_i}^2} 
    = L_t^{[k]}, \quad \forall t \geq h. \label{eqn:e_lb_lagh}
\end{align}
In other words, under any $P \in \calH_0^\sfw(p,q;h)$, $E_t^{[k]}$ is upper-bounded by $L_t^{[k]}$ for each $k$, where $(L_t^{[k]})_{t=0}^\infty$ is a test supermartingale w.r.t.~$\frakG^{[k]}$.
By the supermartingale optional stopping theorem (e.g., Theorem 4.8.4, \citet{durrett2019probability}), we thus have that, for any stopping time $\tau^{[k]}$ w.r.t.~$\frakG^{[k]}$,
\begin{equation}
    \mathbb{E}_P\insquare{E_{\tau^{[k]}}^{[k]}} \leq 1,
\end{equation}
under any $P \in \calH_0^\sfw(p,q;h)$.

Finally, the construction~\eqref{eqn:filtration_k} implies that, for any stopping time $\tau$ w.r.t.~$\frakG$, the mapping $\tau \mapsto \tau^{[k]}$ defined by
\begin{equation}\label{eqn:tau_map}
    \tau^{[k]} := \inparen{\floor{\frac{\tau-k-1}{h}}+1}h + k
\end{equation}
gives a stopping time w.r.t.~$\frakG^{[k]}$~\citep{henzi2021valid}, where $\tau^{[k]} \in \{\tau, \tau+1, \dotsc, \tau+(h-1)\}$.
Therefore, for any stopping time $\tau$ w.r.t.~$\frakG$,
\begin{equation}
    \mathbb{E}_P[\bar{E}_{\tau+h-1}] \leq \frac{1}{h}\sum_{k=1}^h \mathbb{E}_P\insquare{E_{\tau^{[k]}}^{[k]}} \leq 1,
\end{equation}
for any $P \in \calH_0^\sfw(p,q;h)$.

\paragraph{$(\sfp_t^{\sfp\sfw})_{t=0}^\infty$ is a p-process for $\calH_0^{\sfp\sfw}$.}
The key idea here is to first use the fact that $L_t^{[k]}$ is a test supermartingale w.r.t.~$\frakG^{[k]}$ that upper-bounds $E_t^{[k]}$, for each $k \in \{1, \dotsc, h\}$, and then use the time-uniform equivalence lemma for probabilities~\citep{ramdas2020admissible}, along with a p-merging function~\citep{vovk2021evalues}, to obtain a combined p-process.

First, define the following process for each $k = 1, \dotsc, h$: 
\begin{equation}
    \sfq_t^{[k]} := 1 \wedge \inparen{1/\sup_{i\leq t} L_i^{[k]}}, \quad \forall t \geq 1.
\end{equation}
The process involves the running supremum of $(L_t^{[k]})_{t=0}^\infty$, which is a test supermartingale w.r.t.~$\frakG^{[k]}$ as we showed earlier.
In particular, \eqref{eqn:e_lb_lagh} implies that $\sfp_t^{[k]} \geq \sfq_t^{[k]}$ for all $t$ and $k$ under $P \in \calH_0^{\sfp\sfw}$.

Applying \citet{ville1939etude}'s inequality to $(L_t^{[k]})_{t=0}^\infty$, for any $P$,
\begin{equation}
    P\inparen{\exists t \geq 1: \sfq_t^{[k]} \leq \alpha} 
    = P\inparen{\sup_{t \geq 1} L_i^{[k]} \geq \frac{1}{\alpha}} \leq \alpha, \quad \forall \alpha \in (0,1).
\end{equation}
Then, under any $P \in \calH_0^{\sfp\sfw}$, the fact that $\sfp_t^{[k]} \geq \sfq_t^{[k]}$ under $P$ implies
\begin{equation}\label{eqn:lagh_pprocess}
    P\inparen{\exists t \geq 1: \sfp_t^{[k]} \leq \alpha} \leq \alpha, \quad \forall \alpha \in (0,1).
\end{equation}

Now, following an earlier proof in~\eqref{eqn:lagh_term} where we showed that $(L_t^{[k]})_{t=0}^\infty$ is an adapted process w.r.t.~the game filtration $\frakG$, we can analogously show that $(E_t^{[k]})_{t=0}^\infty$ is also an adapted process w.r.t.~$\frakG$, and so is $(\sfp_t^{[k]})_{t=0}^\infty$ by its definition.
Then, by Lemma 2 of \citet{ramdas2020admissible}, (i) $\Rightarrow$ (iii), equation~\eqref{eqn:lagh_pprocess} implies that
\begin{equation}\label{eqn:lagh_pprocess_k}
    P\inparen{\sfp_\tau^{[k]} \leq \alpha} \leq \alpha, \quad \forall \alpha \in (0,1),
\end{equation}
for any stopping time $\tau$ w.r.t.~$\frakG$ and $P \in \calH_0^\mathsf{pw}(h)$.
In other words, $(\sfp_t^{[k]})_{t=1}^\infty$ is a p-process for $\calH_0^{\sfp\sfw}(h)$ w.r.t.~$\frakG$, for each $k \in \{1, \dotsc, h\}$.

Finally, we can merge the p-processes $(\sfp_t^{[k]})_{t=1}^\infty$ at any $\frakG$-stopping times.
For any $\frakG$-stopping time $\tau$, using the harmonic average p-merging function by \citet{vovk2021evalues} combined with \eqref{eqn:lagh_pprocess_k} gives, for any $P \in \calH_0^{\sfp\sfw}$,
\begin{equation}
    P\inparen{\sfp_\tau^{\sfp\sfw} \leq \alpha} \leq \alpha, \quad \forall \alpha \in (0, 1).
\end{equation}

\paragraph{$(E_t^{\sfp\sfw})_{t=0}^\infty$ is an e-process for $\calH_0^{\sfp\sfw}.$}
This follows directly from the validity of a p-to-e calibrator for p-processes (e.g., Proposition 12, \citet{ramdas2020admissible}).
\end{proof}

The statements and proofs for the weak null $\calH_0^\sfw(h)$ are completely analogous, except that instead of taking averages across the $h$ sub-processes we have to take the minimum/maximum for e-/p-processes, because the weak null only implies that there exists some $k$ for which $\Delta_t^{[k]} \leq 0$.

\begin{proposition}[Sequential inference for $\calH_0^\sfw(h)$]\label{ppn:lagh_w}
    Assume the same setup as Proposition~\ref{ppn:lagh_pw}. 
    Then, for each $\lambda \in [0, 1/c)$, the following statements are true:
    \begin{enumerate}
        \item (Minimum sequential e-values.) The process 
            \begin{equation}\label{eqn:seq_evalues_w}
                \bar{E}_t^{\sfw}(\lambda) := \min_{k=1,\dotsc,h} E_t^{[k]}(\lambda)
            \end{equation}
            satisfies $\mathbb{E}_P[\bar{E}_{\tau+h-1}^{\sfp\sfw}(\lambda)] \leq 1$ for any $\frakG$-stopping time $\tau$ and any $P \in \calH_0^{\sfw}(p,q;h)$.
        \item (P-process.) The process $(\sfp_t^{\sfw})_{t=1}^\infty$ defined by
            \begin{equation}\label{eqn:pprocess_w}
                \sfp_t^{\sfw} := \max_{k=1,\dotsc,h} \sfp_t^{[k]}, \quad\text{where}\quad \sfp_t^{[k]} := 1 \wedge \inparen{1/\sup_{i\leq t} E_i^{[k]}(\lambda)},
            \end{equation}
            is an p-process for $\calH_0^{\sfw}(p,q;h)$ w.r.t.~$\frakG$.
        \item (Calibrated e-process.) Let $f: [0,1] \to [0, \infty)$ be any p-to-e calibrator. Then, the process $(E_t^{\sfw})_{t=0}^\infty$ defined by $E_0^{\sfw} = 1$ and
            \begin{equation}
                E_t^{\sfw} := f(\sfp_t^{\sfw}), \quad \forall t \geq 1
            \end{equation}
            is an e-process for $\calH_0^{\sfw}(p,q;h)$ w.r.t.~$\frakG$.
        \end{enumerate}
\end{proposition}

The methods described in Propositions~\ref{ppn:lagh_pw} and~\ref{ppn:lagh_w} both provide valid options for sequentially comparing lag-$h$ forecasters.
While $E_t^{\sfp\sfw}$ may involve a seemingly less intuitive null hypothesis, it upper-bounds $E_t^\sfw$, and it can grow more quickly when either null is false.
Rejecting $\calH_0^{\sfp\sfw}(p,q;h)$ implies that there exists some $k \in \{1, \dotsc, h\}$ such that $\Delta_t^{[k]} > 0$ for some $t$.
For example, if $h=2$, then it implies $p$ outperforms $q$ on average on either odd or even days.
A scenario in which rejecting $\calH_0^{\sfp\sfw}(h)$ would clearly not imply $\calH_0^\sfw(h)$ is when (coincidentally) there is seasonality of period exactly $h$ in the game --- e.g., when comparing 7-day forecasts for a sequence of outcomes that have a different distribution every weekend, $E_t^{\sfw}$ and $E_t^{\sfp\sfw}$ may differ significantly.
A simple way to mitigate this issue is to simply monitor both e-processes (depending on the use case).

In Table~\ref{tbl:weather_evalues_lagh}, we list the sequential e-values for $\calH_0^\sfw$ (Proposition~\ref{ppn:lagh_w}), $\calH_0^{\sfp\sfw}$ (Proposition~\ref{ppn:lagh_pw}), and $\calH_0^\sfs$ (\citet{henzi2021valid}; denoted as $\bar{E}^\sfs$), for the weather comparison tasks in Section~\ref{sec:weather} with lags $h=1, \dotsc, 5$.
As in \citet{henzi2021valid}, no stopping is applied in any of the sequential e-values.
As shown, while $\bar{E}^\sfw$ tends to be overly conservative, $\bar{E}^{\sfp\sfw}$ remains relatively powerful despite testing a substantially weaker null than the strong null (for $\bar{E}^\sfs$).
Across different locations and lags, $\bar{E}^\mathsf{s}$ is generally large ($\geq 20$) whenever $\bar{E}^\mathsf{pw}$ is large, and this is explained by the inclusion relationship between the nulls in~\eqref{eqn:null_inclusions}. 
The comparison of HCLR against HCLR\_ in Zurich is the only case where $\bar{E}^\mathsf{pw}$ exceeds $\bar{E}^\mathsf{s}$.
In this case, the e-values drawn over time (similar to Figure~\ref{fig:weather_plots}) show that there are multiple time periods (2012-2013 and 2014-2015) during which both $\bar{E}^\mathsf{s}$ and $\bar{E}^\mathsf{pw}$ decrease substantially, and it is possible that the choice of the hyperparameter or the variance-adaptivity of our e-values affects how quickly they ``rebound'' after such sharp decreases.

We close with the note that the choice of how aggressively one can bet, either via the choice of the hyperparameter in the mixture distribution $F$ for $\bar{E}^\sfw$ and $\bar{E}^{\sfp\sfw}$ (cf. Section~\ref{sec:eprocess}) or the alternative probability $\pi_1$ for $E^\sfs$, directly affects the power of these e-values. 
Developing powerful strategies for choosing $F$ in the lagged scenario remains a problem deserving of future investigation.

\begin{table}[t]
    \centering
    \small
    \begin{tabular}{c|c|rrr|rrr|rrr}
    \Xhline{1.1pt}
    \multirow{2}{*}{\bf Location} & \multirow{2}{*}{\bf Lag}  &  \multicolumn{3}{c|}{\bf HCLR/IDR}  & \multicolumn{3}{c|}{\bf IDR/HCLR\textsubscript{--}} & \multicolumn{3}{c}{\bf HCLR/HCLR\textsubscript{--}}  \\ 
                     &          & \multicolumn{1}{c}{$\bar{E}^\sfw$} & \multicolumn{1}{c}{$\bar{E}^{\sfp\sfw}$} & \multicolumn{1}{c|}{$\bar{E}^\sfs$}  & \multicolumn{1}{c}{$\bar{E}^\sfw$} & \multicolumn{1}{c}{$\bar{E}^{\sfp\sfw}$} & \multicolumn{1}{c|}{$\bar{E}^\sfs$} & \multicolumn{1}{c}{$\bar{E}^\sfw$} & \multicolumn{1}{c}{$\bar{E}^{\sfp\sfw}$} & \multicolumn{1}{c}{$\bar{E}^\sfs$} \\
    \hline	\multirow{5}{*}{Brussels}
                 & 1    &   0.012 &   0.012 &   0.000 & $>$ 100 & $>$ 100 & $>$ 100 &   1.083 &   1.083 & $>$ 100 \\
                 & 2    &   0.021 &   0.033 &   0.000 &   0.196 &   1.659 & $>$ 100 &   0.510 &   1.196 & $>$ 100 \\
                 & 3    &   0.049 &   0.060 &   0.006 &   0.060 &   0.121 &   1.786 &   0.698 &   2.289 & $>$ 100 \\
                 & 4    &   0.053 &   1.032 &  22.811 &   0.018 &   0.042 &   0.000 &   0.114 &   1.855 & $>$ 100 \\
                 & 5    &   0.145 &   0.714 & $>$ 100 &   0.021 &   0.034 &   0.000 &   0.254 &  19.411 & $>$ 100 \\
    \hline	\multirow{5}{*}{Frankfurt}
                 & 1    &   0.034 &   0.034 &   0.000 &   1.284 &   1.284 & $>$ 100 & $>$ 100 & $>$ 100 & $>$ 100 \\
                 & 2    &   0.022 &   0.029 &   0.000 &   1.573 &   7.223 & $>$ 100 &   1.537 &  69.508 & $>$ 100 \\
                 & 3    &   0.022 &   0.041 &   0.000 &   0.311 &   3.814 & $>$ 100 &   0.836 & $>$ 100 & $>$ 100 \\
                 & 4    &   0.047 &   0.214 &   0.361 &   0.033 &   0.090 &   0.122 &   0.163 &  27.920 & $>$ 100 \\
                 & 5    &   0.037 &   0.334 &   2.468 &   0.023 &   0.104 &   0.001 &   0.173 &   1.781 & $>$ 100 \\
    \hline	\multirow{5}{*}{London}
                 & 1    &   0.041 &   0.041 &   0.029 &   0.277 &   0.277 &   1.351 &   0.285 &   0.285 &   2.845 \\
                 & 2    &   0.038 &   0.038 &   0.021 &   0.289 &   0.321 &   2.002 &   0.164 &   0.200 &   5.178 \\
                 & 3    &   0.037 &   0.061 &   0.185 &   0.087 &   0.367 &   0.203 &   0.141 &   0.241 &   9.613 \\
                 & 4    &   0.077 &   0.121 &   1.751 &   0.051 &   0.108 &   0.018 &   0.077 &   1.714 &   8.428 \\
                 & 5    &   0.070 &   0.208 &   4.949 &   0.032 &   0.066 &   0.002 &   0.113 &   0.279 &   1.427 \\
    \hline	\multirow{5}{*}{Zurich}
                 & 1    &   0.034 &   0.034 &   0.003 &   6.670 &   6.670 &  25.692 & $>$ 100 & $>$ 100 &  61.747 \\
                 & 2    &   0.054 &   0.061 &   0.012 &   0.328 &   0.415 &  19.229 &   2.195 & $>$ 100 &  74.745 \\
                 & 3    &   0.066 &   0.487 &   1.079 &   0.037 &   0.197 &   0.661 &   1.877 &   7.311 &  94.613 \\
                 & 4    &   0.091 &   1.553 &  30.478 &   0.023 &   0.066 &   0.004 &   0.210 &  54.131 &  47.069 \\
                 & 5    &   0.082 &   8.436 & $>$ 100 &   0.026 &   0.053 &   0.000 &   0.192 &   3.964 &  40.648 \\
    \Xhline{1.1pt}
    \end{tabular}
    \caption{Lag-$h$ sequential e-values between pairs of statistical postprocessing methods for ensemble weather forecasts across different locations and lags, where $T$ is the last time step (January 01, 2017). 
    $\bar{E}^\sfw$, $\bar{E}^{\sfp\sfw}$, and $\bar{E}^\sfs$ indicate the lag-$h$ sequential e-values for the lag-$h$ weak, period-wise weak, and strong nulls, respectively.
    All procedures use the Brier score as the scoring rule.
    ``p/q'' indicates the null that ``p is no better than q.''
    Generally speaking, $\bar{E}^\sfw$ is the most conservative, while $\bar{E}^{\sfp\sfw}$ can be powerful against its relatively weak null (compared to the strong null for $\bar{E}^\sfs$).
    }
    \label{tbl:weather_evalues_lagh}
\end{table}

\section{Inference for Predictable Subsequences and Bounds}\label{sec:predictable}

Martingale theory tells us that we can substitute each variable in the exponential supermartingale~\eqref{eqn:expm} with any predictable terms, similar to $(\gamma_i)_{i=1}^\infty$ in Theorem~\ref{thm:main}.
In doing so, we must make sure that the resulting test supermartingale leads to estimating/testing an appropriate quantity of interest.
Here, we illustrate two useful extensions involving this general technique.

\subsection{Inference for Predictable Subsequences}\label{sec:predictable_conditions}

Suppose that each round of our forecast comparison game (Game~\ref{game:general}) happens daily, but we are only interested in comparing the forecasters on weekdays, on every other day, or more interestingly, on days after some specific event happens (e.g., days following market crashes). 
To formalize this, we introduce a predictable $\{0,1\}$-valued process $\xi := (\xi_t)_{t=1}^\infty$ and then estimate/test the average score differential \emph{only} at times when $\xi_t = 1$.
The resulting parameter of interest is expressed as follows:
\begin{equation}
    \Delta_t(\xi_{1:t}) := \frac{\sum_{i=1}^t \xi_i \delta_i}{\sum_{i=1}^t \xi_i} = \frac{1}{\sum_{i=1}^t \xi_i} \sum_{i=1}^t  \xi_i\mathbb{E}_{i-1}\insquare{S(p_i, y_i) - S(q_i, y_i)},
\end{equation}
where $\delta_i = \mathbb{E}_{i-1}[\hat\delta_i] = \mathbb{E}_{i-1}[S(p_i, y_i) - S(q_i, y_i)]$ and $\xi_{1:t} = (\xi_1, \dotsc, \xi_t)$.
$\Delta_t(\xi_{1:t})$ measures the time-varying average score differential \emph{only} for times when $\xi_i = 1$. 
\citet{henzi2021valid} introduce an analogous extension to testing the strong null~\eqref{eqn:strong_null}, where the predictable condition $\xi_t = \indicator{\max\{p_t, q_t\} \geq \frac{1}{2}}$ is used to compare extreme precipitation forecasts.

Because the conditions are predictable, we have the property that $\mathbb{E}_{i-1}[\xi_i\hat\delta_i] = \xi_i\mathbb{E}_{i-1}[\hat\delta_i] = \xi_i\delta_i$, from which the proofs of Theorem~\ref{thm:hoeffding} (assuming sub-Gaussianity), as well as Theorem~\ref{thm:main} and Theorem~\ref{thm:eprocess} (assuming boundedness), straightforwardly follow. 
For example, for each $\lambda \in [0, 1/c)$, consider
\begin{align}
    L_t(\lambda; \xi_{1:t}) &:= \prod_{i: \xi_i = 1} \exp\incurly{ \lambda (\hat\delta_i - \delta_i) - \psi_E(\lambda) (\hat\delta_i - \gamma_i)^2} \\
    &= \prod_{i=1}^t \insquare{ (1-\xi_i) + \xi_i \exp\incurly{ \lambda (\hat\delta_i - \delta_i) - \psi_E(\lambda) (\hat\delta_i - \gamma_i)^2}}. \label{eqn:pred_nsm}
\end{align}
Then, under the same conditions as Proposition~\ref{ppn:subexp_bdd_below}, $L_t(\lambda; \xi_{1:t})$ is a test supermartingale w.r.t.~$\frakG$:
\begin{align}
    \mathbb{E}_{t-1}[L_t(\lambda;\xi_{1:t})] 
    &= L_{t-1}(\lambda; \xi_{1:t-1}) \insquare{ (1-\xi_i) + \xi_i \mathbb{E}_{t-1} \exp\incurly{ \lambda (\hat\delta_i - \delta_i) - \psi_E(\lambda) (\hat\delta_i - \gamma_i)^2}} \label{eqn:predcond_eq} \\
    &\leq L_{t-1}(\lambda; \xi_{1:t-1}),  \label{eqn:predcond_ineq}
\end{align}
for each $t \geq 1$. 
We used the predictability of $(\xi_t)_{t=1}^\infty$ in~\eqref{eqn:predcond_eq} and the boundedness condition (see proof of Proposition~\ref{ppn:subexp_bdd_below}) in~\eqref{eqn:predcond_ineq}.
Applying this to the proof of Theorem~\ref{thm:main} shows that we can construct an EB CS for $(\Delta_t(\xi_{1:t}))_{t=1}^\infty$.

Similarly, we can also derive the corresponding sub-exponential e-process for the null $\calH_0^\sfw(\xi): \Delta_t(\xi_{1:t}) \leq 0 ,\; \forall t$.
This e-process is given by
\begin{equation}
    E_t(\lambda; \xi_{1:t}) := \prod_{i: \xi_i = 1} \exp\incurly{ \lambda \hat\delta_i - \psi_E(\lambda) (\hat\delta_i - \gamma_i)^2},
\end{equation}
for any $\lambda \in [0, 1/c)$.
This is an e-process because, under $\calH_0^\sfw(\xi)$, we have that $\exp(-\lambda \sum_{i=1}^t \xi_i \delta_i) = \prod_{i: c_i=1} \exp(-\lambda \delta_i) \geq 1$, and thus
\begin{align}
    E_t(\lambda; \xi_{1:t}) &\leq \prod_{i: \xi_i = 1} \exp\incurly{ \lambda (\hat\delta_i - \delta_i) - \psi_E(\lambda) (\hat\delta_i - \gamma_i)^2} = L_t(\lambda; \xi_{1:t}).
\end{align}
Since $E_t(\lambda; \xi_{1:t})$ is upper-bounded by the test supermartingale $L_t(\lambda; \xi_{1:t})$ for all $t$ under $\calH_0^\sfw(\xi)$, it follows that $E_t(\lambda; \xi_{1:t})$ is an e-process for $\calH_0^\sfw(\xi)$~\citep{ramdas2020admissible}.

In summary, both the CS and the e-process remain valid under predictable conditions.

\subsection{Inference Under Predictable Bounds}\label{sec:predictable_bounds}

For Theorems~\ref{thm:main} and~\ref{thm:eprocess}, we require that the pointwise score differentials are bounded by some fixed constant, i.e.,  $|\hat\delta_i| \leq \frac{c}{2}$ for all $i$, for some $c \in (0, \infty)$.
In practice, this may be restrictive when the value of $c$ is not known a priori or its range shifts drastically over time.
One way to mitigate this issue is to have a predictable bound $(c_i)_{i=1}^\infty$ at each round, such that
\begin{equation}
    \absval{\hat\delta_i} \leq \frac{c_i}{2},
\end{equation}
for $i \geq 1$, instead of having a uniform bound over all rounds.
Predictable bounds can also be useful in cases where one can guess how bad/good the forecasts can be before each new round begins.

Here, we show that we can extend both Theorem~\ref{thm:main} and Theorem~\ref{thm:eprocess} to work for predictably bounded score differentials.
This result depends on the following facts about the exponential CGF-like function, $\psi_{E,c}(\lambda)$, \emph{as a function of its scale $c$}.
Below, we take $1/0 = \infty$.
\begin{lemma}\label{lem:ecgf_convex}
    For each $\lambda \geq 0$, the function $f_\lambda(c) := \psi_{E,c}(\lambda) = c^{-2}[-c\lambda -\log(1-c\lambda)]$ is non-decreasing and convex on $c \in (0, 1/\lambda)$.
    Furthermore, $f_\lambda$ is strictly increasing and strongly convex on $c \in (0, 1/\lambda)$ if and only if $\lambda > 0$.
\end{lemma}
\begin{proof}
Since $f_\lambda(c)$ is twice differentiable w.r.t.~$c$, it suffices to show that $f_\lambda'(c) \geq 0$ and $f_\lambda''(c) \geq 0$ for all $c$, and also that $f_\lambda'(c) > 0$ and $f_\lambda''(c) > 0$ for all $c$ if and only if $\lambda > 0$.

Given that $0 \leq c\lambda < 1$, we utilize the Taylor series of $x \mapsto -\log(1-x)$ at $x=0$:
\begin{equation}
    -\log(1-c\lambda) = \sum_{t=1}^\infty \frac{(c\lambda)^t}{t} = c\lambda + \frac{c^2\lambda^2}{2} + \frac{c^3\lambda^3}{3} + \dotsb,
\end{equation}
which converges (absolutely).
It then follows that 
\begin{align}
    f_\lambda(c) = \frac{-c\lambda -\log(1-c\lambda)}{c^2} 
    = \frac{\lambda^2}{2} + \frac{c\lambda^3}{3} + \dotsb
    = \lambda^2 \sum_{t=0}^\infty \frac{(c\lambda)^t}{t+2}.
\end{align}
Taking first derivatives term-by-term,
\begin{align}
    f_\lambda'(c) = \lambda^2 \sum_{t=1}^\infty \frac{t\lambda^tc^{t-1}}{t+2}.
\end{align}
Given that $c > 0$, we have that $f_\lambda'(c) \geq 0$ for any $\lambda \geq 0$.
Furthermore, we have that $f_\lambda'(c) > 0$ for $\lambda > 0$ and $f_\lambda'(c) = 0$ for $\lambda = 0$.

Similarly, taking second derivatives term-by-term,
\begin{align}
    f_\lambda''(c) = \lambda^2 \sum_{t=2}^\infty \frac{t(t-1) \lambda^t c^{t-2}}{t+2}.
\end{align}
Given that $c > 0$, we have that $f_\lambda''(c) \geq 0$ for any $\lambda \geq 0$.
Furthermore, we have that $f_\lambda''(c) > 0$ for $\lambda > 0$ and $f_\lambda''(c) = 0$ for $\lambda = 0$.
\end{proof}

\begin{figure}[t]
    \centering
    \includegraphics[width=0.48\textwidth]{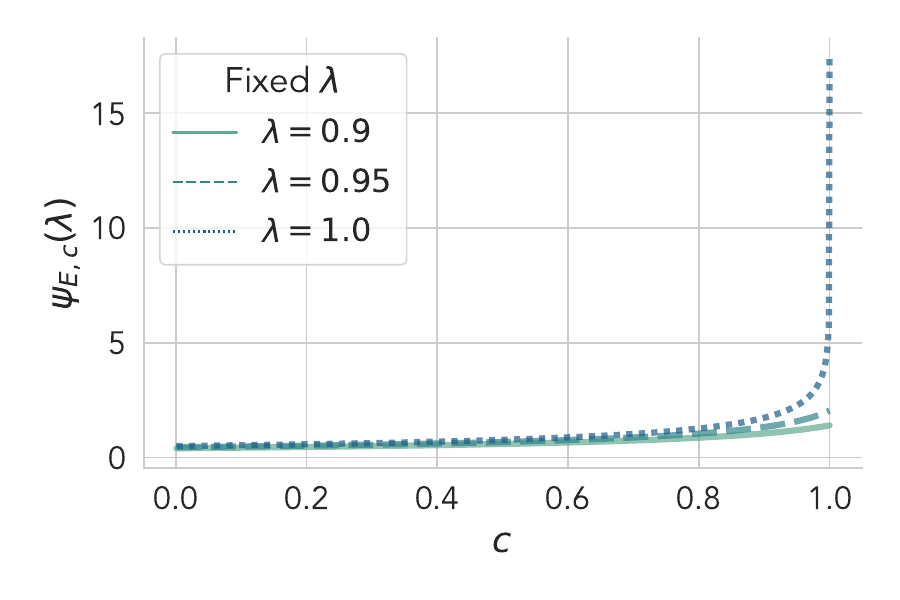}
    \includegraphics[width=0.48\textwidth]{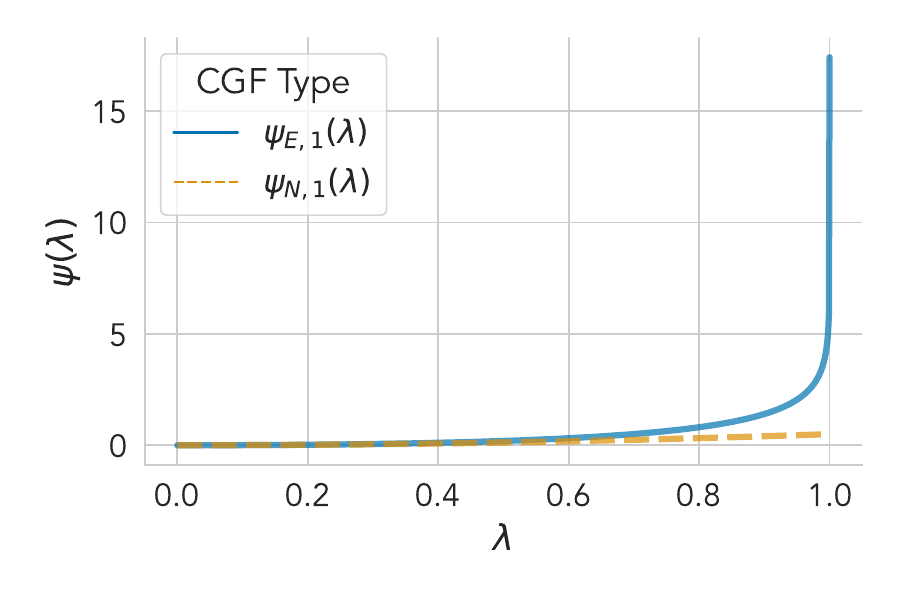}
    \caption{\emph{Left:} Plots of the exponential CGF-like function $f_\lambda(c) = \psi_{E,c}(\lambda)$ against $c \in (0, 1/\lambda)$, for fixed $\lambda$ values of $0.9$, $0.95$, and $1.0$.
    For each $\lambda \geq 0$, $f_\lambda(c)$ is strictly increasing and strongly convex on $c \in (0, 1/\lambda)$.
    \emph{Right:} Comparing $\psi_{E,1}(\lambda)$, as a function of $\lambda \in [0, 1)$, with the Gaussian CGF $\psi_{N,1}(\lambda) = \lambda^2/2$.
    }
    \label{fig:cgfs}
\end{figure}
In Figure~\ref{fig:cgfs}, we plot $\psi_{E,c}(\lambda)$ as a function of $c$, illustrating that it is indeed strictly increasing and strongly convex for different values of $\lambda > 0$, and we also show that $\psi_{E,1}$ as a function of $\lambda$ approximates $\psi_{N,1}(\lambda) = \lambda^2/2$ as $\lambda \to 0^+$.

Now, we derive an e-process that involves predictable bounds and is upper-bounded by a test supermartingale that uses a uniform bound~\eqref{eqn:expm}.
First, let $c_0$ be a (possibly infinite) constant such that $c_i \leq c_0$ for all $i$.
Also, let $\hat{v}_i = (\hat\delta_i - \gamma_i)^2$ where $(\gamma_i)_{i=1}^\infty$ is any predictable sequence as in Theorems~\ref{thm:main} and~\ref{thm:eprocess}.

Now, for each $\lambda \in [0, 1/c_0)$ (as before, we set $1/\infty=0$ and $[0, 0) = \{0\}$), define the following processes: $\undertilde{L}_0(\lambda) = L_0(\lambda) = 1$, and for $t \geq 1$,
\begin{align}
    \undertilde{L}_t(\lambda) &:= \prod_{i=1}^t \exp \incurly{ \lambda \inparen{\hat\delta_i - \delta_i } - \psi_{E, c_0}(\lambda) \inparen{\hat\delta_i - \gamma_i}^2 }; \\
    L_t(\lambda) &:= \prod_{i=1}^t \exp \incurly{ \lambda \inparen{\hat\delta_i - \delta_i } - \psi_{E, c_i}(\lambda) \inparen{\hat\delta_i - \gamma_i}^2 }. 
\end{align}
(If $c_0 = \infty$, then $\psi_{E,c_0}$ is not well-defined, so set $\undertilde{L}_t(\lambda) = 1$ for all $t \geq 1$.)

\begin{proposition}
Suppose that $|\hat\delta_i| \leq \frac{c_i}{2}$, where $(c_i)_{i=1}^\infty$ is a strictly positive predictable sequence.
Also, let $\hat{V}_t = \sum_{i=1}^t (\hat\delta_i - \gamma_i)^2$, where $(\gamma_i)_{i=1}^\infty$ is any $\insquare{-\frac{c_i}{2}, \frac{c_i}{2}}$-valued predictable sequence.
Then, for each $\lambda \in [0, 1/c_0)$, the following statements are true:
\begin{enumerate}
    \item $\undertilde{L}_t(\lambda) \leq L_t(\lambda)$ for all $t \geq 1$;
    \item The process $(L_t(\lambda))_{t=0}^\infty$ is a test supermartingale w.r.t.~$\frakG$;
    \item \emph{(A predictably-bounded e-process.)} The process $(E_t(\lambda))_{t=0}^\infty$, defined as $E_0(\lambda) = 1$ and
    \begin{equation}\label{eqn:predbound_eprocess}
        E_t(\lambda) := \prod_{i=1}^t \exp \incurly{ \lambda \hat\delta_i - \psi_{E, c_i}(\lambda) \inparen{\hat\delta_i - \gamma_i}^2 },\; \forall t \geq 1,
    \end{equation}
    is an e-process for $\calH_0^\sfw(p, q): \Delta_t \leq 0, \; \forall t\geq 1$.
\end{enumerate}
\end{proposition}

\begin{proof}
\begin{enumerate}

\item Using the fact that $c_i \leq c_0$ for each $i$ and that $\psi_{E,c}(\lambda)$ is non-decreasing in $c$ by Lemma~\ref{lem:ecgf_convex}, we obtain
\begin{equation}
    \undertilde{L}_t(\lambda) = \exp\incurly{\lambda S_t - \psi_{E, c_0}(\lambda) \hat{V}_t} \leq L_t(\lambda).
\end{equation}

\item If $c_0 = \infty$, then we must have $\lambda = 0$, so $(L_t(\lambda))_{t=0}^\infty$ always takes the value 1 and is a (trivial) test supermartingale.
Otherwise, Proposition~\ref{ppn:eprocess_bdd_below} directly implies that $(L_t(\lambda))_{t=0}^\infty$ is a test supermartingale w.r.t.~$\frakG$.

\item Because $(c_i)_{i=1}^\infty$ is predictable w.r.t.~$\frakG$, the process $(E_t(\lambda))_{t=0}^\infty$ is adapted w.r.t.~$\frakG$. 
Then, $E_t(\lambda) \leq L_t(\lambda)$ ($P$-a.s.) for all $t$ under any $P \in \calH_0^\sfw(p,q)$, as in the proof of Theorem~\ref{thm:eprocess}, and thus the result follows by Corollary 22 of \citet{ramdas2020admissible}.
\end{enumerate}
\end{proof}

Note that, if a constant bound $c_0 = c > 0$ were known \emph{a priori}, then $\undertilde{L}_t(\lambda)$ coincides with the exponential test supermartingale in Equation~\eqref{eqn:expm}.
The e-process~\eqref{eqn:predbound_eprocess} can be more powerful than using the analogous $(\undertilde{E}_t(\lambda))_{t=0}^\infty$ involving $c_0$ in some cases, although taking the mixture over $\lambda$ (Section~\ref{sec:uniform_boundary}) may not yield a closed form.

\section{Generalizations To Other Outcome and Forecast Types}\label{sec:extensions}

In principle, the game-theoretic approach we describe in Section~\ref{sec:game} can straightforwardly generalize beyond the case of probability forecasts on dichotomous events. 
We briefly discuss two such generalizations and to what extent our methods are applicable in each case.

The first is to the case of $C$-categorical outcomes, for $C \geq 2$.
We can start with the game-theoretic setup (Game~\ref{game:general}) and parameterize the outcome space using $C$-dimensional length-$1$ binary vectors, i.e., $\calY = \{\bfe_c\}_{c=1}^C$ where $\bfe_c = [\indicator{i=c}]_{i=1}^C$, and the set of forecasts as the $C$-dimensional probability simplex, i.e., $\calP = \Delta^{C-1} = \{\bfp \in [0,1]^C : \sum_{c=1}^C p^{(c)} = 1\}$.
Reality also makes its choices from $\Delta^{C-1}$.
Note that, if $C = 2$, we can recover the binary case via the mapping $\bfp = (1-p, p)$, for $p \in [0, 1]$. 
Then, by choosing any bounded scoring rule for categorical outcomes, we can straightforwardly apply Theorems~\ref{thm:main} and~\ref{thm:eprocess} to obtain CSs and e/p-processes (respectively) on the average score differentials.
The $C$-dimensional Brier score, defined as $S(\bfp, \bfy) = 1-\norm{\bfp - \bfy}_2^2$, is bounded within $[0,1]$; the spherical and zero-one scores can be defined analogously~\citep{gneiting2007strictly} and are similarly bounded.
We note that using the normalized Winkler score to utilize unbounded scores, as in Section~\ref{sec:winkler}, is not straightforward.

The next extension is to the case of continuous outcomes.
In this case, we can once again start with the game-theoretic setup 
 (Game~\ref{game:general}) and parameterize the outcome space as $\calY \subseteq \R^d$ for some $d \geq 1$.
At each round $t$, Reality now chooses an arbitrary distribution $r_t$ on $\calY$, from which $y_t$ is sampled.
Depending on the specific forecasting task, the forecasters may either predict (i) certain functional(s) of the outcome distribution, denoted as $\Gamma(P)$ for each $P \in \calP$, or (ii) the CDF (or density) itself.
As an example for (i), each forecaster may predict a level-$\alpha$ (e.g., 95\%) prediction interval $(l_t, u_t)$, in which case the statistician can use the $\alpha$-interval score~\citep{dunsmore1968bayesian}:
\begin{equation}
    S_\alpha((l, u), y) = -(u-l) - (2/\alpha) (l-y) \indicator{y<l} - (2/\alpha) (y-u) \indicator{y>u}, 
\end{equation}
for $(l, u) \subseteq \calY$ and $y \in \calY$. 
As an example for (ii), each forecaster may predict a (Borel-measurable) CDF $F_t$ for $y_t$, in which case the statistician can use the continuously ranked probability score (CRPS)~\citep{matheson1976scoring}:
\begin{equation}
    S(F, y) = -\int_{-\infty}^\infty (F(x) - \indicator{x \geq y})^2 dx = \Ex{Y, Y' \sim F}{|Y - Y'|} - \Ex{Y \sim F}{|Y - y|},
\end{equation}
for any CDF $F$ and outcome $y \in \calY$.
In either case, our main results (Theorems~\ref{thm:main} and~\ref{thm:eprocess}) are applicable when the associated score differentials are bounded.
Specifically, we can allow the choices of $\calY$, $\calP$, and $S$ such that $\calP \subseteq \calP^{(c)}$, where
\begin{equation}\label{eqn:bounded}
    \calP^{(c)} = \incurly{ p \in \Delta(\calY): |S(p, y) - S(q, y)| \leq c/2, \; \forall q \in \Delta(\calY) }, 
\end{equation}
for some $c \in (0, \infty)$.
For instance, if $\calY = [0,1]$, then our main theorems can be used to compare mean, quantile, or interval forecasts on $\calY$, using the corresponding scoring rule in each case~\citep{gneiting2011making}.
If~\eqref{eqn:bounded} is restrictive for the use case, then one may consider using predictable bounds (Section~\ref{sec:predictable_bounds}) or the asymptotic CS (Section~\ref{sec:asympcs}).
Deriving a fully general anytime-valid procedure for unbounded domains and scoring rules remains an open problem.

In Table~\ref{tbl:games}, we summarize these extensions based on the different choices of the outcome space $\calY$ and the forecast type $\calP$ within Game~\ref{game:general}.

\begin{table}[htb]
    \centering
    \begin{tabular}{c|c|c|c}
        \Xhline{1.1pt} 
        \bf Outcome Type   & \bf Categorical & \multicolumn{2}{c}{\bf Continuous} \\ \hline
                Domain     & $\calY = \{\bfe_c\}_{c=1}^C$ & \multicolumn{2}{c}{$\calY \subseteq \R^d$} \\ 
         Reality's Choice &  $r_t \in \Delta^{C-1}$    &  \multicolumn{2}{c}{$r_t \in \Delta(\calY)$ (arbitrary distribution)} \\
        \Xhline{1.1pt}
        \bf Forecast Type  & \bf Probability & \bf Functional & \bf Distribution \\  \hline
                Domain     & $\calP = \Delta^{C-1}$  & $\Gamma(\calP)$  & $\calP \subseteq \Delta(\calY)$ \\ 
         Forecast Examples  & any $C$-dim. probability & mean, prediction interval & CDF \\
         Score Examples & Brier, spherical, 0-1, log scores & quadratic, interval scores & CRPS \\ 
         \Xhline{1.1pt}
         Thms.~\ref{thm:main}~\&~\ref{thm:eprocess} apply & \multicolumn{3}{c}{if $\calP \subseteq \calP^{(c)}$ for some $c \in (0, \infty)$} \\
        \Xhline{1.1pt} 
    \end{tabular}
    \caption{Different specifications of Game~\ref{game:general} based on the outcome space and the forecast type, and the types of scoring rules that can be used in each case.
    In principle, the game-theoretic setup in our main paper (Section~\ref{sec:game}) can straightforwardly extend to these settings; our main approaches (Theorems~\ref{thm:main} and~\ref{thm:eprocess}) extend to cases where the score differentials are bounded.}
    \label{tbl:games}
\end{table}

\section{Comparison with Other Forecast Comparison Methods}\label{sec:comparison} 

\subsection{Methodological Comparison with Henzi and Ziegel (2022)}\label{sec:comparison_hz}

The biggest difference between our approach and \citet{henzi2021valid}'s (HZ) is in the difference between the strong and weak nulls, as described in the main text.
Here, we summarize other methodological differences that are worth noting for practical use cases.
HZ focus on sequentially comparing forecasts on dichotomous events using consistent scoring functions~\citep{gneiting2011making}, which straightforwardly induce proper scoring rules, and they develop e-processes of the form
\begin{equation}\label{eqn:hz}
    E_t^\textsf{HZ}(\lambda_1, \dotsc, \lambda_t) = \prod_{i=1}^t \inparen{1 + \lambda_i \tilde\delta_i}, \quad \text{where}\quad 
    \tilde\delta_i = \frac{S(p_i, y_i) - S(q_i, y_i)}{|S(p_i, \indicator{p_i \geq q_i}) - S(q_i, \indicator{p_i \geq q_i})|},
\end{equation}
for a $[0,1]$-valued predictable sequence $(\lambda_t)_{t=1}^\infty$ and a \emph{negatively oriented} scoring function $S$. 
The form of $\tilde\delta_i$ is exactly that of the Winkler score: by Lemma~\ref{lem:winkler} and reversing the orientation of $S$, we see that $\tilde\delta_i = -w(p_i, q_i, y_i)$, and thus HZ's e-process can be interpreted as betting on the relative forecasting skill as determined by the pointwise empirical Winkler score~\eqref{eqn:winkler}.
In this sense, our e-process for the weak Winkler null in Proposition~\ref{ppn:winkler} is a weak-null counterpart of HZ's e-process.

In terms of the specific form of the e-process, \eqref{eqn:hz} is an example of a \emph{product} form e-process, contrasting with our \emph{exponential} form variant.
The two forms of e-processes are both found the literature, such as the product form in \citet{waudbysmith2020estimating} and the exponential form in \citet{howard2021timeuniform} for estimating bounded means.
Also, while the e-process we derive in~\eqref{eqn:eprocess} explicitly shows its variance-adaptive property and further utilizes the method of mixtures~\citep{robbins1970statistical}, HZ's e-process seeks to optimize its power by optimizing the growth rate of the e-process in the worst case (GROW)~\citep{grunwald2019safe} under a chosen alternative (typically set to a convex combination of $p_t$ and $q_t$).

In terms of use cases, the CSs perform estimation and thus provide information as to exactly \emph{how much} one forecaster is outperforming the other.
The methods in our paper are agnostic to the different types of outcomes (Section~\ref{sec:extensions}), so they can, e.g., be applied to forecasts on categorical outcomes with $C>2$ categories and to forecasts on bounded continuous outcomes.
HZ's approach is applicable to any consistent scoring functions~\citep{gneiting2011making} on binary outcomes and can also test for forecast dominance w.r.t.~all consistent scoring functions.

\subsection{Comparison with DM and GW Tests}\label{sec:dmgw}

As we highlighted in Section~\ref{sec:relatedwork}, the key difference between our work and existing forecast comparison methods, such as \citet{diebold1995comparing,giacomini2006tests,lai2011evaluating,ehm2018forecast}, is whether they have an anytime-valid guarantee.
Here, we present additional experiments to illustrate that (i) the DM and GW tests are \emph{not} valid at arbitrary stopping times, like most other classical tests including \citet{lai2011evaluating}, and (ii) anytime-valid methods need not require larger sample sizes than DM and GW tests for high power.

To recap, the DM test of \emph{unconditional} predictive ability tests 
\begin{equation}
    \calH_0^{\mathsf{DM}}: \mathbb{E}[\hat\delta_n] = 0, \quad \forall n \geq 1,
\end{equation}
where the scoring rule is assumed to depend only on the forecast error, e.g., $S(p_n, y_n) = 1-(p_n - y_n)^2$.
By the DM assumption, the loss differentials are assumed to be covariance stationary, implying that $\mathbb{E}[\hat\delta_n] = \delta$ for some fixed $\delta$ at any $n$.
Given the (stationary) autocovariance function $\gamma(k)$ for score differentials and a consistent estimator $\hat{f}(0)$ of its spectrum at frequency zero, the DM test uses the asymptotic normality under $\calH_0^{\mathsf{DM}}$ given by $\sqrt{n}(\hat\Delta_n - \mu)/\sqrt{2\pi \hat{f}(0)} \rightsquigarrow N(0,1)$.

The GW test, on the other hand, is a test of \emph{conditional} predictive ability that tests
\begin{equation}
    \calH_0^{\mathsf{GW}}: \mathbb{E}_{n-1}[\hat\delta_{m,n}] = 0, \quad \forall n \geq 1.
\end{equation}
Here, $m$ is the maximum window size that each forecaster can look back to, meaning that the test now depends on the forecasting model. 
The GW assumption allows for nonstationarity, although the test statistic involves weights that depend on mixing assumptions~\citep{lai2011evaluating}.

\begin{figure}[t]
    \centering
    \includegraphics[height=0.25\textwidth]{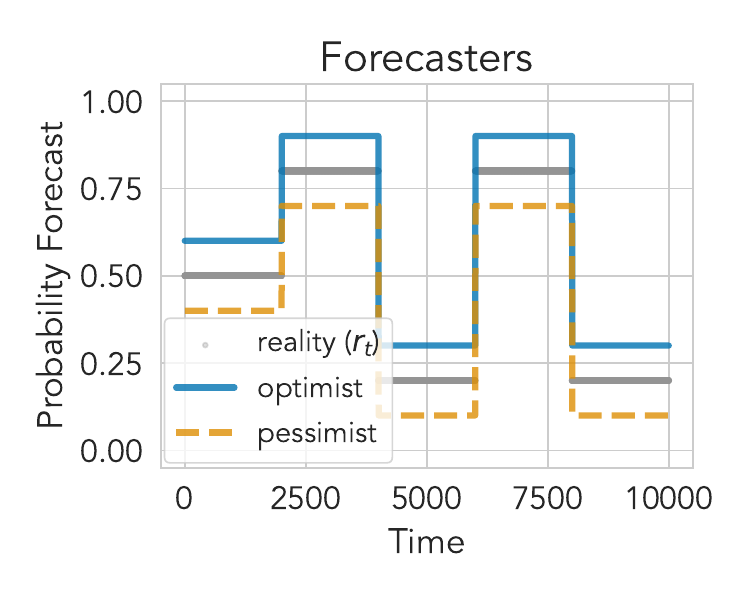}
    \includegraphics[height=0.25\textwidth]{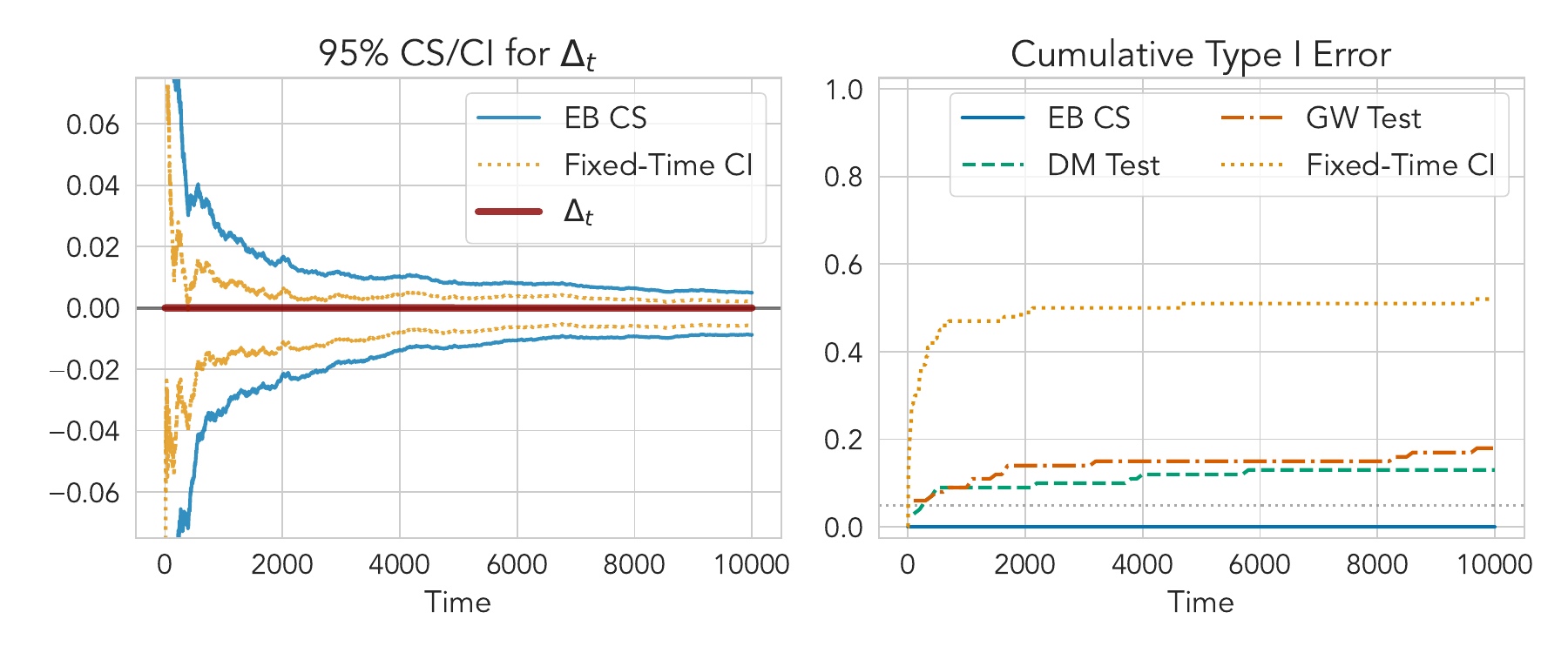}
    \caption{\emph{Left:} Two forecasters, denoted as \texttt{optimist} (blue) and \texttt{pessimist} (orange), on a simulated reality sequence (gray). There is no performance gap between the two in Brier score.
    \emph{Middle:} The true average score differentials $(\Delta_t)_{t=1}^T$ (dark red) along with the 95\% EB CS (blue) and the fixed-time CI (yellow). 
    \emph{Right:} Comparing the cumulative type I error rate for the EB CS (blue), the DM test of unconditional predictive ability (green), the GW test of conditional predictive ability (orange), and \citet{lai2011evaluating}'s asymptotic CIs (yellow).
    All tests are for one-sided nulls of the form ``optimist performs no better than the pessimist.''
    Unlike the EB CS, all classical fixed-time methods, including DM and GW tests, incur a cumulative miscoverage/false decision rate higher than $\alpha = 0.05$.
    }
    \label{fig:dmgw}
\end{figure}

First, we consider a simplistic setting in which $\Delta_t = 0$ for each time $t$ and both the DM and GW assumptions are met.
We compare two forecasters, named \texttt{optimist} ($p_t$) and \texttt{pessimist} ($q_t$), that are equally apart from Reality ($r_t$) in their forecasts (Figure~\ref{fig:dmgw}, left).
For all methods, we test their form of the null that ``the \texttt{optimist} is no better than the \texttt{pessimist}'' under the Brier score.
As expected, both the EB CS (Theorem~\ref{thm:main}) and the fixed-time CI~\citep{lai2011evaluating} to quickly shrink to zero (Figure~\ref{fig:dmgw}, middle), and also neither the DM nor GW test falsely rejects the null at $T=10,000$.

Now, we can also compute the cumulative type I error rate, which for p-values $(\sfp_t)$ is given by $\alpha_t = P\inparen{\exists i \leq t: \sfp_i \leq \alpha}$.
For CS/CIs $(C_t)$, this is equivalent in this case to the cumulative miscoverage rate $\alpha_t = P\inparen{\exists i \leq t: 0 \notin C_i}$ that we used earlier in Section~\ref{sec:simulated}, because $\Delta_t = 0$ under any $P \in \calH_0$.
The quantity is estimated over a repeated sampling of the data under $P$.
We expect that an anytime-valid procedure satisfies $\alpha_t \leq \alpha$ for any $t$ by definition, whereas classical fixed-time tests such as the DM and GW tests do not. 
As shown Figure~\ref{fig:dmgw} (right), the cumulative type I errors of both the DM and GW tests exceed the significance level of $\alpha = 0.05$ after roughly $100$ and $1000$ steps, respectively, and they continue to trend upward in log-scale.
This confirms that the p-values obtained by DM or GW tests, much like the fixed-time CI, are overconfident under continuous monitoring and thus at data-dependent stopping times, even when their assumptions are met.
In other words, the DM and GW tests, along with fixed-time CIs, do not have an anytime-valid guarantee.

\begin{figure}[t]
    \centering
    \includegraphics[width=0.32\textwidth]{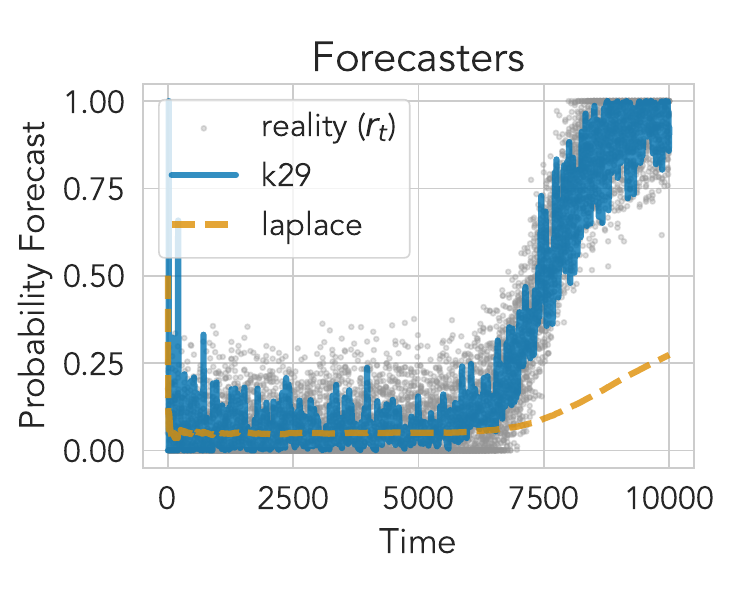}
    \includegraphics[width=0.32\textwidth]{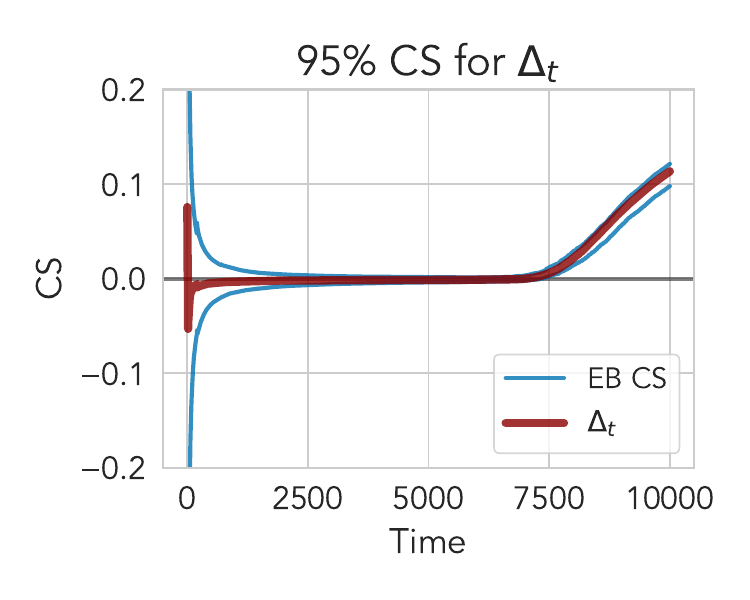}
    \includegraphics[width=0.32\textwidth]{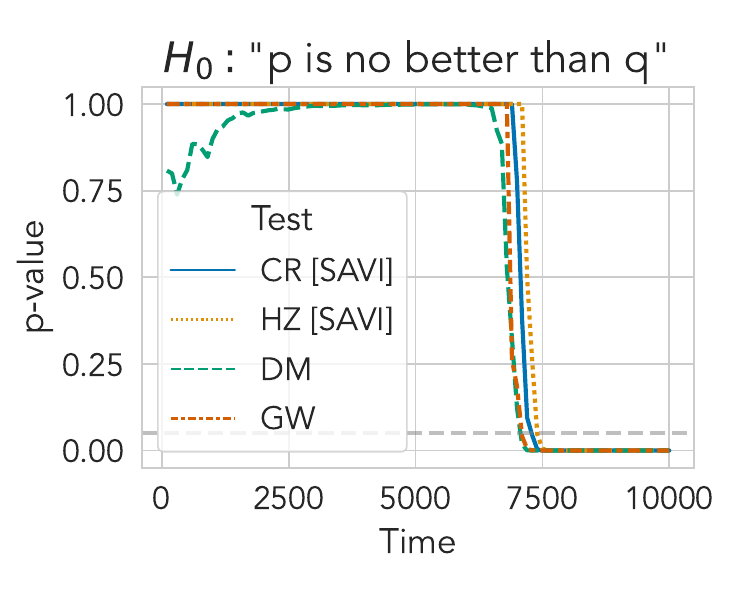}
    \caption{\emph{Left:} Two forecasters, \texttt{k29} (blue) and \texttt{laplace} (orange), on a simulated reality sequence (gray) that induces a changepoint in the loss differentials later in the time horizon.
    \emph{Middle:} The 95\% EB CS for $(\Delta_t)_{t=1}^T$ using the Brier score.
    $\Delta_t$ stays zero initially but trends positive later.
    \emph{Right:} P-values for the null ``\texttt{k29} is no better than \texttt{laplace}'' at each sample size $t$.
    CR (ours; blue) and HZ (yellow) are anytime-valid (SAVI), whereas DM (green) and GW (orange) are not.
    When $\Delta_t$ quickly trends positive ($t \approx 7300$), all p-values shrink to zero, and neither CR nor HZ requires substantially many extra samples to get to zero compared to DM and GW. 
    }
    \label{fig:dmgw_power}
\end{figure}

Next, we show that the anytime-validity of SAVI methods (CSs, e-processes, and p-processes), do not necessarily require larger sample sizes than the classical tests.
We compare two forecasters, \texttt{k29} with a 3-degree polynomial kernel ($p_t$) and \texttt{laplace} ($q_t$), whose average and pointwise score differentials stay close to zero for a while ($t \leq 7000$) until a sharp changepoint in the data is introduced and $\Delta_t$ trends positive afterwards (Figure~\ref{fig:dmgw_power}, left).
Note that this invalidates the covariance stationarity assumption of the DM test.
The EB CS for $\Delta_t$ is drawn in the middle plot of Figure~\ref{fig:dmgw_power}, which shows that the CS uniformly covers the time-varying average as expected.

To illustrate that SAVI approaches do not necessarily require larger sample sizes for ``detecting'' this changepoint, we compare SAVI and non-SAVI p-values for the null that ``\texttt{k29} is no better than \texttt{laplace}'' under the Brier score.
First, we plot the p-process, $\sfp_t = 1/\sup_{i\leq t}E_i$ given by~\eqref{eqn:pprocess}, where $(E_t)_{t=0}^\infty$ is the sub-exponential e-process~\eqref{eqn:eprocess} that corresponds to the LCB of the CS.
This is denoted in the right plot of Figure~\ref{fig:dmgw_power} (denoted as ``CR'').
We also plot the p-process constructed from \citet{henzi2021valid}'s e-process $(E_t^{\mathsf{HZ}})_{t=0}^\infty$ via the same mapping, i.e., $\sfp_t^{\mathsf{HZ}} = 1/\sup_{i\leq t}E_i^{\mathsf{HZ}}$.
As shown in the plot, when compared against the DM and GW p-values, both our and HZ's p-processes shrink to zero nearly as quickly, indicating that they require comparable amounts of data to reject the null when $\Delta_t$ trends positive.


\section{Additional Experiment Details and Results}\label{sec:additional_experiments}

\subsection{Additional Details \& Results from Numerical Simulations}\label{sec:simulated_app}

\subsubsection{Data Generation}\label{sec:simulated_details}

The reality sequence $(r_t)_{t=1}^T$ is specifically chosen to be non-IID and contain sharp changepoints, as drawn with gray dots in Figure~\ref{fig:forecasters}:
\begin{equation*}
    r_t = \insquare{0.8 \cdot \theta_t + 0.2 \cdot (1 - \theta_t)} + \epsilon_t ,
\end{equation*}
where
\begin{equation*}
\theta_t = \begin{cases}
    0.5 & \mbox{\indent for } t \in [1, 2000] \\
    1   & \mbox{\indent for } t \in [2001, 4000] \\
    0   & \mbox{\indent for } t \in [4001, 6000] \\
    1   & \mbox{\indent for } t \in [6001, 8000] \\
    0   & \mbox{\indent for } t \in [8001, 10000] \\
\end{cases}
\end{equation*}
and $\epsilon_t \sim \calN(0, 0.1^2)$ is an independent Gaussian noise for each $t$.

\subsubsection{All Pairwise Comparisons in Numerical Simulations}\label{sec:pairwise}

\begin{figure}[t]
    \centering
    \includegraphics[width=\textwidth]{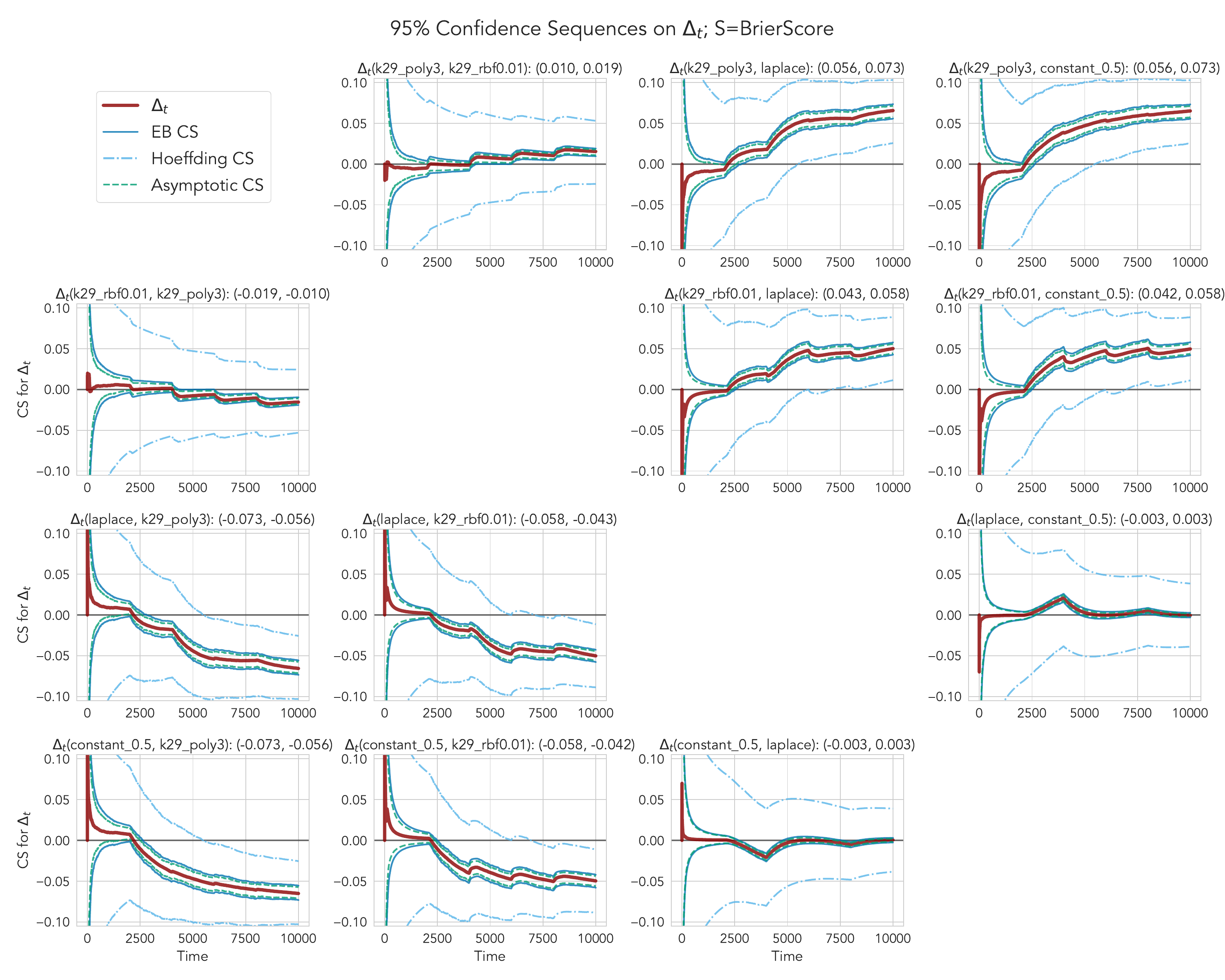}
    \caption{95\% EB (blue), Hoeffding-style (skyblue), and asymptotic (green) CSs on $\Delta_t$ between four different forecasters (\texttt{k29\_poly3}, \texttt{k29\_rbf0.01}, \texttt{laplace}, and \texttt{constant\_0.5}) plotted in Figure~\ref{fig:forecasters}.
    Scoring rule is the Brier score, and positive values of $\Delta_t$ indicate that the first forecaster is better than the second.
    In all comparisons, both CSs cover $\Delta_t$ uniformly, and the width of the EB CS approaches that of the asymptotic CS as time grows large.}
    \label{fig:pairwise}
\end{figure}

In Figure~\ref{fig:pairwise}, we plot the 95\% EB, Hoeffding-style, and asymptotic CSs for all pairwise comparisons between the constant baseline (\texttt{constant\_0.5}), the Laplace forecaster (\texttt{laplace}), and the K29 forecasters with the 3-degree polynomial kernel and the Gaussian RBF kernel with bandwidth $0.01$ (\texttt{k29\_poly3} and \texttt{k29\_rbf0.01}, respectively).
The Brier score is used.
Across all pairwise comparisons, both CSs uniformly cover the true score differentials across all times, regardless of whether the score differentials contain sharp changepoints and contain specific trends.

\clearpage

\subsection{Additional Details \& Results from the MLB Experiment}

For all MLB-related experiments, we choose $v_\text{opt} = 100$, given the longer time horizon considered (compared to other experiments in this paper).

\subsubsection{Details on the MLB Forecasters}\label{sec:mlb_forecasters}

\begin{figure}[t]
    \centering
    \includegraphics[width=\textwidth]{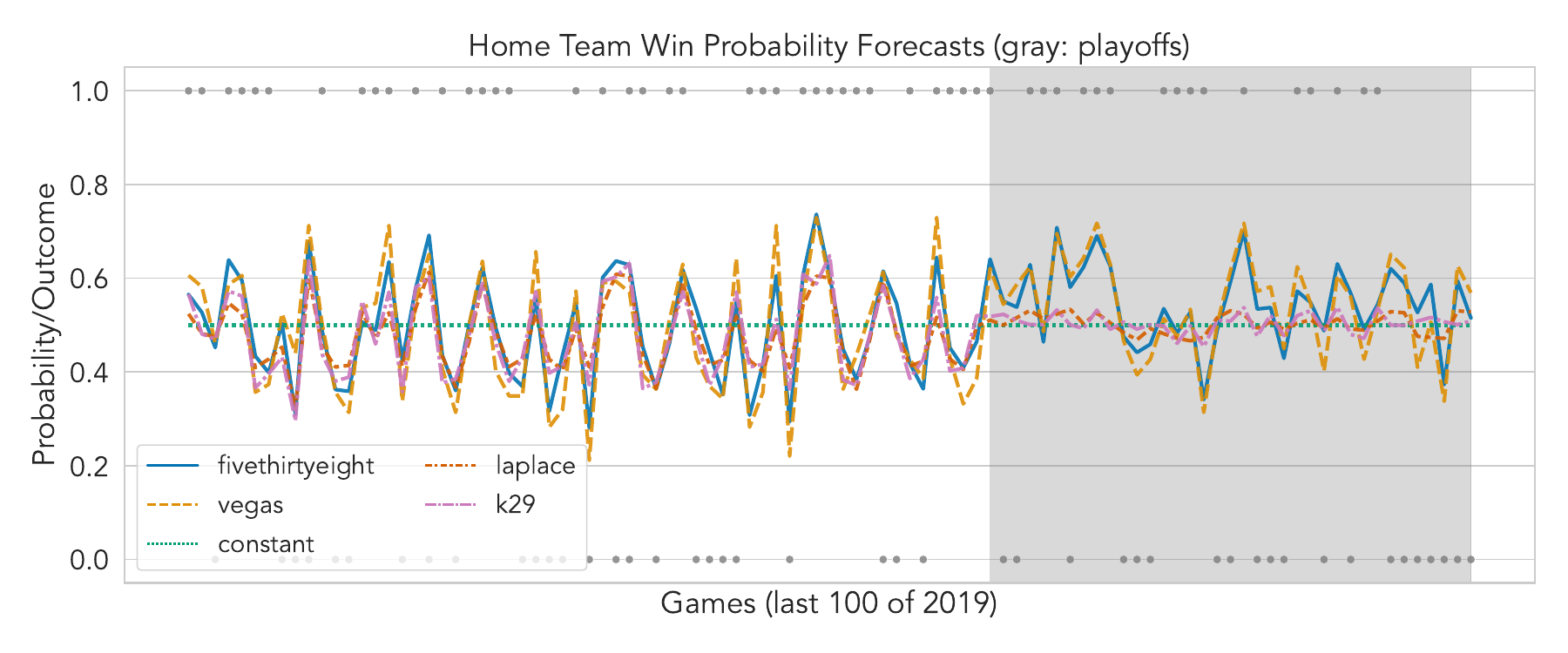}
    \caption{Various forecasters on the last 100 MLB games played in 2019 (including regular season and postseason). FiveThirtyEight and Vegas forecasts are publicly available forecasts online; Laplace and K29 forecasts are made using historical outcomes as data without external information. \emph{Note that the forecasts are computed using data from a 10-year window (2010 to 2019), but we only show the last 100 games here for visualization purposes.} The shaded region highlights the playoffs (the last seven being the World Series games).}
    \label{fig:mlb_forecasts}
\end{figure}

Here, we describe in detail the five Major League Baseball (MLB) forecasters that are compared in Section~\ref{sec:2019ws}.
Figure~\ref{fig:mlb_forecasts} illustrate their forecasts on the last 100 games of 2019.
\begin{itemize}
    \item \texttt{538}: Game-by-game probability forecasts on every MLB game since 1871, available at \url{https://data.fivethirtyeight.com/#mlb-elo}. According to the methodology report at \url{https://fivethirtyeight.com/features/how-our-mlb-predictions-work/}, the probabilities are calculated using an ELO-based rating system for each team, and game-specific adjustments are made for the starting pitcher as well as other external factors (travel, rest, home field advantage, etc.).
    Before each new season, team ratings are reverted to the mean by one-third and combined with preseason projections from other sources (Baseball Prospectus's PECOTA, FanGraphs' depth charts, and Clay Davenport’s predictions).
    \item \texttt{vegas}: Pre-game closing odds made on each game by online sports bettors, as reported by \url{https://Vegas-Odds.com}. (Download source: \url{https://sports-statistics.com/sports-data/mlb-historical-odds-scores-datasets/}.)
    The betting odds are given in the American format, so each odds $o$ is converted to its implied probability $p$ via $p = \indicator{o \geq 0} \frac{100}{100+o} + \indicator{o < 0}\frac{-o}{100-o}$.
    Then, for each matchup, the pair of implied probabilities for each team is rescaled to sum to 1. For example, given a matchup between team $A$ and team $B$ with betting odds $o_A = -140$ and $o_B = +120$, the implied probabilities are $\tilde{p}_A = 0.58$ and $\tilde{p}_B = 0.45$, and the rescaled probabilities are $p_A = 0.56$ and $p_B = 0.44$.
    \item \texttt{constant}: a constant baseline predicting $p_t = 0.5$ for each $t$.
    \item \texttt{laplace}: A seasonally adjusted Laplace algorithm, representing the season win percentage for each team. 
    Mathematically, it is given by $p_t = \frac{k_t + c_t}{n_t + 1}$, where $k_t$ is the number of wins so far in the season, $n_t$ is the number of games played in this season, and $c_t \in [0, 1]$ is a baseline that represents the final probability forecast from the previous season, reverted to the mean by one-third.
    For example, if the previous season ended after round $t_0$, then $k_t = \sum_{i=t_0}^{t-1} \indicator{y_i = 1}$, $n_t = t - t_0$, and $c_t = \frac{2}{3} \cdot p_{t_0} + \frac{1}{3} \cdot \frac{1}{2}$ (with $c_0 = \frac{1}{2})$. 
    The final probability forecast for a game between two teams is rescaled to sum to 1.
    \item \texttt{k29}: The K29 algorithm applied to each team, using the Gaussian kernel with bandwidth $0.1$, computed using data from the current season only. 
    The final probability forecast for a game between two teams is rescaled to sum to 1.
\end{itemize}

\begin{figure}[t]
    \centering
    \includegraphics[width=\textwidth]{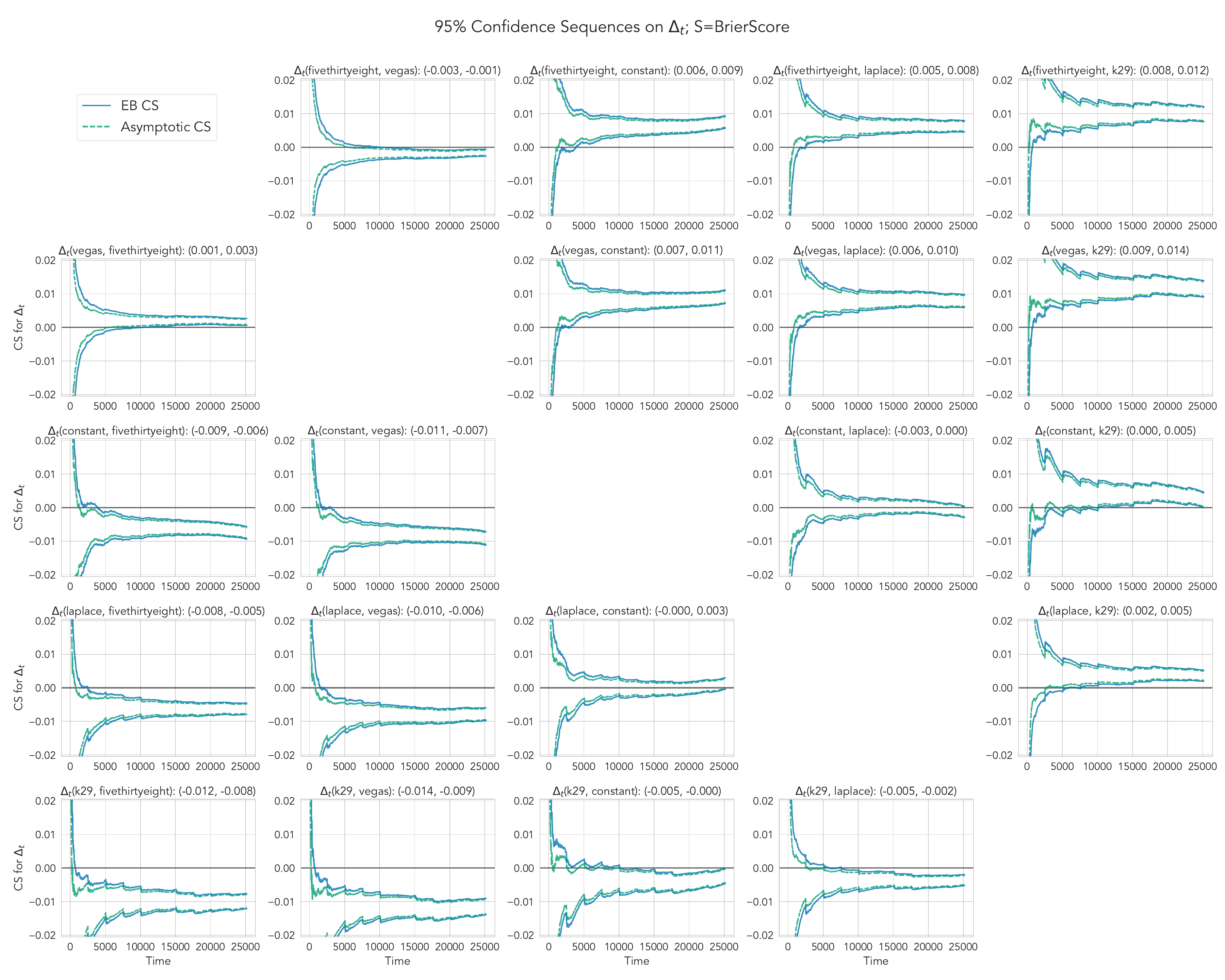}
    \caption{Comparing MLB win probability forecasts from 2010 to 2019, using the EB and Hoeffding-style CSs at significance level $\alpha = 0.05$. 
    $T = 25,165$ corresponds to the final game of the 2019 World Series.
    The Brier score is used.
    We find that, over time, the five forecasters are found to achieve significantly different predictive performance from each other (except \texttt{laplace} and \texttt{constant}), with the \texttt{vegas} forecaster achieving the best performance, followed by \texttt{fivethirtyeight}, \texttt{laplace} $\approx$ \texttt{constant}, and \texttt{k29}.
    The title of each subplot includes the 95\% EB CS at $T = 25,165$.
    }
    \label{fig:mlb_comparisons}
\end{figure}

\clearpage

\subsubsection{All Pairwise Comparisons of MLB Forecasters}\label{sec:mlb_comparisons}

Figure~\ref{fig:mlb_comparisons} includes all pairwise comparisons between the five MLB forecasters considered in our experiment.
See main text from Section~\ref{sec:2019ws} for further details.

\subsection{Additional Details \& Results from the Weather Experiment}\label{sec:weather_forecast}

\begin{figure}[t]
    \centering
    \includegraphics[width=\textwidth]{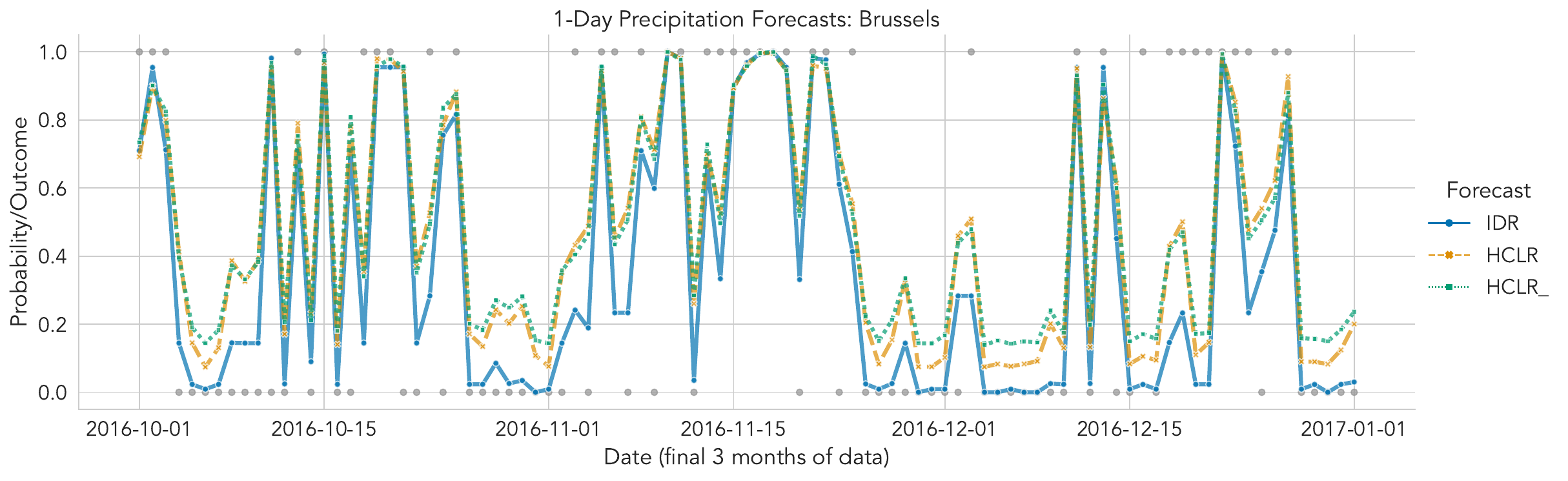}
    \caption{Comparing three statistical postprocessing methods (IDR, HCLR, HCLR\_) for 1-day ensemble weather forecasts on the Probability of Precipitation (PoP). The binary outcome is drawn as gray dots. 
    \emph{For visualization purposes, we plot the data and the forecasts only for the final 3 months (October 01, 2016 to January 01, 2017) and at one airport location (Brussels).}}
    \label{fig:weather_forecast}
\end{figure}

The setup closely follows the comparison experiment by \citet{henzi2021valid}, who compare statistical postprocessing methods for predicting the probability of precipitation (PoP) using the ensemble forecast data from the European Centre for Medium-Range Weather Forecasts (ECMWF; \citet{molteni1996ecmwf}).
The dataset includes the observed 24-hour precipitation from January 06, 2007 to January 01, 2017 at four airport locations (Brussels, Frankfurt, London Heathrow, and Zurich), and for each location and date it also includes 1- to 5-day ensemble forecasts, consisting of a higher resolution forecast, 50 perturbed ensemble forecasts at a lower resolution, and a control run for the perturbed forecasts.
They consider three statistical postprocessing methods in their experiments: isotonic distributional regression (IDR; \citet{henzi2021isotonic}), heteroscedastic censored logistic regression (HCLR; \citet{messner2014extending}), and a variant of HCLR without its scale parameter (HCLR\_).
Each method is applied to the first half of the data, separately for each airport location and lag $h = 1, \dotsc, 5$, and the second-half data is used to make sequential comparisons of the postprocessing methods.
Note that each location has a different number of observations: 3,406 for Brussels, 3,617 for Frankfurt, 2,256 for London, and 3,241 for Frankfurt.
See Section 5 in \citet{henzi2021isotonic} and Section 5.1 in \citet{henzi2021valid} for further details about the dataset and the postprocessing methods.

In Figure~\ref{fig:weather_forecast}, we plot the three forecasters (1-day) on the PoP for the final year (2016-2017) in Brussels.

\subsection{Fine-Tuning the CS Width Using Simulated IID Mean Differentials}\label{sec:iid_mean}

The uniform boundaries we use in our CSs come with hyperparameter(s) that one can choose to optimize the CS widths at specific intrinsic times (i.e., values that the non-decreasing sequence $(\hat{V}_t)_{t=1}^\infty$ can take).
As explained in Section~\ref{sec:boundary_details}, this choice can be thought of as an additional fine-tuning step and is secondary to choosing the type of uniform boundary.
Nevertheless, since it is a hyperparameter, we seek to find a reasonable default that can be used for typical scenarios of forecast comparison without an a priori knowledge of how large the intrinsic time can get.

To achieve this, we compare the widths of various CSs for the mean differential between two independent and identically distributed (IID) random variables. 
The main reason for using IID data is so that we can compare the width of our CSs with other CSs developed in previous work~\citep{howard2021timeuniform,waudbysmith2020estimating,waudbysmith2021doubly}, including ones that only apply to IID means.

\begin{figure}[t]
    \centering
    \includegraphics[width=\textwidth]{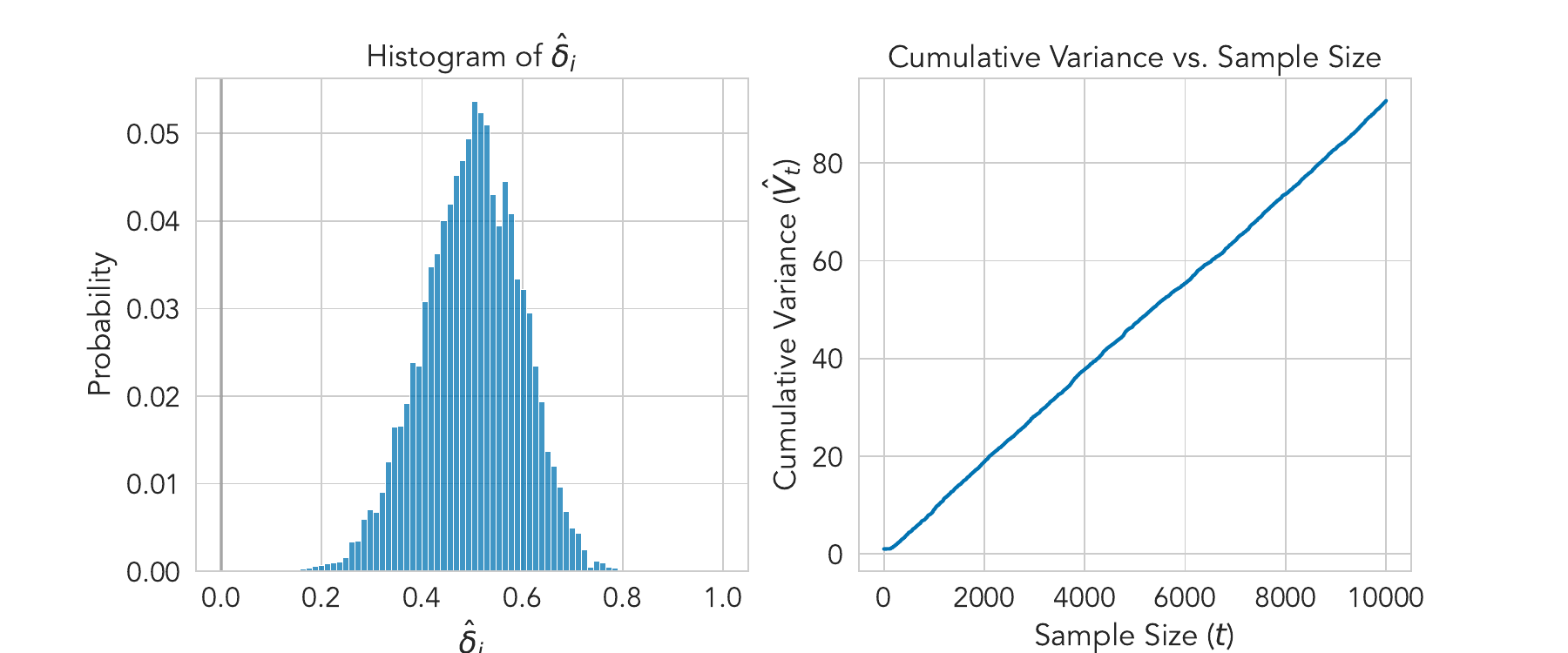}
    \caption{(Left) Histogram of $\hat\delta_i \overset{\text{IID}}{\sim} \mathrm{Beta}(30, 10) - \mathrm{Beta}(10, 30)$ for $i = 1, \dotsc, 10,000$. (Right) Plot of the cumulative variance (intrinsic time) $\hat{V}_t = \sum_{i=1}^t (\hat\delta_i - \hat\Delta_{i-1})^2$, where $\hat\Delta_{i-1} = \sum_{j=1}^{i-1} \hat\delta_j$. 
    Note that the hyperparameter $v_\text{opt}$, which we discuss below, determines the intrinsic time $\hat{V}_t$ at which the uniform boundary is the tightest.}
    \label{fig:iid_mean_data}
\end{figure}

To begin, we simulate score differences by sampling two IID Beta random variables and taking their differences: 
\begin{equation}\label{eqn:iid_delta}
    \hat\delta_i \overset{\text{IID}}{\sim} \mathrm{Beta}(30, 10) - \mathrm{Beta}(10, 30), \quad \forall i = 1, \dotsc, 10,000 .
\end{equation}
Note that $-1 \leq \hat\delta_i \leq 1$ a.s. and that $\mathbb{E}[\hat\delta_i] = \frac{30}{30+10} - \frac{10}{10+30} = \frac{1}{2}$.
Figure~\ref{fig:iid_mean_data} illustrates the data sampled according to~\eqref{eqn:iid_delta} (left) as well as the cumulative variance (intrinsic time) $\hat{V}_t = \sum_{i=1}^t (\hat\delta_i - \hat\Delta_{i-1})^2$, where $\hat\Delta_{i-1} = \sum_{j=1}^{i-1} \hat\delta_j$, over the sample size $t$ (right).

\begin{figure}[t]
    \centering
    \includegraphics[width=\textwidth]{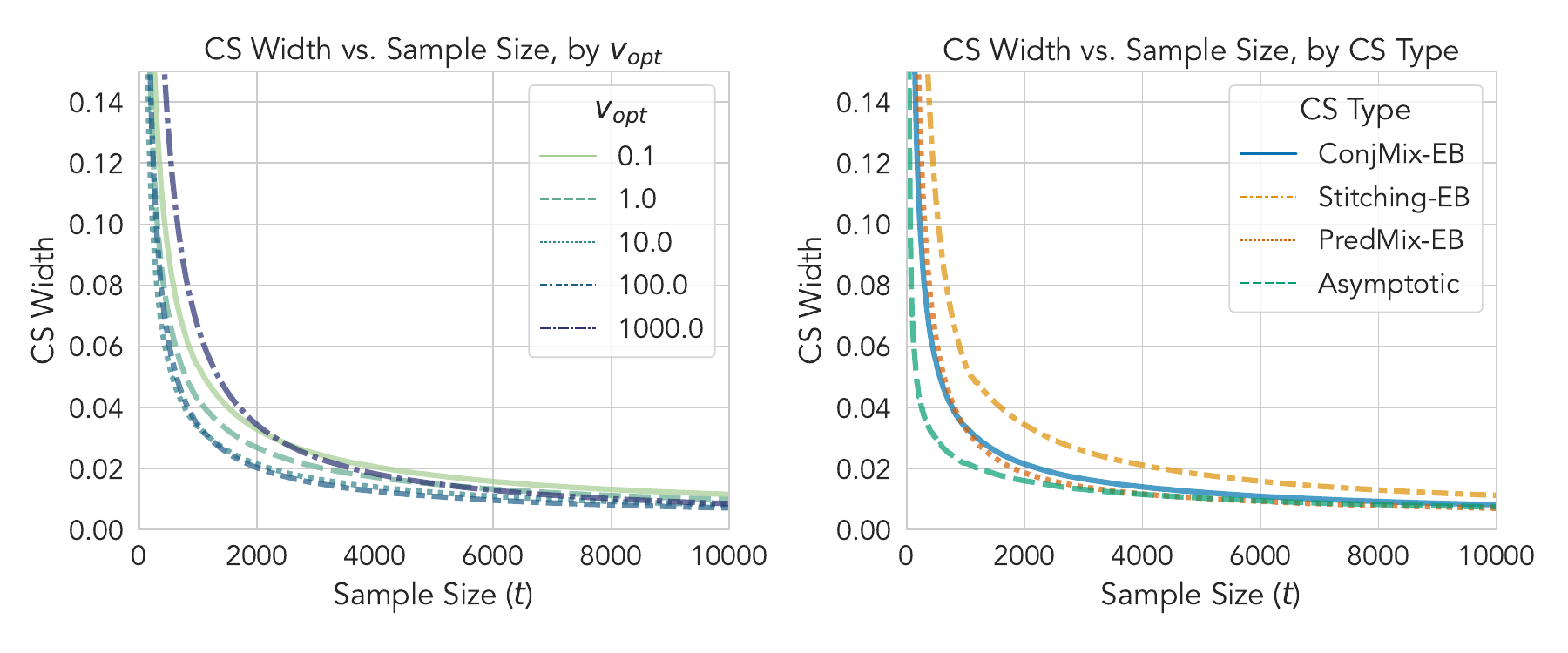}
    \caption{\emph{Left:} Widths of conjugate-mixture EB CSs per sample sizes ($t$), across different values of the hyperparameter $v_\text{opt}$ (optimal intrinsic time). 
    The choices $v_\text{opt} = 10$ and $v_\text{opt} = 100$ give the smallest widths overall, with the former being tighter early on and the latter later on.
    \emph{Right:} Widths of EB CSs using different uniform boundaries, including the conjugate-mixture (``ConjMix'') and predictable-mixture (``PredMix'') boundaries, and also the asymptotic CS.
    Overall, the asymptotic CS is the tightest, although the mixture EB CSs achieve similar widths for large sample sizes. The stitching EB CS is considerably wider than the mixture variants.
    }
    \label{fig:iid_mean}
\end{figure}

Given the data, we now compare different configurations of the EB CS (Theorem~\ref{thm:main}) for the mean score differential.
Using the EB CS with the conjugate-mixture uniform boundary (Section~\ref{sec:uniform_boundary}), we first show how we choose a default value for $v_\text{opt}$, the hyperparameter for the uniform boundary that specifies the intrinsic time at which the CS width is optimized (defined in Section~\ref{sec:boundary_details}).
Recall that, in our previous plot, we showed the values of intrinsic times across sample sizes for this data.
In Figure~\ref{fig:iid_mean} (left), we plot the widths of the 95\% EB CS against different choices of $v_\text{opt}$.
Comparing the values of $v_\text{opt} \in \{0.1, 1, 10, 100, 1000\}$, we find that the EB CS is generally the tightest across time for $v_\text{opt} = 10$ or $v_\text{opt} = 100$.
Based on the result, we use a default value of $v_\mathrm{opt} = 10$ for all our experiments involving the EB CS in the paper, unless specified otherwise. 

We now compare EB CSs constructed using different types of uniform boundaries, including the conjugate-mixture (``ConjMix'') boundary and the polynomial stitching boundary (Section~\ref{sec:stitching}).
In this comparison, we additionally include EB CSs constructed using the predictable-mixture (``PredMix'') boundary~\citep{waudbysmith2020estimating}, which is an efficient alternative that works specifically for bounded IID means.
Finally, we include the asymptotic CSs that we described in Section~\ref{sec:asympcs} as a reference.

In Figure~\ref{fig:iid_mean} (right), we plot the widths of all CS variants at the coverage level of 95\%, optimized for the intrinsic time $v_\mathrm{opt} = 10$ when applicable.
Generally speaking, we observe that the asymptotic CS achieves the tightest width, although the (non-asymptotic) EB CS variants using mixture boundaries approach that width for large sample sizes.
This is consistent with our intuition, as the asymptotic CS is the large-sample ``limit'' of the EB CS in terms of width~\citep{waudbysmith2021doubly}.
Among the EB CS variants, the conjugate-mixture variant is tighter towards the beginning ($t < 10^3$) while the predictable-mixture becomes slightly tighter afterwards; the stitching CS is not as tight as the other two. 
This is also as expected, as both mixture CSs are known to have similar widths (up to differences determined by hyperparameters)~\citep{waudbysmith2020estimating}, while the stitching CS tends to be looser in practice~\citep{howard2021timeuniform}. 
We close with the note that any of these (EB or asymptotic) CSs are substantially tighter than Hoeffding-style CSs (Theorem~\ref{thm:hoeffding}) in most cases, regardless of the uniform boundary choice.
This is evident from our earlier experiments in Section~\ref{sec:simulated}.

\end{document}